\documentclass[12pt]{article}
\usepackage{latexsym}
\usepackage{amsmath}
\usepackage{amssymb}
\usepackage{epsfig,graphics}
\usepackage{graphicx}
\usepackage{booktabs}
\usepackage{multirow}
\usepackage{caption}
\usepackage{subcaption}

\newcommand{\be}{\begin{equation}}
\newcommand{\ee}{\end{equation}}
\newcommand{\beq}{\begin{eqnarray}}
\newcommand{\eeq}{\end{eqnarray}}

\begin{document}

\begin{titlepage}


\begin{center}
  {\large\bf SU(N) gauge theories in 3+1 dimensions: \\
  glueball spectrum, string tensions and topology} \\
  \vspace*{.35in}
{Andreas Athenodorou$^{a,b}$ and Michael Teper$^{c,d}$\\
  \vspace*{.35in}
  $^{a}$ Dipartimento di Fisica, Universit\`a di Pisa and INFN, Sezione di Pisa, \\
  Largo Pontecorvo 3, 56127 Pisa, Italy. \\
\vspace*{.10in}
$^{b}$ Computation-based Science and Technology Research Center, The Cyprus Institute,\\
20 Kavafi Str., Nicosia 2121, Cyprus \\
\vspace*{.10in}
$^{c}$Rudolf Peierls Centre for Theoretical Physics, University of Oxford,\\
Parks Road, Oxford OX1 3PU, UK\\
\vspace*{.10in}
$^{d}$All Souls College, University of Oxford,\\
High Street, Oxford OX1 4AL, UK}
\end{center}

\vspace*{0.2in}

\begin{center}
{\bf Abstract}
\end{center}

We calculate the low-lying glueball spectrum, several string tensions and some properties of
topology and the running coupling for $SU(N)$ lattice gauge theories in $3+1$ dimensions. We
do so for $2 \leq N \leq 12$, using lattice simulations with the Wilson plaquette action,
and for glueball states in all the representations of the cubic rotation group, for
both values of parity and charge conjugation. We extrapolate these results to
the continuum limit of each theory and then to $N=\infty$. For a number of these states
we are able to identify their continuum spins with very little ambiguity.
We calculate the fundamental string tension and $k=2$ string tension and investigate
the $N$ dependence of the ratio. Using the string tension as the scale, we calculate
the running of a lattice coupling and confirm that $g^2(a)\propto 1/N$  for
constant physics as $N\to\infty$. We fit our calculated values of $a\surd\sigma$
with the 3-loop $\beta$-function, and extract a value for $\Lambda_{\overline{MS}}$,
in units of the string tension, for all our values of $N$, including $SU(3)$. We use these
fits to provide analytic formulae for estimating the string tension at a given lattice coupling.
We calculate the topological charge $Q$ for $N\leq 6$ where it fluctuates
sufficiently for a plausible estimate of the continuum topological susceptibility.
We also calculate the renormalisation of the lattice topological charge, $Z_Q(\beta)$, for
all our $SU(N)$ gauge theories, using a standard definition of the charge, and we provide
interpolating formulae, which may be useful in estimating the renormalisation of
the lattice $\theta$ parameter. We provide quantitative results for how
the topological charge `freezes' with decreasing lattice spacing and with increasing $N$.
Although we are able to show that within our typical errors our glueball and string tension
results are insensitive to the freezing of $Q$ at larger $N$ and $\beta$, we choose to
perform our calculations with a typical distribution of $Q$ imposed upon the fields
so as to further reduce any potential systematic errors.

\vspace*{0.15in}

\leftline{{\it E-mail:} a.athenodorou@cyi.ac.cy, mike.teper@physics.ox.ac.uk}

\end{titlepage}

\setcounter{page}{1}
\newpage
\pagestyle{plain}

\tableofcontents

\section{Introduction}
\label{section_intro}

In this paper we calculate various physical properties of $SU(N)$ 
gauge theories in $3+1$ dimensions. We do so by performing calculations 
in the corresponding lattice gauge theories over a sufficient range of lattice
spacings, $a$, and with enough precision that we can obtain plausible
continuum extrapolations. We also wish to be able to extrapolate to the
theoretically interesting $N=\infty$ limit and to compare this to the
physically interesting $SU(3)$ theory. To do so we have performed
our calculations for $N=2,3,4,5,6,8,10,12$ gauge theories. 

Our main aim is to provide a calculation of the low-lying `glueball' mass spectrum
for all quantum numbers and all $N$. This means calculating the lowest states in all 
the irreducible representions, $R$, of the rotation group of a cubic lattice, and for
both values of parity $P$ and charge conjugation parity $C$. We also calculate the
confining string tension, $\sigma$, so as to provide a useful scale for the glueball
masses. In addition we calculate a number of other interesting quantities.
However these are side-products of our glueball calculations rather than being
dedicated calculations of these quantities. The first is the $k=2$ string tension, 
$\sigma_{k=2}$ and the nature of its approach to $N=\infty$. The second is an estimate 
of the scale parameter $\Lambda_{\overline{MS}}$ for all our $SU(N)$ gauge theories. 
Thirdly, in order to monitor the topological `freezing' as $a$ becomes small
at fixed $N$, or $N$ becomes large at fixed $a$, we calculate the interval (in units
of our Monte Carlo steps) between changes in the topological charge, as well
as the renormalisation of that charge, and also the topological susceptibility.

Our calculations of the glueball spectrum are intended to make necessary improvements to
previous work. We recall that the pioneering glueball calculations in
\cite{BLMT_N,BLMTUW_N}
were restricted to obtaining the continuum masses of the lightest and first excited 
$J^{PC}=0^{++}$ glueballs and the lightest $2^{++}$ glueball, 
in units of the string tension, with the continuum spin assignments being based on plausible
assumptions. This sufficed to provide the first fully non-perturbative demonstration that 
the basic physical quantities of the  $SU(3)$ gauge theory are `close to' those of $SU(\infty)$, 
as had long been hoped, but it did not provide the kind of detailed $N=\infty$ glueball spectrum 
that would be useful in testing theoretical approaches. Such a detailed spectrum was
subsequently provided in 
\cite{BLARER_N}
where the lightest masses in all the  irreducible representations of the rotation group of 
the cubic lattice were calculated for $N\in [2,8]$ and extrapolated to $N=\infty$. In addition 
a serious effort was made in that work to identify those states that might be multi-glueball 
`scattering' states or finite volume artifacts (based on conjugate pairs of winding flux tubes), 
rather than being the single glueballs that one would obtain in an infinite volume.  
The important drawback of this calculation is that it was made at a fixed value of 
the lattice spacing $a$ (`fixed' in units of the deconfinement temperature), corresponding 
to a value of $\beta \simeq 5.895$ in $SU(3)$. This corresponds to a coarse lattice 
discretisation, typically between the two largest lattice spacings used in our present work.
This has two important adverse consequences. The first is that it leaves uncertain
the values of the masses in the desired continuum limit. (Although the calculations in
\cite{BLARER_N}
and earlier $SU(3)$ calculations left room for being optimistic about the limited
size of any lattice corrections.) The second problem is that since the lattice 
masses $aM$ are large when $a$ is large, as in that paper, it makes the extraction
of heavier glueball masses much more ambiguous and the statistical errors much larger
than they would be at smaller values of $a$. A corollary is that it renders
continuum spin assignments more ambiguous, except for the very lightest glueballs,
since the spin assignment depends on observing near-degeneracies amongst states 
in appropriate irreducible representations of the lattice rotation group, and for this 
it is clearly essential to achieve an adequate precision in the various mass estimates. 
In addition to this earlier work, there recently appeared a potentially more relevant paper 
\cite{SpN_Lucini} 
when our work was largely completed.
This paper provides a pioneering calculation of the masses of the 
lightest glueballs in all the irreducible representations, of a number of continuum 
$Sp(2N)$ gauge theories with an extrapolation to $N=\infty$. In principle this extrapolation 
is equally relevant to the $N\to\infty$ limit of $SU(N)$ gauge theories, as pointed out in
\cite{SpN_Lucini},
since the  $Sp(2N)$ and $SU(N)$ gauge theories share a common (perturbative) planar limit.
In practice, however, the usefulness for $SU(N\to\infty)$ turns out to be very limited. Firstly, 
$Sp(2N)$ has no $C=-$ sector, so there are no predictions for half the $SU(N\to\infty)$
spectrum. Secondly the limited range of $N$ in
\cite{SpN_Lucini},
coupled with the fact that in $Sp(2N)$ the leading
correction to $N=\infty$ is $O(1/N)$ renders the large-$N$ extrapolations less convincing 
and much less precise when compared to  $SU(N)$ where the leading correction is $O(1/N^2)$.
The net result is that the errors on the $SU(N\to\infty)$ glueball masses obtained in 
\cite{SpN_Lucini},
are larger by 
a factor of $\sim 5 - 20$ compared to the values obtained in this paper, and the mass
estimates are mainly for just the ground states in each channel. This means that
the evidence for the assignment of a continuum spin $J$ only appears plausible for
the $J^P=0^+,0^-,2^+,2^-$ ground states. Of course, none of these comments detract
from the success of
\cite{SpN_Lucini} 
in their primary aim of elucidating the mass spectrum of $Sp(2N)$  gauge theories.

In this paper we provide a calculation of the masses of the ground states and some excited
states in all the irreducible representations $R$ of the rotation group of our cubic lattice 
with an extrapolation of these masses (in units of the string tension) to the continuum limit.
To make this extrapolation  more reliable and more precise we extend the range of
our calculations to much smaller lattice spacings than earlier work. This enables
us to extract the masses of the heavier ground states and most of the first few
excited states much more reliably than earlier calculations.
As an important by-product, all this will allow us to make a significant number 
of continuum spin assignments after the extrapolation to the continuum limit. 
We recall that the assignment of a continuum spin typically depends on observing 
near-degeneracies amongst both ground and excited states in the various irreducible 
representations of the lattice rotation group, and this requires both precision 
in the mass estimates and small enough lattice spacings for the masses of the
heavier excited states to be plausibly estimated. We also extend the range 
of our calculations to larger values of $N$. This is not only to make the 
extrapolation to $N=\infty$ somewhat more reliable and precise but also, and perhaps 
more importantly, to help in excluding from our $N\to\infty$ glueball spectrum any 
multiglueball scattering states and  finite volume states, since these states 
will decouple from our single trace operators as $N\uparrow$.

This kind of calculation is of course standard in the case of $SU(3)$, see for example
\cite{CMMT-1989,CM-UKQCD-1993,MP-1999,HM_Thesis,HMMT-2004,MP-2005,AAMT-2020},
and what we have attempted to do is to bring the $SU(N)$ calculations towards a similar level
of sophistication. The computational cost of performing calculations at larger
$N$ means that further improvement to our work is still desirable. 
(It would also be very interesting to see the existing calculations at very large $N$ using 
space-time reduction
\cite{MGPAGAMO-2020}
extended to calculations of the glueball spectrum, just as they have been in
$2+1$ dimensions
\cite{MGPAGAMKMO-2018}.)
Nonetheless we are able in this paper to provide the first calculation of the masses 
of the ground states in all the $R^{PC}$ channels of the continuum $SU(N\to\infty)$ gauge 
theory,  as well as some excited states in most channels.

The plan of the paper is as follows. In the next section we introduce our lattice
setup and describe how we calculate energies from correlators. We discuss some of
the main systematic errors affecting these calculations and how we deal with them,
with a particular focus on the rapid loss of tunneling between sectors
of differing topology as $N\uparrow$. In Section~\ref{section_strings} we describe in 
detail our calculation
of the confining string tension, $\sigma$, which we will later use as the physical scale in
which to express our glueball masses. As a side product we also calculate the string tension, 
$\sigma_{k=2}$, of the lightest flux tubes carrying $k=2$ units of fundamental flux.
In Section~\ref{subsection_kstringlargeN} we study the approach
of the ratio $\sigma_{k=2}/\sigma$ to the $N=\infty$ limit so as to address the
old controversy concerning the power of  $1/N$ of the leading correction. In
Section~\ref{section_coupling} we show how our precise calculation of the fundamental 
string tension, $a^2\sigma$, as a function of the lattice coupling 
enables us to confirm the expected scaling of $g^2$ with $N$, and motivates 
(lattice improved) perturbative fits that allow us, in Section~\ref{subsection_Lambda}, 
to estimate a value for $\Lambda_{\overline{MS}}$ as a function of $N$. We also provide, 
in Section~\ref{subsection_interpol}, some analytic interpolation/extrapolation formulae 
for the variation of the string tension
with the coupling, which may be of use in other calculations.
We then turn, in Section~\ref{section_glueballs}, to our main
calculation in this paper, which is that of the low-lying glueball spectrum.
We calculate the masses on the lattice, extrapolate to the continuum limit, and
then extrapolate to $N=\infty$. Although these states are classified according to
the representations of the rotation group of our cubic lattice, we are able to
identify the continuum spins in many cases, as described in Section~\ref{subsection_spins}.
In doing all this we need to address the
problem of the extra states that winding modes introduce into the glueball spectrum and
also the possible presence of multi-glueball 'scattering' states.
We complete Section~\ref{section_glueballs} with a brief comparison of our results with
those of some earlier calculations. We then return,
in Section~\ref{section_topology}, to some of the properties of the topological
fluctuations in our lattice fields. After illustrating in Section~\ref{subsection_Qcooling}
how our cooling algorithm reproduces the topological charge of a lattice field, we
calculate in Section~\ref{subsection_Qtunneling} the rate of topological freezing with increasing
$N$ and with decreasing $a(\beta)$ and compare our results to the simplest theoretical expectations.
In Section~\ref{subsection_Qsusc} we provide our results for the topological
susceptibility in the continuum limit of $SU(N\leq 6)$ gauge theories and in the $N\to\infty$
limit. Then, in Section~\ref{subsection_QZ}, we calculate the multiplicative renormalisation of
our lattice topological charge for each $SU(N)$ group and provide interpolating formulae
which may be useful in calculations with a $\theta$ parameter in the lattice action.
Section~\ref{section_conclusion} summarises our main results.

Finally we remark that in parallel with the present calculations most of our $SU(3)$
calculations, which are of particular physical interest, have recently been published
separately 
\cite{AAMT-2020}.
\section{Calculating on a lattice}
\label{section_lattice} 

\subsection{lattice setup}
\label{subsection_lattice_setup}

We work on hypercubic lattices of size $L_s^3L_t$ with lattice spacing $a$ and with
periodic boundary conditions on the fields. The Euclidean time extent, $aL_t$, is
always chosen large enough that we are in the confining phase of the theory, at a
temperature that is well below the deconfining phase transition
\cite{BLMTUW_Tc}.
Our fields are $SU(N)$ matrices, $U_l$,
assigned to the links $l$ of the lattice. The Euclidean path integral is 
\begin{equation}
Z=\int {\cal{D}}U \exp\{- \beta S[U]\},
\label{eqn_Z}
\end{equation}
where ${\cal{D}}U$ is the Haar measure and we use the standard plaquette action,
\begin{equation}
\beta S = \beta \sum_p \left\{1-\frac{1}{N} {\text{ReTr}} U_p\right\}  
\quad ; \quad \beta=\frac{2N}{g^2_L}.
\label{eqn_S}
\end{equation}
Here $U_p$ is the ordered product of link matrices around the plaquette $p$. We write
$\beta=2N/g^2_L$ since in this way we recover the usual continuum action when we
take the continuum limit of the lattice theory and replace $g^2_L\to g^2$.
The subscript $L$ reminds us that this coupling is defined in a specific coupling
scheme  corresponding to the lattice and the plaquette action. Since $g^2_L$ is the
bare coupling corresponding to a short distance cut-off $a$, it provides a definition
of the running coupling $g^2_L(a)=g^2_L$ on the length scale $a$. Since the theory is
asymptotically free $g^2_L(a)\to 0$ as $a\to 0$ and hence $\beta\to \infty$ as $a\to 0$
and so we can decrease the cut-off
$a$ and so approach the desired continuum limit of the theory by increasing $\beta$. Our
calculations of this lattice path integral are carried out via a standard Monte Carlo using a
mixture of Cabibbo-Marinari heat bath and over-relaxation sweeps through the lattice.
We typically perform $\sim 2\times 10^6$ sweeps at each value of $\beta$ at each lattice size,
and we typically calculate correlators every $\sim 25$ sweeps and the topological charge
every $50$ or $100$ sweeps. Naturally we choose values of $\beta$ that place us on the
weak coupling branch of the lattice theory. We discuss the `bulk' transition that separates weak
coupling from strong coupling, and becomes a first order transition for $N\geq 5$
\cite{BLMTUW05},
in Section~\ref{subsection_bulk}. There are of course other possible choices for the action,
and we briefly comment upon our choice in Section~\ref{subsection_comparisons}, where
we compare our results to those of other recent calculations.

We simulate $SU(N)$ lattice gauge theories for $N=2,3,4,5,6,8,10,12$ over a range of values
of $\beta$ so as to be able to plausibly obtain, by extrapolation, the glueball spectra and
string tensions of the corresponding continuum gauge theories. A summary of the basic
parameters of our calculations is given in Tables~\ref{table_param_SU2}-\ref{table_param_SU12}.
For each of our $SU(N)$ calculations we show the values of $\beta$, the lattice sizes, the
average plaquette, the string tension, $a^2\sigma$, and the mass gap, $am_G$. In the header of
each table we show the approximate spatial size in units of the string tension. Since
finite volume corrections are expected to decrease with increasing $N$ we decrease the
lattice volume as we increase $N$. This expectation needs to be confirmed by explicit
calculations which we shall provide in Section~\ref{subsection_massV}.

\subsection{energies and correlators}
\label{subsection_lattice_energies}

We calculate energies from correlation functions in the standard way. Suppose we wish to
calculate glueball masses in some representation $R^{PC}$. If one picks an operator $\phi(t)$
with quantum numbers $R^{PC}$ and momentum $p=0$ then the correlator will provide us with
the energies $E_i$ of states with those quantum numbers 
\begin{equation}
  C(t) = \langle \phi^{\dagger}(t)\phi(0)\rangle
  = \sum_{n=0} |c_n|^2 \exp\{-E_nt\} \stackrel{t\to\infty}{\longrightarrow}
  |c_0|^2 \exp\{-E_0t\},
\label{eqn_C}
\end{equation}
where $E_n\leq E_{n+1}$ and $E_0$ is the lightest state with $R^{PC}$ quantum numbers --
which will often be the lightest glueball in that sector. (If $\phi$ has vacuum quantum
numbers then one uses the vacuum subtracted operator.)
Since on the lattice time is measured in lattice units, $t=an_t$, 
what we obtain is the value, $aE_0$, of the mass in lattice units. If we calculate two masses,
$aM$ and $a\mu$, in this way then the lattice spacing drops out of the ratio and if we
calculate the ratio for several values of $a(\beta)$ we can extrapolate to the continuum
limit in the standard way, using
\begin{equation}
 \frac{aM(a)}{a\mu(a)} = \frac{M(a)}{\mu(a)} \simeq \frac{M(0)}{\mu(0)} + ca^2\mu^2 
\label{eqn_cont}
\end{equation}
once  $a(\beta)$ is small enough. (This standard tree level extrapolation could be improved with
perturbative corrections
\cite{Sommer-cutoff}
and with higher order power corrections but our calculations are not so extensive and precise as to
motivate such modifications.)

The starting point for an operator $\phi$ is the trace of a closed loop on the lattice.
For an operator that projects onto glueballs the loop should be contractible.
A non-contractible loop that closes across the periodic spatial boundary
will project onto a confining flux tube that winds around that spatial torus.
For the contractible glueball loop one takes a suitable linear combination of rotations
of that loop for it to be in the desired representation $R$ of the lattice rotation
group and together with the parity inverse this allows us to form an operator
with parity $P$. The real and imaginary parts of the original loop will correspond to
$C=+$ and $C=-$ respectively. Summing this linear combination over all spatial sites
at time $t$ gives us the operator with $p=0$.

The statistical errors on $C(t)$ are roughly independent of $t$ while its value decreases
exponentially with $t$, so if we are to estimate $aE_0$ from $C(t)$ in eqn(\ref{eqn_C})
we need to be able to do this at small $t$. (The fluctuations of $\phi^{\dagger}(t)\phi(0)$
involve the higher order correlator $(\phi^{\dagger}(t)\phi(0))^2$ which typically has a vacuum
channel and this is independent of $t$.) For this to be possible we need the operator
$\phi$ to have a large overlap onto the state $|n=0\rangle$ that corresponds to $E_0$,
i.e. that $|c_0|^2/\sum_n |c_n|^2 \sim 1$. One can achieve this using iteratively `blocked'
link matrices and loops
\cite{MT-block1,MT-block2}
as described in detail in, for example,
\cite{BLMTUW_N}.
To monitor the approach of  $C(t)$ to the asymptotic exponential decay in eqn(\ref{eqn_C})
it is useful to define an effective energy
\begin{equation}
  \frac{C(an_t)}{C(a(n_t-1))} = \exp\{-aE_{eff}(n_t)\}.
\label{eqn_Eeff}
\end{equation}
(In practice we use a cosh modification of this definition to take into account the
periodicity in the temporal direction.)
If, within errors,  $aE_{eff}(n_t) = \mathrm{const}$ for $n_t\geq \tilde{n}_t$ then
we can use $aE_{eff}(\tilde{n}_t)$ as an estimate for $aE_0$, or we can do a simple
exponential fit to $C(t)$ for $n_t\geq \tilde{n}_t-1$ to estimate $aE_0$.

To calculate not just the ground state energy but also some usefully precise excited state
energies from $C(t)$ in eqn(\ref{eqn_C}) would require a precision that is at present
unachievable. Instead the standard strategy is to use a variational calculation. One chooses
some basis of $n_0$ operators $\{\phi_i(t): i=1,..,n_0\}$, calculates the cross-correlators
$C_{ij}(t)= \langle \phi^{\dagger}_i(t)\phi_j(0)\rangle$ and finds the linear combination,
$\Phi = \Phi_0$, that maximises $C(t)= \langle \Phi^{\dagger}(t)\Phi(0)\rangle$
for some suitable small $t=t_0$. This is then our best estimate of the wave-functional
of the ground state $|n=0\rangle$. We then extract $aE_0$ from the asymptotic
exponential decay of $C(t)= \langle \Phi^{\dagger}_0(t)\Phi_0(0)\rangle$. To calculate
the first excited energy $aE_1$ we consider the subspace of $\{\phi_i: i=1,..,n_0\}$
that is orthogonal to $\Phi_0$ and repeat the above steps. This gives our best
estimate $\Phi_1$ for  $|n=1\rangle$ and we estimate $aE_1$ from the asymptotic
exponential decay of the correlator of $\Phi_1(t)$. Similarly by considering
the subspace of the operator vector space that is orthogonal to $\Phi_0$ and $\Phi_1$ 
we can obtain estimates for the next excited state and so on. As with any variational
calculation, the accuracy of such estimates will depend on having a large enough
basis of operators.

\subsection{systematic errors}
\label{subsection_lattice_systematics}

In addition to the statistical errors in the calculation of energies and masses, which
can be estimated quite reliably, there are systematic errors that are harder to control.
We will now briefly point to some of these, leaving a more detailed discussion till later on.

Since the error to signal ratio on $C(n_t)$ increases at least as fast as
$\propto \exp\{+aE_0n_t\}$, the range of $n_t$ which is useful rapidly decreases
when the energy $aE_0$ increases, and an obvious systematic error is that we begin our
fit of the asymptotic exponential decay at too small a value of $t=an_t$.
Given the positivity properties of our correlators, this means that as the true
value of $E_0$ increases, so does our overestimate of its value. For any particular
state this problem will become less severe as we decrease $a$, and this provides
some check on this error.

The correlator of an excited state, using our best variational choice of operator,
may contain small contributions from lighter excitations since our basis is far
from complete. So at large enough $n_t$ these will dominate the correlator and
the effective mass will drop below that of the mass of the state of interest.
This does not appear to be a severe problem for us because the pattern of our statistical
errors, which grow rapidly with $n_t$, prevents us from obtaining values at large $n_t$.
In any case we attempt to identify at most an intermediate effective mass plateau
from which we extract the desired mass.

If a state has too small an overlap onto our basis of operators then, given the
exponential growth with $t$ of the statistical error on the effective energy, the
state will be, at best, assigned a mass that is much too high and so it will typically not
appear in its correct order in the mass spectrum. If so this means that there will be
a missing state in our calculated glueball spectrum and this will not be helpful if,
for example, one wishes to check theoretical models of the spectrum against the lattice
spectrum. To reduce the possibility of this occurring one needs to use a large basis of
operators, just as in any variational calculation.

The states we are primarily interested in are single particle glueball states rather
than multi-glueball scattering states. (In our finite spatial volume multiglueball states
are interacting at all times but for simplicity we refer to them as scattering states.)
We use single trace operators which should ensure that at large $N$ our correlators
receive no multi-glueball contributions but this is not necessarily the case
at smaller $N$. This is of course related to the fact that at smaller $N$ 
heavier glueball states may have substantial decay widths. Later in the paper we will
provide an exploratory analysis of this issue by combining single and double trace
operators.

The spectrum in a finite spatial volume will differ from the spectrum in an infinite
volume. One type of correction to a glueball propagator comes, for example, from a virtual
glueball loop where one line of the two in the loop encircles the spatial torus. Such
corrections are exponentially small in, typically, $M_GL$ where $M_G$ is the mass gap
(the lightest scalar glueball) and $L$ is the spatial size
\cite{Luscher-V}.
These corrections will be very small in our calculations. For the intermediate volumes
on which we work a much more relevant finite volume correction comes from a confining
flux tube that winds once around one of the spatial tori. Such a flux tube is often
referred to as a torelon. By itself a torelon has zero overlap onto local particle states.
However a state composed of two torelons, where one is conjugated, will in general
have a nonzero overlap. Such a `ditorelon' will not only shift the energies of
the usual particle glueballs but will introduce extra states into the glueball spectrum.
These extra states can be identified through their volume dependence and in some
other ways as we shall see later on.

In addition to the usual gradual loss of ergodicity as the lattice spacing decreases,
the tunnelling between sectors of differing topological charge $Q$ suffers a much more
rapid supression not only as $a\to 0$ at fixed $N$ but also as $N$ increases at fixed $a$.
This affects a significant part of our calculations and we discuss next how we deal with it
in practice.

\subsection{topological freezing}
\label{subsection_Qfreezing}

We begin by briefly summarising the reason that one loses ergodicity in the
topology of lattice gauge fields much faster than in other typical quantities 
as $a\to 0$ at fixed $N$ and as $N\to\infty$ at fixed $a$, when using a local 
Monte Carlo algorithm,. We then describe where this impacts on our particular
calculations and why this inpact should be minor on theoretical grounds. 
We provide explicit calculations to confirm this expectation and, finally, 
we describe the additional procedure that we follow to minimise any remaining
systematic errors. It is important to note that all these arguments are for pure 
gauge theories and would have to be reconsidered in the presence of light fermions
as in $QCD$.

In a periodic four-volume a continuum gauge field of topological charge $Q$ cannot be smoothly
deformed into a gauge field of charge $Q^\prime\neq Q$. So in a sequence of lattice fields
that have been generated using a Monte Carlo algorithm that is local any change in
the value of $Q$ should be increasingly suppressed as we approach the continuum limit.
This statement can be made quantitative as we do below. But before that we note that
there is a separate suppression that occurs at any fixed small value of $a$
when we increase $N$. To smoothly deform a lattice field from, say, $Q=1$ to $Q=0$
in a finite volume with periodic gauge potentials, the core of an original
extended instanton must gradually shrink until it disappears within a hypercube,
leaving a simple gauge singularity centered on that hypercube.
Before it does so it will be a small instanton (assuming $a$ is small) which
we can describe using standard semiclassical methods. So its effective action 
will be $S_I(\rho) \sim 8\pi^2/g^2(\rho)$ where $\rho$ is the size of the core
and provides the relevant scale, and the probability of it existing
in a field will be $\propto \exp\{-S_I(\rho)\} \sim \exp\{- 8\pi^2/g^2(\rho)\}$
per unit volume up to a $1/\rho^4$ volume factor and some less important factors
\cite{tHooft_Q,Coleman_Q}.
Now at large $N$ the 't Hooft coupling $\lambda = g^2N$ needs to be kept constant
for a smooth limit
\cite{tHooft_N,Coleman_N}.
In terms of the running coupling this means $\lambda(l) = g^2(l)N$
is independent of $N$ up to corrections
that are powers of $1/N$, where $l$ is measured in some physical units, e.g.
the mass gap. So the weight of a small instanton vanishes as
$\propto \exp\{- 8\pi^2N/\lambda(\rho)\}$, i.e. it vanishes exponentially
with $N$
\cite{Witten_Q,Teper_Q}.
Since the process of going `smoothly' from a field with $Q=1$ to a field with
$Q=0$ necessarily involves at an intermediate stage of the process the presence of
fields containing such small instantons, we immediately infer that the change in
topology must be at least exponentially suppressed with increasing $N$ once $a$ is small.
Such a strong suppression has indeed been observed in earlier lattice calculations
\cite{BLMT_N,LDD_K2}.
Returning to the $a\to 0$ limit at fixed $N$ we note that one can use the one-loop expression
for $g^2(\rho)$ and this leads to a suppression of the tunnelling between
topological sectors that is the appropriate power of $\rho$ and hence of $a$, once
$\rho \sim O(a)$.

In our calculations we normally start a Monte Carlo sequence with the trivial $Q=0$
gauge field $U_l=I, \forall l$, and `thermalise' with some tens of thousands of sweeps
to reach the `equilibriated' lattice gauge fields which we can begin to use for our
calculations. As described above, for larger $N$ and smaller $a$ the topological 
tunnelling can become
sufficiently rare that the fields remain at $Q=0$ even after our attempted
equilibriation. In practice this proves not to be an issue for any of our $SU(2)$, $SU(3)$,
$SU(4)$ and $SU(5)$ calculations. For $SU(6)$ it becomes a problem for the smallest values
of $a$ and for $SU(N\geq 8)$ it is a problem for all but our largest values of $a$.
That is to say, within our statistics and our range of $a$ topological freezing
is largely an issue to do with $N\uparrow$ rather than with $a\downarrow$.
This distinction is important since the currently accepted theoretical picture
\cite{Witten_98}
tells us that we are sitting in a vacuum with periodicity in the $\theta$ parameter of
$2\pi N$ rather than $2\pi$. (It is other interlaced $\theta$-vacua that provide the
expected $2\pi$ periodicity.) This implies that higher moments in $Q$ disappear as inverse
powers of $N$, and so does the correlation between the topological charge and gluonic 
operators. Lattice calculations 
\cite{LDDGMEV_Q}
have confirmed this expectation. This tells us that at large enough $N$ the loss of
ergodicity should not affect physical quantities such as glueball masses. 

Of course we do not know in advance what value of $N$ is `large enough' and so it
is useful to perform
an explicit check of how such freezing affects the physics we are interested in.
For this we choose to work in $SU(8)$ at $\beta=45.50$ where changes in topology
are extremely rare, but which is close to the value $N=6$ where we begin to encounter
a serious loss of topological ergodicity in our calculations.
We begin with a comparison of two sets of about $2\times 10^6$ fields
on a $14^320$ lattice, with glueball and flux tube `measurements' taken every
25'th update. One set of fields has $Q=0$ throughout while the other has what one expects
to be the correct distribution of $Q$ imposed through about 40 different starting
configurations as described above. We calculate glueball and flux tube correlators
in exactly the same way in both ensembles of fields and perform the usual variational
calculation for each representation $R^{PC}$. Note that here we include all the states
that we obtain and make no attempt to identify any of the finite volume ditorelon states
alluded to in Section~\ref{subsection_lattice_systematics}.
In most, but not all, cases the effective
energy plateau in the effective energies, $aE_{eff}(t)$, defined in eqn(\ref{eqn_Eeff})
begin at $t=2a$. So we show in Table~\ref{table_GKvsQ_SU8_l14} the values of 
$aE_{eff}(t=2a)$ that we obtain in each of the two sets of fields, for the ground
states and some of the lowest lying excited states. We see that the two sets of values
are remarkably consistent within errors, with only a small handful of heavier states
differing by slightly more than $2\sigma$. In addition to the glueball effective
energies we also show those for the lightest two states of a flux tube that  winds
around the spatial torus. We do so for the flux in the fundamental representation
($k=1$) and for the flux in the $(k=2)=(k=1)\otimes (k=1)$ representation.
Here too we see consistency within statistical errors. In addition to this study
on the $14^320$ lattice we perform a similar one on a smaller $12^320$ lattices at
the same $\beta$. This is the lattice size, in physical units, that is used in
our calculations later on in this paper, and we recall that in general the local
effects of constraining the topological charge should increase with decreasing volume
(see e.g.
\cite{Aoki}).
%
Again we have consistency within errors for almost all the states.
The conclusion of these explicit calculations is that within our typical statistical
errors the physical quantities we calculate in this paper are not affected by
constraining the total topological charge to $Q=0$. 

Although both the theory and our above explicit checks suggest that the loss of
topological ergodicity in our calculations will have very little effect on
our measured glueball spectrum, we implement the following additional procedure that
is designed to reduce any remaining biases. Suppose we are confronted with 
topological freezing on a lattice of size $L_s^3L_t$ at some value of $\beta$ in some
$SU(N)$ theory. Now instead of starting with a $Q=0$ field with $U_l=I, \forall l$,
we produce a set of sequences in parallel which start with  fields that have various
values of $Q$. We choose the ensemble of these starting fields to roughly reproduce
the expected distribution. What is `expected' we infer from what we observe at the lower
values of $N$ where there is no topological freezing and also from the values obtained
at the same value of $N$ but at the larger values of $a(\beta)$ (if any) where the freezing
has not yet set in. This is a plausible strategy since we find that in those cases
where the distribution of $Q$ can be determined there is little variation
in the distribution of $Q$ with either $N$ or $a$ on an equal physical volume.
To produce such a starting field with some
$Q$ we generate a sequence of $SU(3)$ fields on the same size lattice as our $SU(N)$ one
and we pick out a field with the desired charge $Q$. We `cool' this field
to produce a field with minimal fluctuations apart from the net topological charge.
(See Section~\ref{section_topology} for details.)
We then embed this lattice field of $3\times 3$ complex
matrices into a suitable corner of our set of $N\times N$ unit matrices, so as to produce
a set of $SU(N)$ matrices with charge $Q$. If we have topological freezing then this
starting distribution of $Q$ will be maintained in our set of Monte Carlo sequences.
This method has some disadvantages but also some advantages. One advantage is that
the lack of topological tunneling is a better representation of the continuum
theory: we are sampling disconnected sectors. After all, the field configurations near the
point of topological tunnelling have no continuum correspondence and are a lattice
artifact. The major disadvantage is that we necessarily have a rather limited set of
parallel sequences (typically 20 or 40) and hence values of $Q$, and so only the most
probable values of $Q$ are properly represented. Lattice fields with values of $Q$ that have
a low probability simply do not appear. That is to say we are missing the
tail of the $Q$ distribution, and so any physics that depends on 
higher moments of $Q$ will be something we cannot address here.
However theory tells us that the correlation between such higher moments of $Q$
and our gluonic operators will be suppressed as a correpondingly higher power of $N$,
so any bias should be a small fraction of what, as we have seen above, will be at most
a very small shift to the glueball masses.

In conclusion, our explicit calculations, the theoretical arguments, and our
additional procedure of imposing upon our fields something quite close to the 
expected distribution of $Q$ makes us confident that within our typical statistical 
errors any systematic error from the freezing of topology is negligible in our 
glueball calculations.

\section{String tensions}
\label{section_strings}

Let us label the sites of the hypercubic lattice by the integers $n_x,n_y,n_z,n_t$.
To project onto a fundamental ($k=1$) flux tube the simplest operator we can use is
$\mathrm{Tr}(l_f)$ where $l_f$ is the product of link matrices along a minimal path
encircling  the spatial torus. This is just the usual spatial Polyakov loop. For example,
if the spatial lattice size is $L$ then the Polyakov loop in the $x$ direction is
\begin{equation}
 l_f(n_y,n_z,n_t) = \prod_{n_x=1}^{n_x=L} U_x(n_x,n_y,n_z,n_t).
  \label{eqn_Poly}
\end{equation}
We also use such Polyakov loops composed of blocked links and we translate the operator
along the $x$ direction.
Clearly the parities of such states are positive. This loop is translationally
invariant along the flux tube direction and so the momentum along the flux tube is
zero; $p_{\parallel}=0$. It is rotationally invariant around its axis, so $J=0$.
Because of the usual centre symmetry (see below) the $C=+$ and $C=-$ flux tubes are
degenerate. We then sum $l_f(n_y,n_z,n_t)$ over $n_y$ and $n_z$ so that the transverse
momentum $p_{\perp}=0$. Thus the lightest state onto which this final operator projects
should be the ground state of the winding flux tube that carries fundamental flux.
To project onto a $k=2$ flux tube we do the same using linear combinations of the
operators $\mathrm{Tr}(l_fl_f)$ and $\mathrm{Tr}(l_f)\mathrm{Tr}(l_f)$. For $k=3$ we use linear
combinations of $\mathrm{Tr}(l_fl_fl_f)$, $\mathrm{Tr}(l_fl_f)\mathrm{Tr}(l_f)$ and
$\mathrm{Tr}(l_f)\mathrm{Tr}(l_f)\mathrm{Tr}(l_f)$, and so on for higher $k$.
The defining property of a $k$-string that winds around the spatial torus in say the
$x$ direction is that under a transformation of the link matrices
$U_x(n_x,n_y,n_z,n_t) \to \exp\{i2\pi /N\} U_x(n_x,n_y,n_z,n_t)$ for all $n_y,n_z,n_t$
and at some fixed value of $n_x$, it will acquire a factor $\exp\{i2k\pi /N\}$.
Since this transformation is a symmetry of the theory (the action and measure are
unchanged) and at low temperatures the symmetry is not spontaneously broken,
any two operators for which this factor of $\exp\{i2k\pi /N\}$ differ will
have zero mutual overlap. So we have separate sectors of flux tubes labelled
by $k$ with the fundamental $k=1$ strings existing for $N\geq 2$,
the $k=2$ strings for $N\geq 4$, the $k=3$ strings for $N\geq 6$, and so on.
We note that our basis of operators for a $k$-string could easily be extended by
multiplying the operators by factors such as $\mathrm{Tr}(l_f)\mathrm{Tr}(l_f^\dagger)$,
but while this might be useful to improve the overlap onto massive excitations
it would be very surprising if it improved the overlap onto the ground state
which is what we are interested in here.

\subsection{finite volume corrections}
\label{subsection_KcorrnV}

As described above, we calculate the ground state energy of a flux tube that winds once around
a spatial torus. If the lattice has size $l^3l_t$, this flux tube has length $l$ and we generically
denote the energy by $E_k(l)$, where $k=1$ corresponds to a flux tube carrying fundamental flux
and $k$ to a $k$-string (carrying $k$ units of fundamental flux). The string tension $\sigma_k$
is the energy per unit length of a very long flux tube
\begin{equation}
\sigma_k = \lim_{l\to\infty}\frac{E_k(l)}{l}.
  \label{eqn_K}
\end{equation}
In practice our tori are finite and so we expect that there will be corrections to the
leading linear dependence of $E_k(l)$.
In our calculations we will normally estimate the asymptotic string tension
$\sigma_k$ using the `Nambu-Goto' formula
\begin{equation}
E_k(l) = \sigma_k l\left( 1 - \frac{2\pi}{3\sigma_k l^2}\right)^{\frac{1}{2}}.
  \label{eqn_NG}
\end{equation}
When expanded in powers of $1/\sigma_kl^2$ this formula generates all the known universal terms
\cite{string_1,string_2,string_3,string_4}
and, at least for fundamental flux, proves to accurately describe numerical calculations for the range
of $l$ relevant to our calculations
\cite{AABBMT_K}.
However, strictly speaking, these results hold for flux tubes that are effectively in an infinite
transverse volume, while in our case, in order to maintain the rotational symmetry needed for our glueball
calculations, the transverse size is $l\times l$ and so decreases when we decrease $l$.
We therefore need to perform some explicit checks, i.e. to check whether eqn(\ref{eqn_NG}) encodes all
the finite-$l$ corrections that are visible within our typical statistical errors.

We see from eqn(\ref{eqn_NG}) that the relevant scale is $l$ in units of the string tension,
i.e. $l\surd\sigma$. These scales are listed in the headers of
Tables~\ref{table_param_SU2}-\ref{table_param_SU12} and they vary from $l\surd\sigma\sim 4$
for lower $N$ to $l\surd\sigma\sim 2.6$ for our highest values of $N$. As far as glueball
masses are concerned the theoretical justification for this decrease in the volume
is that we expect finite volume corrections to decrease as $N$ grows and the practical
reason is that the expense of the $SU(N)$ matrix product calculations grows as $\propto N^3$. 
In order to check whether there are significant corrections to eqn(\ref{eqn_NG}) at the values
of $l$ we use, we calculate $\sigma$ on larger (sometimes smaller) lattices at selected values of
$\beta$. These calculations are summarised in Table~\ref{table_V_k1_SUN}. For $N=8,10,12$
the scale $l\surd\sigma\sim 2.6$ corresponds to $l=12$ in the Table, and we see that the
value of $a\surd\sigma$ is the same for the longer $l=14$ flux tube, suggesting that there are no
significant finite $l$ corrections to eqn(\ref{eqn_NG}) for these values of $N$. The same
conclusion holds for the other groups in Table~\ref{table_V_k1_SUN}. All this provides evidence
that the fundamental string tensions quoted in Tables~\ref{table_param_SU2}-\ref{table_param_SU12}
do not suffer significant finite volume errors, at least within our typical statistical errors.

The energies of $k\geq 2$ flux tubes are larger and therefore so are the statistical
and systematic errors in estimating them. Since the energies grow with $k$, we have
only attempted a finite volume study for $k=2$ flux tubes. This is presented in
Table~\ref{table_V_k2_SUN}. The message here is more nuanced than for $k=1$: while any finite $l$
corrections to eqn(\ref{eqn_NG}) appear to be insignificant when $l\surd\sigma \gtrsim 3$,
and hence for our $N \leq 6$ string tension calculations, there is strong evidence that
there is a correction of $\simeq 1.9(5)\%$ to the values obtained with $l\surd\sigma \sim 2.6$,
which are the values we emply for $N\geq 8$. This is something
we will need to consider when we estimate the errors on our $k=2$ string tensions. Presumably
the $k=3$ and $k=4$ string tensions show corrections at least as large, but we do not
have any finite $l$ study of these.

\subsection{fundamental ($k=1$) string tensions}
\label{subsection_k1strings}

\begin{figure}[htb]
\begin	{center}
\leavevmode
\input	{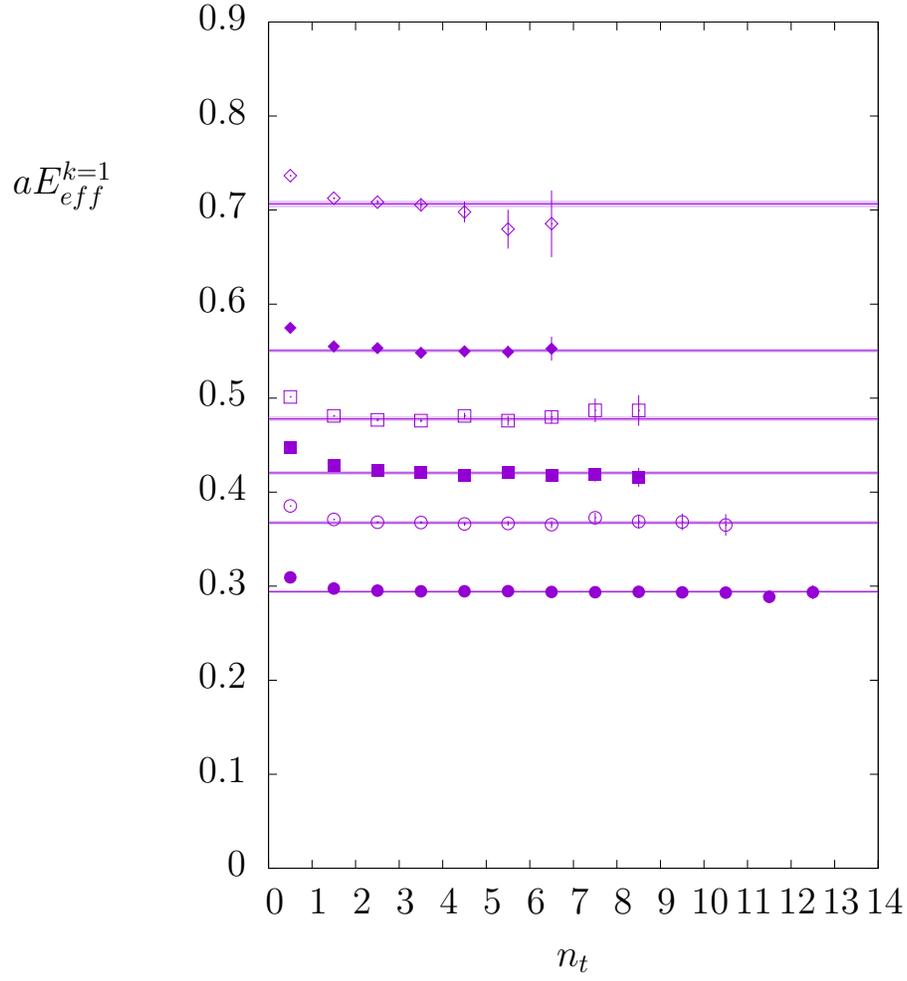}
\end	{center}
\caption{Effective energies of the ground state of a fundamental $k=1$ flux tube winding
  around a spatial torus, extracted from
  the best correlator $C(t)$ between $t=an_t$ and $t=a(n_t+1)$. For $SU(8)$ and at
  $\beta=44.10, 44.85, 45.50, 46.10, 46.70, 47.75$ in descending order. Lines are our estimates of
  the $t\to\infty$ asymptotic energies. The nearly invisible bands around those lines denote the errors
  on those estimates.}
\label{fig_EeffK1_SU8}
\end{figure}

\begin{figure}[htb]
\begin	{center}
\leavevmode
\input	{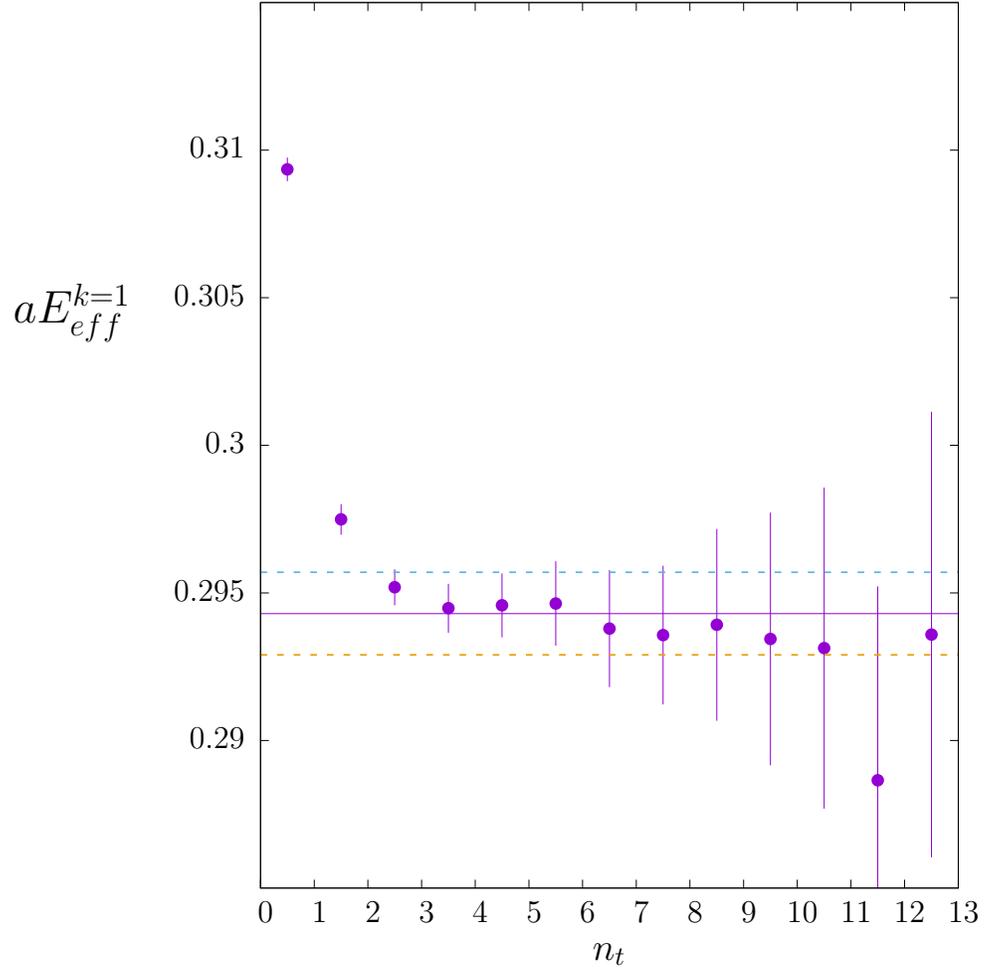}
\end	{center}
\caption{
  Effective energy of the ground state of a fundamental $k=1$ flux tube winding
  around a spatial torus, as in Fig.\ref{fig_EeffK1_SU8}, for $\beta=47.75$ in $SU(8)$,
  with a rescaling sufficient to expose the errors. The solid line is the best
  estimate from a fit to the correlation function, and the dashed lines show
  the $\pm 1$ standard deviation fits.}
\label{fig_EeffK1b_SU8}
\end{figure}

We calculate the ground state energies of the fundamental flux tube for the lattices, couplings
and gauge groups listed in Tables~\ref{table_param_SU2}-\ref{table_param_SU12}.
We then use eqn(\ref{eqn_NG}) to extract our estimates of the infinite volume string tensions.
The energies are obtained by identifying effective energy plateaux as described in
Section~\ref{subsection_lattice_energies}.
As an example we show in Fig.\ref{fig_EeffK1_SU8} the effective energies
that lead to the $SU(8)$ string tension estimates in Table~\ref{table_param_SU8}.
As one increases $\beta$ and hence decreases $a$, the energy in lattice units decreases
so that we can obtain precise effective energies over a larger range of $n_t$
which in turn makes it easier to identify a plateau and so estimate the corresponding
value of $E_{eff}(n_t\to\infty)$. Although we attempt to show in Fig.\ref{fig_EeffK1_SU8}
the error bands on our estimates of the corresponding flux tube energies, these are
almost invisible because the errors are too small on the scale used in the plot.
We therefore replot, in Fig.\ref{fig_EeffK1b_SU8}, the effective energies for $\beta=47.75$,
which is the calculation that is the closest to the desired continuum limit,
with a sufficient rescaling to expose
the errors on the effective masses. On the plot we show the best estimate for the
energy obtained from a fit to the correlation function, together with the
energies corresponding to $\pm 1$ standard deviations.

For the corresponding $SU(3)$ plots we refer the reader
to Fig.1 of
\cite{AAMT-2020}.
The fact that the $SU(3)$ lattices are (much) larger than the $SU(8)$ ones at similar
values of $a\surd\sigma$ means that the flux tubes are more massive, so the correlators
decrease faster with $n_t$ and the extraction of $E_{eff}(n_t\to\infty)$ is less
compelling -- despite the fact that the calculations extend closer to the continuum
limit in the case of $SU(3)$. These $SU(3)$ plots are typical of our  $N \leq 4$
calculations, where the lattices are chosen to be particularly large, while the $SU(8)$
plot is typical of  $N \geq 8$ where the spatial volumes used are smaller.

In the case of $SU(2)$ and $SU(3)$ we have also calculated the string tension on
smaller volumes which are still large enough for the flux tube calculations even if too
small for reliable glueball calculations.
This enables us to extend the calculations to smaller values of
$a\surd\sigma$ at modest computational cost. The purpose of these calculations
is to feed into our analysis of the running coupling later on in this paper. We list
the results of the calculations in Tables~\ref{table_Ksmall_SU2},\ref{table_Ksmall_SU3}.
%
%

%
%
\subsection{$k=2$ string tensions}
\label{subsection_k2strings}

One expects that as $N\to\infty$ the $k=2$ flux tube will become two non-interacting $k=1$
flux tubes and we will have $\sigma_{k=2}\to 2\sigma_{k=1}$. However, as
earlier calculations have shown
\cite{BLMT_K2,AABBMT_K,LDD_K2},
at lower values of $N$ one finds that $\sigma_{k=2}$ is substantially less than $2\sigma_{k=1}$,
so that one can think of it as behaving like a bound state of two fundamental flux tubes.
This suggests that at lower values of $N$ the $k=2$ flux tube can be treated as a single
`string' with finite volume corrections well described by eqn(\ref{eqn_NG}). As we increase
$N$ the $k=2$ flux tube should increasingly look like two fundamental flux tubes that
are loosely bound with the binding energy vanishing as $N\to\infty$. Here one might
expect each of these two flux tubes to have finite volume corrections given by
eqn(\ref{eqn_NG}) so that the overall finite volume behaviour of the energy of a
$k=2$ flux tube becomes significantly different:
\begin{equation}
  E_{k=2}(l) \stackrel{\mathrm{small}\, N}{=}
  \sigma_{k=2} l\left( 1 - \frac{2\pi}{3\sigma_{k=2} l^2}\right)^{\frac{1}{2}}
  \stackrel{N\to\infty}{\longrightarrow}
  \sigma_{k=2} l\left( 1 - \frac{4\pi}{3\sigma_{k=2} l^2}\right)^{\frac{1}{2}}
  \label{eqn_NGk2}
\end{equation}
with $\sigma_{k=2}=2\sigma_{k=1}$. One can expect a smooth transition between these
two behaviours which clearly creates ambiguities in extracting the string tension
$a^2\sigma_{k=2}$ from the flux tube energy $aE_{k=2}(l)$.

\begin{figure}[htb]
\begin	{center}
\leavevmode
\input	{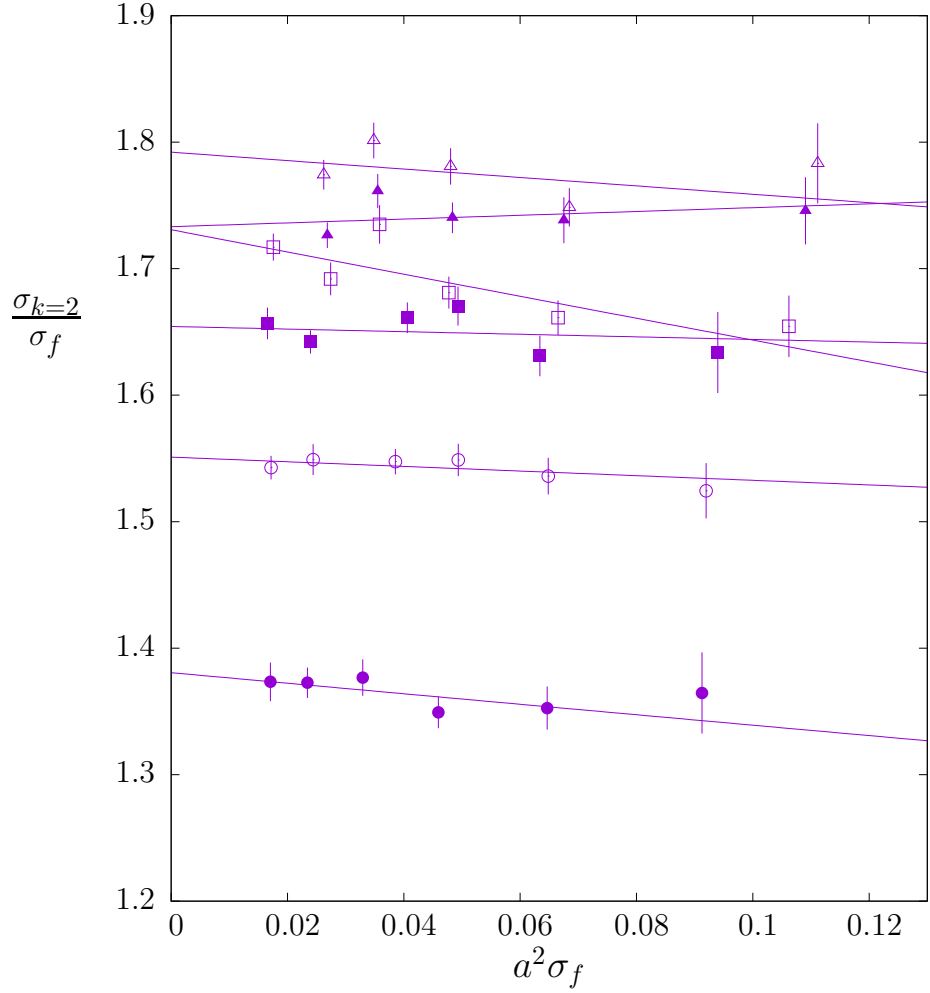}
\end	{center}
\caption{$k=2$ string tensions, $\sigma_k$, in $SU(N)$ gauge theories
  for $N=4,5,6,8,10,12$ in ascending order, in units of the $k=1$ fundamental string
  tension, $\sigma_f$. Lines are extrapolations to the continuum limits.}
\label{fig_k2k1_cont}
\end{figure}

Assuming for now that we can treat the $k=2$ flux tube as a single string with
finite volume corrections as given by  eqn(\ref{eqn_NG}) we calculate $\sigma_{k=2}$
on our various ensembles of lattice fields. Using the previously calculated
values of $\sigma_f \equiv \sigma_{k=1}$ we form the dimensionless ratio
$\sigma_{k=2}/\sigma_f$ which we then extrapolate to the continuum limit using
eqn(\ref{eqn_cont}). We show the extrapolations in Fig.~\ref{fig_k2k1_cont}.
For $N\leq 6$ these appear to be under good control, but for larger $N$ that
is less clear. One would expect the slope of the extrapolation to vary smoothly
with $N$, and that is certainly what one sees for $N=4,5,6$. But then
there is a violent break in the behaviour between $N=6$ and $N=8$
and wild oscillations when comparing  the $N=8,10,12$ slopes.
It may be that this is due to our use of smaller spatial volumes for $N\geq 8$.
In any case, we calculate the continuum limits from these fits and the
results for our various gauge groups are presented
in the second column of Table~\ref{table_sigmak2}. As remarked earlier, there
is good evidence from Table~\ref{table_V_k2_SUN} that for the smaller lattices
that we have used for $N\geq 8$ we should apply an additional finite volume correction of
$\simeq 1.9(5)\%$. Doing this leads to the values in the third column of
Table~\ref{table_sigmak2}, which are the values we consider to be more reliable.
There is a caveat here: at our largest values of $N$ the values of $\sigma_{k=2}/\sigma_f$
are quite close to the asymptotic value of two, and so we might expect that we are in
the range of $N$ where the stronger finite volume behaviour displayed in eqn(\ref{eqn_NGk2}) 
is setting in. To settle this question would require a dedicated calculation that is
beyond the scope of this paper.

We have also obtained continuum extrapolations of the $k=3$ and $k=4$ string tensions.
These have been obtained from the corresponding flux tube energies using eqn(\ref{eqn_NG})
and this undoubtedly underestimates the finite volume corrections. Indeed as $N\to\infty$
we expect $\sigma_k\to k \sigma_f$ as the $k$-string becomes $k$ non-interacting
fundamental strings, and then we expect a version of eqn(\ref{eqn_NGk2}),
\begin{equation}
  E_k(l) \stackrel{N\to\infty}{\longrightarrow}
  k\sigma_f l\left( 1 + \frac{2\pi}{3\sigma_f l^2}\right)^{\frac{1}{2}}
  =
  \sigma_k l\left( 1 + \frac{2k\pi}{3\sigma_k l^2}\right)^{\frac{1}{2}}.
  \label{eqn_NGk}
\end{equation}
Given this uncertainty and the increasing ambiguity with increasing $k$ of extracting
an effective energy plateau (because of their increasing energies), we do not
discuss these values any further here.

\subsection{$N\to\infty$ extrapolations}
\label{subsection_kstringlargeN} 

\begin{figure}[htb]
\begin	{center}
\leavevmode
\input	{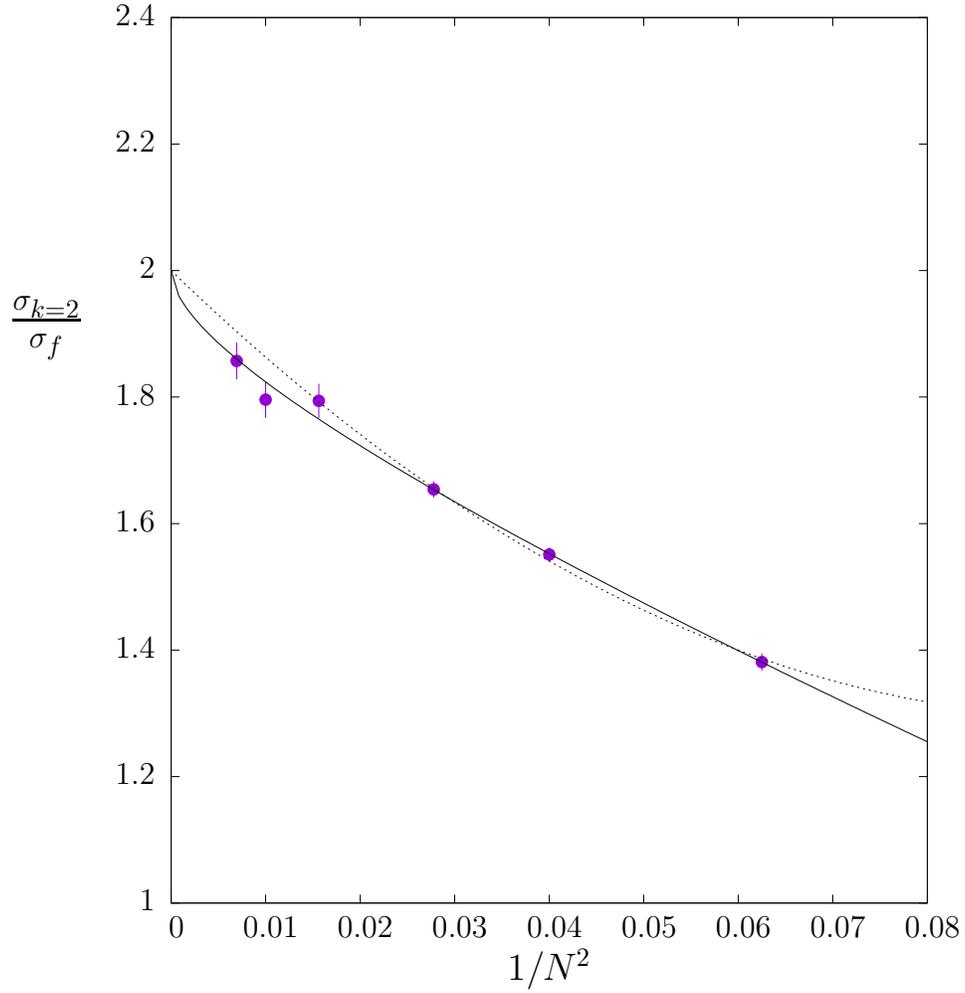}
\end	{center}
\caption{Continuum limit of $k=2$ string tension, $\sigma_k$, in units of
  the $k=1$ fundamental string tension, $\sigma_f$, for our $SU(N)$ gauge
  theories. Solid line is the best fit in powers of $1/N$ and dashed line
  is the best fit in powers of $1/N^2$, with the constraint that the ratio
  is 2 at $N=\infty$.}
\label{fig_k2k1_N}
\end{figure}

An interesting question about the ratio $\sigma_{k=2}/\sigma_f$ concerns its dependence
on $N$. On general grounds we expect the leading correction to the asymptotic value
to be $O(1/N^2)$. However there are old ideas under the label of `Casimir Scaling',
that the (string) tension of a flux tube should be proportional to the smallest quadratic
Casimir of the representations that contribute to that flux, as though the flux tube
joining two sources behaved, in this respect, just like one gluon exchange. For
the $k$ flux tube the relevant representation is the totally antisymmetric one and this
predicts
\begin{equation}
 \frac{\sigma_k}{\sigma_f}
 \stackrel{CS}{=}
 \frac{k(N-k)}{N-1}
 \stackrel{N\to\infty}{=}
  k-\frac{k(k-1)}{N}-\frac{k(k-1)}{N^2} +O\left(\frac{1}{N^3}\right)
  \label{eqn_sigkCS}
\end{equation}
so that the leading correction is $O(1/N)$ rather than $O(1/N^2)$. Previous lattice
calculations have often favoured Casimir scaling as a good approximation
\cite{CSlattice}
so it is interesting to test this idea against our $k=2$ values listed in Table~\ref{table_sigmak2}.
In Fig.\ref{fig_k2k1_N} we plot our continuum values of  $\sigma_{k=2}/\sigma_f$ against $1/N^2$.
In addition we know that $\sigma_{k=2}/\sigma_f=2$ at $N=\infty$ so we impose this as a constraint
in our fits. In Fig.\ref{fig_k2k1_N} we show our best fit in powers of $1/N^2$
\begin{equation}
 \frac{\sigma_{k=2}}{\sigma_f}
 =
 2.0-\frac{1.28(19)}{N}-\frac{4.78(90)}{N^2}
  \label{eqn_k2k1N}
\end{equation}
with a $\chi^2$ per degree of freedom of $\sim 0.5$ and an alternative best fit in powers of $1/N^2$
\begin{equation}
 \frac{\sigma_{k=2}}{\sigma_f}
 =
 2.0-\frac{14.43(60)}{N^2}+\frac{73.8(12.1)}{N^4}
  \label{eqn_k2k1NN}
\end{equation}
with a $\chi^2$ per degree of freedom of $\sim 2.2$. The former fit is clearly better, but the
latter cannot be entirely excluded. On the other hand while the fit in powers of $1/N$ is very good,
the coefficients are very different from those obtained if we expand the Casimir scaling
prediction, as in eqn(\ref{eqn_sigkCS}). One may speculate that for the range
of $N$ where the $k=2$ flux tube is a bound state the $N$ dependence of the
string tension is best described in powers of $1/N$ while once $N$ is large enough
that it has become a weakly interacting pair of fundamental flux tubes, the dependance
will be best described in powers of $1/N^2$. At which $N$ this transition occurs will depend 
on the length $l$ of the flux tube. If we use the formula in eqn(\ref{eqn_NG}) for the
fundamental and $k=2$ flux tubes and if we assume that $\sigma_{k=2}=2\sigma - O(1/N)$,
then one finds that the lightest state will be the weakly interacting pair of fundamental 
flux tubes for $l\leq l_c$ where $l_c\surd\sigma \propto N^{\frac{1}{2}}$. (This essentially 
reflects a competition between the leading linear terms and the $O(1/l)$ Luscher corrections.)
That is to say, for any fixed length $l$,  the (very) asymptotic leading
correction is $O(1/N^2)$ as expected by the usual large-$N$ counting. However one has to
be careful in which order one takes the $l\to\infty$ and $N\to\infty$ limits.

\section{Running coupling}
\label{section_coupling}

A question of theoretical interest is whether our calculations show that  we should keep the
\mbox{'t Hooft} coupling $g^2N$ fixed as $N\to \infty$ in order to have a smooth large-$N$ limit.
A question of phenomenological interest is whether we can estimate the scale $\Lambda$
of our $SU(N)$ gauge theories, from the calculated running of the gauge coupling, and
in particular whether we can do so for $SU(3)$. Finally something that is often useful in
lattice calculations is to have an interpolation function for $a(\beta)$. These are
the three issues we address in this section. A summary of the technical background to the
perturbative calculations has been placed in Appendix~\ref{section_appendix_couplings}.

\subsection{scaling with N}
\label{subsection_couplingN}

Since the lattice coupling $g^2_L$ defined in eqn(\ref{eqn_S}) provides a definition of the running coupling
on the scale $a$, and our above calculation of the (fundamental) string tension at various $\beta$ enables
us to express $a$ in physical units, i.e. as $a\surd\sigma$, we can use these calculations to address
some questions about the properties of the running coupling in $SU(N)$ gauge theories. Before doing so
we recall that this lattice coupling, corresponding to the particular coupling scheme defined by the lattice
and the plaquette action, is well-known to be a `poor' definition of a running coupling in the sense that
higher order corrections will typically be very large. This is indicated, for example, by the relationship
between the scale parameters $\Lambda_L$ and $\Lambda_{\overline{MS}}$ in this scheme and the
standard ${\overline{MS}}$ scheme 
\cite{Hasenfratz,Dashen-Gross}:
\begin{equation}
\frac{\Lambda_{\overline{MS}}}{\Lambda_L}
=
38.853 \exp\left\{-\frac{3\pi^2}{11N^2}\right\}.
\label{eqn_lamLlamMS}
\end{equation}
This is a long-standing issue that has led to the formulation of a number of improved couplings.
(For a review see
\cite{gimp_review}.)
Here we shall use the `mean-field' improved coupling of Parisi 
\cite{MF_Parisi},
\be
\frac{1}{g^2_I} = \frac{1}{g^2_L}
\langle \frac{1}{N}\mathrm{Tr}U_p \rangle,
\label{eqn_gI}
\ee
which has a nice physical motivation as the effective coupling experienced by a background field
(in a simple approximation). Denoting the corresponding scale by $\Lambda_I$, one finds
\be
\frac{\Lambda_{\overline{MS}}}{\Lambda_I}
=
\frac{\Lambda_L}{\Lambda_I}
\frac{\Lambda_{\overline{MS}}}{\Lambda_L}
=
\exp\left\{ -\frac{w_1}{2b_0} \right\}
\times
38.853 \exp\left\{-\frac{3\pi^2}{11N^2}\right\}
\simeq
2.633
\label{eqn_lamIlamMS}
\ee
using eqn(\ref{eqn_lamLlamMS}) and eqns(\ref{eqn_b0b1},\ref{eqn_plaq_pert}), with the
value of $\Lambda_L/\Lambda_I$ being obtained using eqn(\ref{eqn_aa0f}) for each $\Lambda$,
with $g^2_I$ and $g^2_L$ related by eqns(\ref{eqn_gIgL},\ref{eqn_plaq_pert}), and then
taking $g^2\to 0$. This already suggests that $g^2_I$ has the potential to be more-or-less
as good as $g^2_{\overline{MS}}$. One can also show
\cite{CAMTAT}
that this coupling tracks quite accurately the Schrodinger-functional coupling
\cite{SF1,SF2}
over a very wide range of scales. (See
\cite{CAMTAT}
for a detailed discussion.) So from now on, in this section, we shall use
$g^2_I(a)$ as our lattice running coupling.

\begin{figure}[htb]
\begin	{center}
\leavevmode
\input	{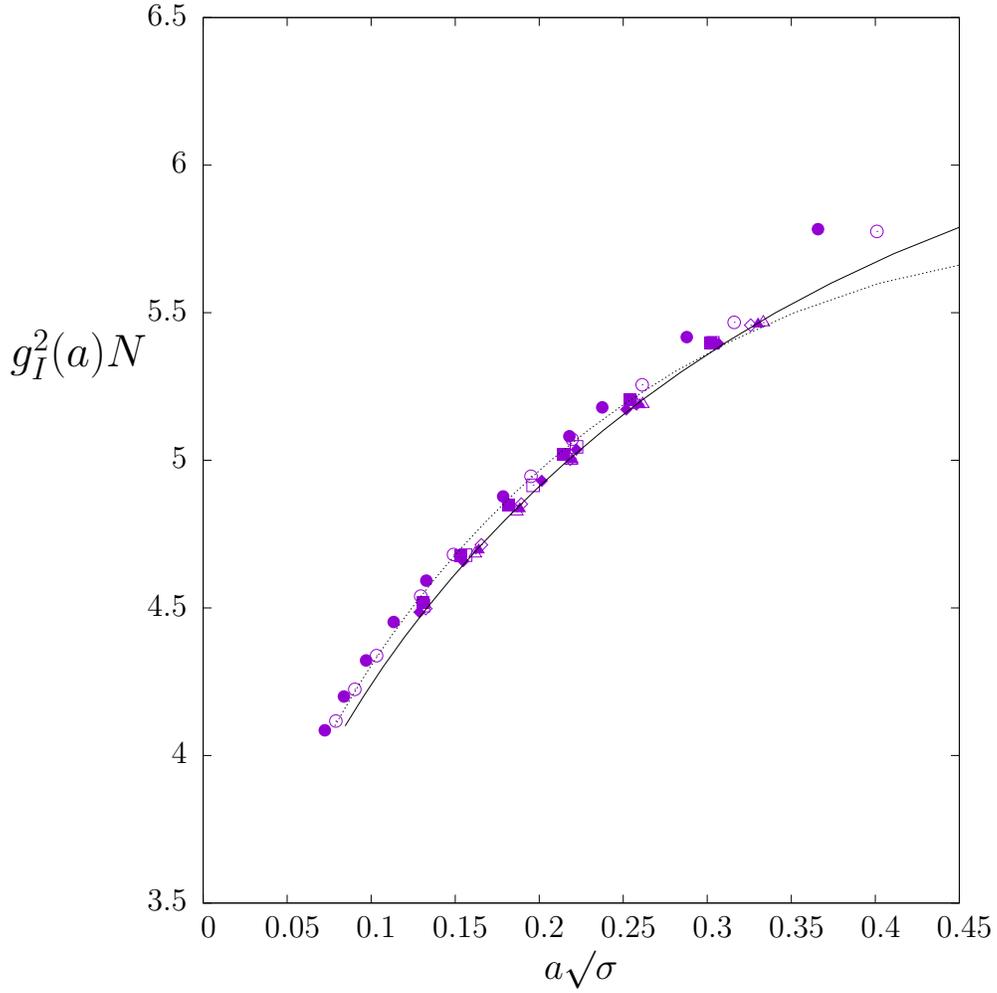}
\end	{center}
\caption{Running (mean-field improved) 't Hooft coupling on the lattice scale $a$, expressed in
  units of the string tension, for
  $SU(2)$, $\bullet$, $SU(3)$, $\circ$, $SU(4)$, $\blacksquare$,
  $SU(5)$, $\square$, $SU(6)$,  $\blacklozenge$, $SU(8)$, $\lozenge$,
  $SU(10)$, $\blacktriangle$, $SU(12)$, $\vartriangle$.
  Solid and dashed lines are (improved) perturbative fits to $SU(8)$ and $SU(3)$
  respectively.}
\label{fig_ggINK_suN}
\end{figure}

The main question we address concerns the $N$ dependence of $g^2_I(a)$. The usual large $N$
counting tells us that we expect to approach constant physics as $N\to\infty$ if we
keep $g^2N$ fixed. For the running coupling this means that if we plot $g^2_I(a)N$
against the calculated values of $a\surd\sigma$ in our $SU(N)$ gauge theories, they
should approach a common envelope as $N\to\infty$. The interesting question is whether
the approach is slow or fast, indicating that our values of $N$ are `far from' or `near to'
$N=\infty$ respectively. In Fig.~\ref{fig_ggINK_suN} we plot our calculated values of the
running 't Hooft coupling $g^2_I(a)N$ against $a\surd\sigma$ for all our values of $N$.
The results are quite remarkable: even $SU(2)$ is very close to $SU(\infty)$ in this respect,
and it is only when the lattice spacing becomes large that appreciable differences appear.
It is also interesting to see what happens if
one uses the poor $g^2_L(a)$ coupling instead of $g^2_I(a)$, and one finds that
while the convergence to a large $N$ limit is still evident,
the corrections at lower $N$ are substantial indicating that, not surprisingly, higher order
non-planar contributions are important using this coupling scheme. Finally we comment
that these conclusions are not unexpected or novel: similar analyses have appeared in
for example 
\cite{BLMT_N,CAMTAT},
albeit usually with less accuracy and over more limited ranges of $N$ and $a\surd\sigma$.

\subsection{perturbative running and ${\mathrm{\Lambda_{\overline{MS}}}}$}
\label{subsection_Lambda}

The second question we address is whether the running coupling dependence displayed in 
Fig.~\ref{fig_ggINK_suN} can be described by the usual perturbative $\beta$-function
for at least some $N$ once $a$ is small. If so we can extract a value of $\Lambda_I$
for each such $N$ and a corresponding value of $\Lambda_{\overline{MS}}$ using
eqn(\ref{eqn_lamIlamMS}). Our analysis will broadly follow that of
\cite{CAMTAT}
and we refer the reader to that paper for background and context; here we merely outline
the calculation, with more details in Appendix~\ref{section_appendix_couplings}. To fix
our notation we begin with the standard $\beta$-function for the lattice bare coupling $g_I^2$:
\be
\beta(g_I) = -\frac{\partial g^2_I}{\partial\log a^2}
=
- b_0  g^4_I - b_1  g^6_I - b^I_2  g^8_I + ...
+ O(a^2),
\label{eqn_bfunction}
\ee
where the scheme independent coefficients $b_0,b_1$ and the scheme dependent $b^I_2$
are given in eqn(\ref{eqn_b0b1}) and eqn(\ref{eqn_b2I}) of
Appendix~\ref{section_appendix_couplings}.
As shown in Appendix~\ref{section_appendix_couplings} this motivates the following
3-loop  expression for $a$:
\be
a \sqrt\sigma(a)
\stackrel{3 loop}{=} 
\frac{\sqrt\sigma(0)}{\Lambda_I}
\left( 1 + c_{\sigma} a^2\sigma \right) 
\left(b_0g_I^2(a)\right)^{-\frac{b_1}{2b^2_0}}
e^{-\frac{1}{2b_0g_I^2(a)}}
e^{-\frac{1}{2} \int^{g_I^2(a)}_0 dg^2
\left(\frac{b_0b^I_2-b^2_1-b_1b^I_2g^2}{b^3_0+b^2_0b_1g^2+b^2_0b^I_2g^4}
\right) }.
\label{eqn_agI3loop}
\ee
Here the first factor on the right after the coefficient is a non-perturbative tree-level lattice
spacing correction.
The motivation is that we could just as well use some physical mass, $\mu(a)$, as a scale in place of
$\surd\sigma(a)$ and since $\mu(a)/\surd\sigma(a)=\mu(a=0)/\surd\sigma(a=0)(1+O(a^2))$ it
must be the case that we have, in general, a factor $(1+O(a^2))$ multiplying the perturbative expression
\cite{CAMTAT,CAMTAT2}.
Here we choose to use $\sigma$ as our scale for $a^2$: using some other $\mu^2$ would only change
the $O(a^4)$ term and we will assume that $a$ is small enough that we can neglect any $O(a^4)$ terms.
We evaluate numerically the integral in the exponential for any given value of $g_I(a)$:
for such a smooth integrand any simple technique will be able to give accurate results.
We now describe the result of fitting this function to our values of $a\surd\sigma$, as listed in
Tables~\ref{table_Ksmall_SU2},\ref{table_Ksmall_SU3} for $SU(2)$ and $SU(3)$ and in
Tables~\ref{table_param_SU4}--\ref{table_param_SU12} for $SU(4)$ to $SU(12)$.
In performing these fits one can either use the measured values of $a^2\sigma$ in the
$(1+c_{\sigma}a^2\sigma)$ lattice correction factor, or the value calculated from the
formula itself. In the latter case eqn(\ref{eqn_agI3loop}) becomes a quadratic equation
for $a\surd\sigma$. Not surprisingly, for the range of $g_I^2(a)$ where  eqn(\ref{eqn_agI3loop})
provides a good fit to our calculated string tensions, the difference between the
two methods is insignificant, and we choose here to use the measured value in the correction term.

The results of our fits using eqn(\ref{eqn_agI3loop}) are shown in Table~\ref{table_Lambda_fitN}.
For each value of $N$ we fit our lattice values of $a\surd\sigma$ discarding the largest
values until we obtain a reasonably acceptable fit. The range of lattice spacings for the fits
is listed for each $N$, together with the $\chi^2$ per degree of freedom, $\chi^2/n_{df}$.
In the few cases labelled by $\ast$ in  Table~\ref{table_Lambda_fitN} the fits are very poor
with $\chi^2/n_{df} \geq 3$, and so we quadruple the stated
error in those cases in the hope that this encodes the increased uncertainty.
Referring to  Tables~\ref{table_param_SU2}--\ref{table_param_SU12}
we see that the number of values used within a given fit varies from 6 in $SU(3)$ to 4 in
$SU(10)$ and $SU(12)$. Since we have 2 parameters in our fit, 4 points is a very small number
for a fit. Moreover, for $SU(10)$ and $SU(12)$ the lattice spacings do not extend to values as small 
as for lower $N$. Hence one should treat the resulting fits with extra caution.

In Table~\ref{table_Lambda_fitN}
we also list the fitted values of the perturbative scale parameter, $\Lambda_I$, in units of the
continuum string tension. We then convert this to the corresponding $\Lambda_{\overline{MS}}$
scale in the widely used $\overline{MS}$ coupling scheme, using the $N$-independent relation
$\Lambda_{\overline{MS}}/\Lambda_I \simeq 2.633$ from eqn(\ref{eqn_lamIlamMS})
and the results are listed in the last column of Table~\ref{table_Lambda_fitN}, again
in units of the string tension. The quoted errors on our values of $\Lambda_I$ are very small,
but they are purely statistical and their smallness reflects the very small errors on the
string tensions that are fitted. The systematic errors involved in, for example, the
truncation of the beta-function may be much larger. To get some measure of this
error we also perform an extra fit using the exact 2-loop running,
\be
a \sqrt\sigma(a) \stackrel{2 loop}{=}  c_I \left( 1 + c_{\sigma,I} a^2\sigma \right)
e^{-\frac{1}{2b_0 g_I^2}}
\left(\frac{b_1}{b_0^2}+\frac{1}{b_0 g_I^2}
\right)^{\frac{b_1}{2b_0^2}}.
\label{eqn_agI2loop}
\ee
The resulting 2-loop values of $\Lambda_I$ are also listed in  Table~\ref{table_Lambda_fitN}.
We see that $\Lambda^{2loop}_I$ is typically about $10\%$ smaller than the $\Lambda^{3loop}_I$.
We have decided to use half of this difference as a measure of the error associated
with dropping 4-loop and higher coefficients from the 3-loop calculation. This is
added as a second error, within square brackets, to the resulting 3-loop values for 
$\Lambda_{\overline{MS}}$ that are listed in the last right-hand column of
Table~\ref{table_Lambda_fitN}. Unlike the statistical errors, this error will affect
the results for all values of $N$ in the same direction. 

\begin{figure}[htb]
\begin	{center}
\leavevmode
\input	{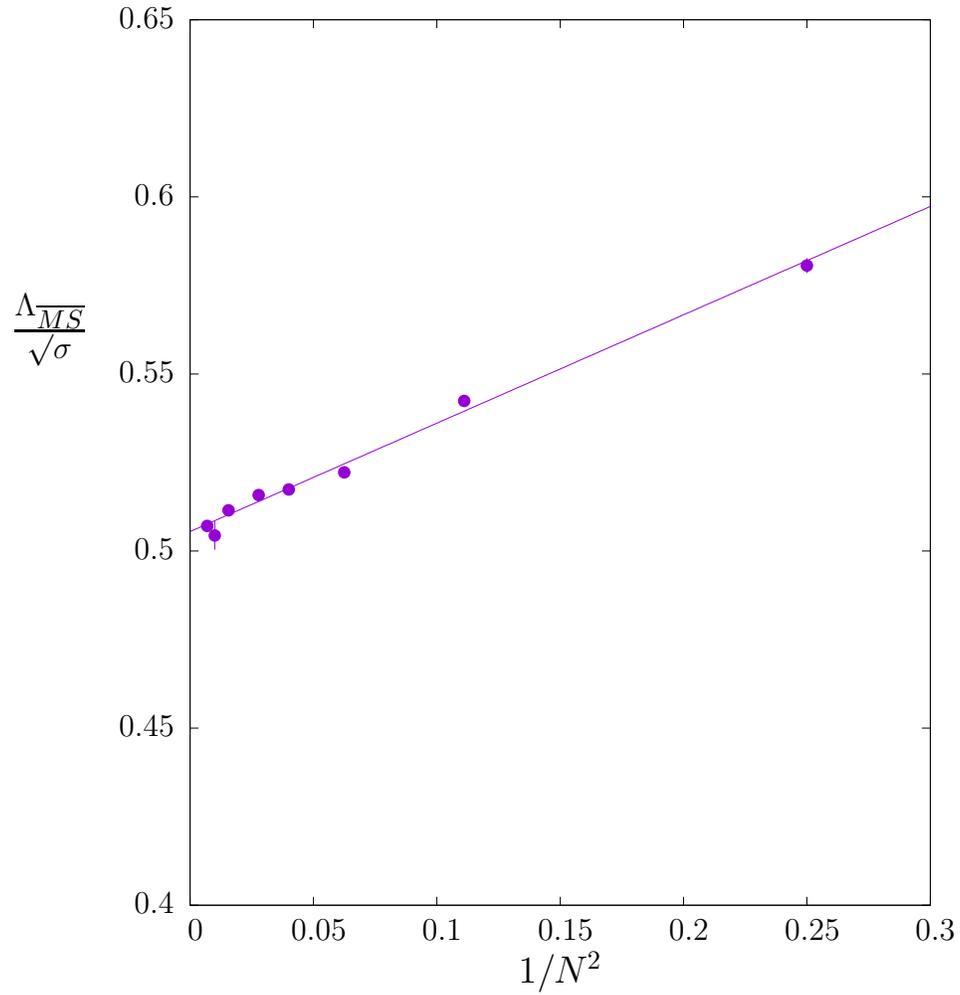}
\end	{center}
\caption{Values of the scale parameter $\Lambda_{\overline{MS}}$
  in units of the string tension in our $SU(N)$ gauge theories.}
\label{fig_LamMS_N}
\end{figure}

In Fig.~\ref{fig_LamMS_N} we display the values that we obtain from our 3-loop fits for
$\Lambda_{\overline{MS}}/\surd\sigma$ as a function of $N$. (We use only the statistical
errors in these fits.) It is clear that the variation
with $N$ is very weak. If we fit all our values for $N \leq 12$ we find the fit,
\be
\frac{\Lambda_{\overline{MS}}}{\surd\sigma}
  = 
  0.5055(7)[250] + \frac{0.306(12)}{N^2}, \quad \chi^2/n_{df}=2.70, \qquad N\in[2,12],  
\label{eqn_LamMS_N}
\ee
which is displayed in Fig.~\ref{fig_LamMS_N}. Although the $\chi^2/n_{df}$ is not good,
the calculated values appear to be scattered around the fit in a random pattern.
Other possible fits include
\beq
\frac{\Lambda_{\overline{MS}}}{\surd\sigma}
  & = &
  0.5067(11)[250] + \frac{0.258(12)}{N^2}, \quad \chi^2/n_{df}=1.65, \qquad N\in[4,12], \nonumber \\
\frac{\Lambda_{\overline{MS}}}{\surd\sigma}
  & = &
  0.5060(8)[250] + \frac{0.303(13)}{N^2}, \quad \chi^2/n_{df}=3.81, \qquad N\in[2,8], \nonumber \\
\frac{\Lambda_{\overline{MS}}}{\surd\sigma}
  & = &
  0.5093(16)[250] + \frac{0.206(38)}{N^2}, \quad \chi^2/n_{df}=0.44, \qquad N\in[4,8].
\label{eqn_LamMS_N_B}
\eeq
In all the above fits we have included within square brackets an estimate of $\sim 5\%$ for
the systematic error associated with the truncation of the perturbative expansion. We note
that this systematic error is much larger than the differences between the various fits above,
and is the dominant source of uncertainty.

To finish we turn briefly to the case of $SU(3)$ where the value of  $\Lambda_{\overline{MS}}$
has phenomenological interest. Here we can attempt to transform our value into $\mathrm{MeV}$
units as follows. We begin with its value in units of the string tension, as listed in
Table~\ref{table_Lambda_fitN}, and we then transform this into a value in terms of the Sommer
length scale $r_0$
\cite{Sommer-r0}
using a recent calculation
\cite{AAMT-2020}
that gives $r_0\surd\sigma = 1.160(6)$. A recent review
\cite{Sommer-r0b}
of calculations in lattice QCD with light quarks concludes that
$r_0=0.472(5)\mathrm{fm}=1/418(5) \mathrm{MeV}^{-1}$.
The usual expectation is that the value of $r_0$ is not very sensitive to the inclusion of light
quarks, so we use this $\mathrm{MeV}$ value in the pure gauge theory. (This is of course the
arguable step.) Doing so we arrive at
\be
\left.\frac{\Lambda_{\overline{MS}}}{\surd\sigma}\right|_{SU3}
= 0.5424(13)[185]
\Longrightarrow
r_0 \Lambda_{\overline{MS}} = 0.629(4)[22]
\Longrightarrow
\Lambda_{\overline{MS}} \stackrel{SU3}{=} 263(4)[9] \mathrm{MeV},
\label{eqn_LamMS_SU3}
\ee
where the first error is statistical and the second is systematic.
This value is consistent with the values recently obtained in the dedicated calculations
\cite{LamMS_SU3a,LamMS_SU3b}
that use very different methods, and our value has a similar accuracy. This adds
confidence that our calculations of $\Lambda_{\overline{MS}}$ for the other $SU(N)$
groups are also reliable.

\subsection{interpolating and extrapolating functions for $a(\beta)$}
\label{subsection_interpol}

It can often be useful to know the value of $a(\beta)$, in physical  units, at some value
of $\beta$. This can be provided, for example, by the value of $a\surd\sigma$. However
calculations of $a\surd\sigma$ are obtained at a number of discrete values of $\beta$
within some finite range $\beta_0\leq\beta\leq\beta_1$ and the $\beta$ value of interest
may lie outside this range, or may be within this range but not at one of the discrete
values where  $a\surd\sigma$ has been calculated. In the latter case one needs to find an
interpolating function that will work in the range $\beta \in [\beta_0,\beta_1]$ and since
$a\surd\sigma$ typically varies smoothly in this range
this is easy to do: a few sensibly chosen terms from almost any complete set of functions
will work adequately. However if the $\beta$ value of interest lies outside the
range $[\beta_0,\beta_1]$ and, in particular, if it is at some weaker coupling, then
extrapolating such a `random' interpolating fit will invariably work badly
unless the original interpolating function has a form motivated by weak coupling
perturbation theory, in which case it should (in principle) be reasonably accurate.
In the section above we have used precisely such functions. We will here
present a fit in an explicit form so that the reader can readily employ it.

Our fit will be in terms of the mean-field improved coupling $g^2_I$ defined
in eqn(\ref{eqn_gI}). This requires that one know the value of the average plaquette;
this is always calculated in a Monte Carlo but one needs to record the
average, which is normally done so as to provide a first check on the calculation.
(And if not, one can obtain a value with adequate accuracy very quickly
using small lattices and modest statistics.)
The interpolating function we will use is a variation on the ones used above
and is as follows:
\be
a \sqrt\sigma(a) =  c_0 \left( 1 + c_{\sigma} a^2\sigma(a) \right) F_{2l+}(g_I),
\label{eqn_agI1}
\ee
where
\be
F_{2l+}(g_I)
=
e^{-\frac{1}{2b_0 g_I^2}}
\left(\frac{b_1}{b_0^2}+\frac{1}{b_0 g_I^2}
\right)^{\frac{b_1}{2b_0^2}} 
e^{-\frac{b^I_2}{2b_0^2}g_I^2}.
\label{eqn_agI2}
\ee
The first two factors of $F_{2l+}(g_I)$ constitute the exact dependence when $\beta(g)$ is
truncated to the first two terms. The last factor on the right is the extra dependence
if one keeps in the exponent the leading $O(b^I_2)$ term to $O(g^2_I)$.
That is to say, it is `more' than 2 loops, but `less' than 3 loops. Hence the subscript
$2l+$ on $F(g)$. This is of course an arbitrary truncation with no guarantee that it
does indeed do better than the 2-loop one; however we use it because, as we shall see
below, it turns out to be very close to the 3-loop result, but without the need to
perform any numerical integrations. Note that here we choose to use in the $\propto a^2\sigma$
correction term the value given by fitting the formula to our data since this is the
mode in which it needs to be used in an extrapolation.

To make use of eqn(\ref{eqn_agI2}) we need to fit the constants $c_0$ and $c_{\sigma}$ to our
calculated values of $a\surd\sigma$, at each $N$. Once the constants $c_0$ and $c_{\sigma}$
have been fitted we can solve eqn(\ref{eqn_agI1}), which is quadratic in $a\surd\sigma$, at
any value of $\beta_I$:
\be
a \sqrt\sigma(a) =  \frac{1}{2c_0c_{\sigma}F_{2l+}(\beta_I)}
  \left(1-\left[1 - 4c^2_0c_{\sigma}F_{2l+}(\beta_I)^2\right]\right).
\label{eqn_aKgI}
\ee
In  eqn(\ref{eqn_agI2}) the values of $b_0,b_1,b^I_2$ need to be specified. The values of
$b_0$ and $b_1$ are  universal and are given in eqn(\ref{eqn_b0b1}). The value of $b^I_2$
is as given in eqn(\ref{eqn_b2I}) 
\be
b^I_2=b^L_2+w_2b_0-w_1b_1,
\label{eqn_b2Ib}
\ee
where we use eqns(\ref{eqn_b2MS},\ref{eqn_b2Lb}) in eqn(\ref{eqn_b2La}) to obtain the
explicit expression for $b^L_2$ and then inserting that together with the functions in
eqn(\ref{eqn_plaq_pert}) and eqn(\ref{eqn_b0b1}) into eqn(\ref{eqn_b2Ib}) we
obtain the explicit expression for $b^I_2$ for any $N$. 

It only remains now to give our fitted values of $c_0$ and $c_{\sigma}$ for each of the
$SU(N)$ groups for which we have calculated the string tension. This we do in
Table~\ref{table_interp_gI}. There we show for each $N$ our best fits to these
two parameters as well as the fitted range and the $\chi^2$ per degree of freedom
of the fit. In performing these fits we systematically drop the largest values of
$a\surd\sigma$ from the fit until we obtain an acceptable $\chi^2$. This is
appropriate since our interpolating function is based on weak-coupling
perturbation theory. Note that being based on weak coupling the resulting function
is not designed to work for values of $a$ or $g^2$ greater than the range
within which it provides a good fit.

As an aside it needs to be emphasised that all the above fits are only relevant
if one uses the Wilson plaquette action.
As a second aside, we note that if one wants $a(\beta)$ in terms of some physical
scale $\mu$ instead of $\surd\sigma$ then one can straighforwardly modify the
above formula for $a\surd\sigma$ to one for $a\mu$ by using the relation
$\mu/\surd\sigma = d_0 + d_1 a^2\sigma$ if it accurately holds in the
range of couplings of interest.

\section{Glueball masses}
\label{section_glueballs} 

\subsection{quantum numbers}
\label{subsection_quantumG}

The glueballs are colour singlets and so our glueball operator is obtained by taking the
ordered product of $SU(N)$ link matrices around a contractible loop and then taking the trace.
To retain the exact positivity of the correlators we use loops that contain only spatial links.
The real part of the trace projects on $C=+$ and the imaginary part on $C=-$.
We sum all spatial translations of the loop so as to obtain an operator with momentum $p=0$.
We take all rotations of the loop and construct the linear combinations that transform
according to the irreducible representations, $R$, of the rotational symmetry group  of our cubic
spatial lattice. We always choose to use a cubic lattice volume that respects these symmetries.
For each loop we also construct its parity inverse so that taking linear combinations
we can construct operators of both parities, $P=\pm$. The correlators of such operators will
project onto glueballs with $p=0$ and the $R^{PC}$ quantum numbers of the operators concerned.

The irreducible representations $R$ of our subgroup of the full rotation group are usually
labelled as $A_1,A_2,E,T_1,T_2$. The $A_1$ is a singlet and rotationally symmetric, so it
will contain the $J=0$ state in the continuum limit. The $A_2$ is also a singlet, while the
$E$ is a doublet and $T_1$ and $T_2$ are both triplets. In Section~\ref{subsection_spins} we
will outline the detailed relationship between these representations and the continuum spin $J$.
Since, for example, the three states transforming as the triplet of $T_2$ are degenerate on
the lattice, we average their values and treat them as one state in our tables of glueball
masses and we do the same with the $T_1$ triplets and the $E$ doublets. (Just as we would
treat the 5 states of a continuum $J=2$ glueball as one entry.)

\subsection{finite volume effects}
\label{subsection_massV} 

For reasons of computational economy we wish to calculate on lattice sizes that are small 
but, at the same time, large enough that any finite volume corrections remain smaller than
our typical statistical errors. Since the computional cost of calculating in $SU(N)$
gauge theories grows roughly $\propto N^3$ (the multiplication of two $N \times N$
matrices) and since finite volume corrections are expected to decrease as powers of $1/N$,
we reduce the size in physical units of our lattices as we increase $N$, as shown
in Tables~\ref{table_param_SU2}-\ref{table_param_SU12}. 

There are two important types of finite volume corrections. The first can be thought of as arising
when the propagating glueball emits a virtual glueball which propagates around the spatial
torus. The resulting shift in the mass of the propagating glueball decreases exponentially
in $m_Gl$ where $m_G$ is the mass gap and $l$ is the length of the spatial torus
\cite{Luscher-V}.
As we see from Tables~\ref{table_param_SU2}-\ref{table_param_SU12} the value of $am_G \times l/a$
is quite large in all of our calculations, so we can expect this correction to be small.

The second type of finite volume correction consists of states composed of multiple flux
tubes winding around a spatial torus in a (centre) singlet state. The lightest of these will be
a state composed of one winding flux tube together with a conjugate winding flux tube,
which we refer to as a `ditorelon'. (A single winding flux tube is usually
referred to as a `torelon'.) Since it can have a non-zero overlap onto the contractible
loops that we use as our glueball operators, it can appear as a state in our calculated
glueball spectrum. Neglecting interactions, the lightest ditorelon will consist
of each flux tube in its ground state with zero momentum and will have an energy, $E_d$, that
is twice that of the flux tube ground state, $E_d=2E_f$. Interactions will shift the energy
but this shift should be small on our volumes so we shall use $E_d\simeq 2E_f$ as a rough guide
in searching for these states. This ground state ditorelon has simple rotational properties
and only contribute to the $A_1^{++}$ and $E^{++}$ representations. If we allow one or
both of the component flux tubes to be excited and/or to have non-zero equal and opposite
transverse momenta we can populate other representations and produce towers of states.
However these excited ditorelon states will be considerably heavier on the lattice volumes
we employ and so we will not consider them any further in this paper, although they certainly
warrant further study.

The first of the above corrections leads to small shifts in the masses of the glueballs.
The second leads to extra states in the glueball spectrum. The signature of such an extra
ditorelon state is that its mass grows roughly linearly with the lattice size:
$aE_D \simeq 2aE_f \simeq 2 a^2\sigma_f L$ where $L$ is the relevant spatial size in lattice
units and  $\sigma_f$ is the (fundamental) string tension. So to test for finite volume effects
we perform calculations at the same value of $\beta$ on different lattice sizes and compare the
glueball spectra. To identify any ditorelon states we look for extra states in the $A_1^{++}$
and $E^{++}$ spectra whose masses increase roughly linearly with the volume. Since the mass
shift associated with the first kind of correction decreases exponentially with the lattice
size any shift on a significantly larger volume should be much smaller than on the smaller volume;
so to check that it is negligible compared to our statistical errors we simply compare the masses
of the states that are not ditorelons on the different volumes. As we see
from Tables~\ref{table_param_SU2}-\ref{table_param_SU12} the lattice sizes we use
fall into three groups: $l\surd\sigma \sim 4.0$ for $SU(2),SU(3),SU(4)$;
 $l\surd\sigma \sim 3.1$ for $SU(5),SU(6)$; $l\surd\sigma \sim 2.6$ for $SU(8),SU(10),SU(12)$.
We will exhibit a finite volume analysis for a representative of each of these three groups.

We begin with $SU(2)$. In Table~\ref{table_GvsV_SU2} we list the low-lying glueball spectra
that we obtain on 3 lattice sizes, together with the energies of the ground and
first excited states of the winding flux tubes. In physical units the $L=12,14,20$ spatial
lattice sizes correspond to $l\surd\sigma \simeq 2.9,3.4,4.8$ respectively and we recall
that the typical lattice size we use in $SU(2)$ is $l\surd\sigma \simeq 4$.
For the glueball states we list the
effective energy at $t=2a$ or, where applicable, the energy of the fit to the effective
energy plateau when that begins at $t=2a$. (These two measures differ very little in practice.)
These are obtained from the correlators of our variationally selected best operators.
For most of our lighter states the value of $aE_{eff}(t=2a)$ is very close to our best estimate
of the mass. For these finite volume comparisons we prefer this measure to the mass itself because
it serves to minimise the statistical errors and makes any finite volume corrections more visible.
Putting aside the $R^P=A_1^{+}$ and the $E^{+}$ spectra for the moment, we see that most of the glueball
energies on the $L=14$ and $L=20$ lattices are consistent within one standard deviation and all within
two standard deviations. The energies from the $L=12$ lattice are broadly consistent but there are
now some examples, such as the $A_2^+$ and $E^-$ ground states, where there appear to be significant
differences. We now return to the $A_1^{+}$ and the $E^{+}$ spectra listed in Table~\ref{table_GvsV_SU2}
to see if there is any evidence of the extra ditorelon states. We do indeed see these on the
$L=12$ and $L=14$ lattices in both the $A_1^{+}$ and the $E^{+}$ spectra: these states are displayed
in the Table with no corresponding entries at other lattice sizes. Their
effective energies increase with
$L$ and are just a little heavier than twice the flux tube energy. For the $L=20$ lattice
any ditorelon state would be much heavier than any of the states shown: at such energies the
spectrum is denser, the errors are larger, and so identifying an `extra' state becomes ambiguous.
The remaining states in the $A_1^+$ and the $E^{+}$ spectra are broadly consistent across the
lattice sizes, except for the lightest $E^+$ on $L=12$ and the first excited $A_1^+$ on $L=14$,
both of which are quite close to their respective ditorelons and possibly mixing with them.
Also the 4th state in the $L=12$ $A_1^+$ spectrum shows a shift. Since the typical lattice size we
use in $SU(2)$ is $l\surd\sigma \sim 4$ (except for the relatively unimportant calculations
at the largest values of $a(\beta)$) we can estimate the ditorelon states to have energies
$E_D \sim 2\sigma l \sim 2 (l\surd\sigma)\surd\sigma \sim 8\surd\sigma$ and we shall therefore
only perform continuum extrapolations of $A_1^{+}$ and $E^{+}$ states that are lighter than
$\sim 8\surd\sigma$, although it is still the case that the heaviest states in these channels
may be perturbed by ditorelon contributions. As for the other channels, since the size
$l\surd\sigma \sim 4$ falls between our $L=14$ and $L=20$ lattices
at $\beta=2.427$, we can conclude from the above that there should be no finite volume
corrections that would be visible outside our statistical errors.
We have performed similar finite volume checks in $SU(3)$ in our earlier paper
\cite{AAMT-2020}
and we refer the reader to that paper for details, in particular for a check of the $C=-$ states.
In the case of $SU(4)$ we have performed a finite volume analysis comparing the spectra on
$18^320$ and $22^4$ lattices at $\beta=11.02$, which has helped us identify the
positions of the ditorelon states in the spectra at other values of $\beta$, and to
remove these states from our spectra. With all these checks we have some confidence
that our $SU(2)$, $SU(3)$ and $SU(4)$ continuum spectra will not be afflicted by significant
finite volume corrections.

Our $SU(5)$ and $SU(6)$ calculations are on significantly smaller volumes, typically with
$l\surd\sigma \sim 3.1$, and therefore deserve a separate finite volume study to the one above.
To do so we compare the low-lying glueball spectra that we obtain in $SU(5)$ on $14^320$ and $18^318$
lattices at $\beta=17.46$, together with the energies of the ground and first excited states of
the fundamental and $k=2$ winding flux tubes. In physical units these $L=14$ and $L=18$ spatial
lattice sizes correspond to $l\surd\sigma \simeq 3.06$ and $3.93$ respectively and we recall
that the typical lattice size we use in $SU(5)$ and $SU(6)$ is $l\surd\sigma \simeq 3.1$ i.e. close
to that of the smaller of our two lattices. As in $SU(2)$
we readily identify the extra ditorelon states in the $A_1^{++}$ and $E^{++}$ representations,
which have energies very close to twice that of the winding fundamental flux tube.
(The ditorelon on the $L=18$ lattice will lie outside our energy range.) They appear
to be well separated from neighbouring glueball states and so we can readily identify and
exclude them from our $SU(5)$ and $SU(6)$ glueball spectra and exclude them from our
continuum limits. The other energies
on our two lattice sizes are mostly within errors of each other, with a few out by up to two
standard deviations, and two states a little more than that. Given the large number of states
being compared, such rare discrepancies are inevitable and we can  claim that
we see no significant finite volume corrections to any of our listed states.

We now turn to our largest $N$ calculations where we use the even smaller spatial lattice size
$l\surd\sigma \simeq 2.6$. Our finite volume study is in $SU(12)$ on $12^320$ and $14^320$ lattices
at $\beta=103.03$, where the $12^320$ lattice represents our typical physical size, and the
energies are listed in Table~\ref{table_GvsV_SU12}. In contrast to the previous tables, we
list under the $A_1^{++}$ and $E^{++}$ representations our best estimate of the true glueball
spectrum with the ditorelons removed. We will return to the identification of the latter shortly.
We observe that most of the energies are the same within errors and all are within two standard
deviations. That is to say, any finite volume shifts in the energies are within our statistical
errors at these values of $N$, despite the fact that the $L=12$ lattice volume is quite small.

The situation with respect to the ditorelons is more complex. The first complication is that on
the  $12^320$ lattice the $A_1^{++}$ ditorelon is nearly degenerate with the first excited  $A_1^{++}$
glueball, so that these states may well mix even if the overlaps are small due to the large-$N$
suppression. Moreover the same occurs in the $E^{++}$ representation where the
ditorelon is nearly degenerate with the ground state $E^{++}$ glueball.
The second complication arises from the fact that the overlap of the ditorelon double trace
operators onto the single trace operators that we used to calculate the glueball spectrum
is suppressed since $N$ is large. This should
be an advantage and indeed if the overlap is small enough then the ditorelon will not appear
in the spectrum obtained from the single trace operators so that there is no issue.
However it may well appear as a minor component of
a state that at first sight is quite massive but then its $aE_{eff}(t)$ drops towards the
ditorelon energy as $t$ increases and the more massive components die away. This 
can lead to ambiguities and indeed does on both our lattice sizes.

Since all our $SU(12)$ glueball calculations, and also those in $SU(10)$ and $SU(8)$,
are on lattices of roughly the same physical size as our $12^320$ one at $\beta=103.03$, we need
to address these problems with the ditorelons. We do so as follows. First we find that
if in constructing our glueball operators we use blocked links whose extent is smaller than
the lattice size then the overlap of ditorelons onto these operators is small enough that
we do not see any ditorelon state in the $A_1^{++}$ spectrum but we do see it embedded as a
small component in more massive states in the $E^{++}$ spectrum. Since a link at blocking level
$bl$ joins sites that are $2^{bl-1}a$ apart, this means keeping to blocking levels $bl=1-4$
on our $l=12a$ and $l=14a$ lattices. If we now do something that may appear less reasonable and
include $bl=5$ blocked links, which join lattice sites $16a$ apart, so that the operators
formed out of these links wrap multiply around the spatial torus in all spatial directions,
we find that the ditorelon overlaps are much larger and the ditorelon states now appear quite
clearly in the resulting spectrum. The results of these various calculations
are shown in Table~\ref{table_GvsV_SU12B}. As usual the energies shown are $aE_{eff}(t=2a)$
and so are often slightly higher than the value of the effective energy plateau.
Consider first the $A_1^{++}$ spectra on the $12^320$ lattice. We see that the second and third
states in the $bl=1-5$ spectrum are nearly degenerate, with masses close to what one might
expect for the ditorelon, i.e. $2aE_f\sim 0.96$, but only one of them appears in the $bl=1-4$
spectrum. Thus we infer that one of those two (or a mixture of the two) is a ditorelon.
On the $14^320$ lattice we also see an extra state in the $bl=1-5$ spectrum, as compared
to the $bl=1-4$ spectrum, but now the state is heavier, as we would expect for a ditorelon 
since  $2aE_f\sim 1.2$ on this lattice. Turning to the $E^{++}$ representation on the $14^320$
lattice we find an extra state using the $bl=1-5$ basis, as compared to using the $bl=1-4$ basis,
with roughly the expected mass of a ditorelon.
On the  $12^320$ lattice things are a bit different: the second state in the $bl=1-5$ spectrum
has no obvious partner in the $bl=1-4$ spectrum, but there is a state in the latter spectrum
where $aE_{ff}(t)$ decreases rapidly with increasing $t$ to a comparable value. This is
the second entry in the table and the mass from the `plateau', indicated in square brackets,
is similar to the energy of the second state in the $bl=1-5$ spectrum. We
interpret this as follows: the $E^{++}$ ditorelon has a substantial overlap onto the $bl=1-5$
basis and only a small overlap onto the $bl=1-4$ basis, but large enough to appear within
what appears to be a higher excited state. This mass is somewhat smaller than that of the
state we identified as a ditorelon on the  $14^320$ lattice which is what one expects.
We note that unlike the states we identify as ditorelons, most of the other states
have nearly the same energies on the two lattice sizes, as one expects for states that
are not ditorelons. Recalling that the lattice sizes we use in our $N\geq 8$ calculations
are of the same physical size as our $12^320$ lattice at  $\beta=103.03$, we can use
the above results in all those cases to identify ditorelon states and remove them from
our glueball spectra. The same type of technique is useful for $N=5,6$.

A final comment is that the rotational symmetries of the ditorelon states
-- and indeed multitorelon states -- remain those of our finite volume, i.e. $\pi/2$ rotations,
even in the continuum limit. That is to say, even in that limit
they will fall into the representations of the cubic subgroup of the continuum rotation group.
The genuine glueballs, on the other hand, will fall into representations of the full continuum
rotation group in the continuum limit once the volume is large enough, up to corrections
that are exponentially small in the spatial size, and this difference can also be useful
in distinguishing ditorelons from genuine glueballs.

\subsection{lattice masses}
\label{subsection_latticemass} 

As described earlier, we calculate glueball masses from the correlators of suitable $p=0$ operators.
These operators are chosen to have the desired $R^{PC}$ quantum numbers, where
$R \in \{A_1,A_2,E,T_1,T_2\}$ labels the irreducible representation of the cubic subgroup of the
rotation group and $P=\pm$ and $C=\pm$ label parity and charge conjugation. We typically
start with a set of 12 different closed loops on the lattice. (For $N=2,3$ we used 27, but
we then observed that we can get almost equally good spectra by using a suitable subset of 12 loops.)
We calculate all 24 rotations of the loop and construct linear combinations of the traces that
tranform as $R$. We do so separately using the real and imaginary parts of the traces,
which gives us operators with $C=\pm$ respectively. We also calculate the parity inverses of
each of these twelve loops, and of their rotations, and by adding and subtracting appropriate
operators from these two sets we form operators for each $R$ with $P=\pm$. To ensure that
we have non-trivial operators for each set of quantum numbers it is useful to include
some loops that have no symmetry under rotations and parity inversion. With our particular
choice of 12 (or 27) closed loops we are able to construct the number of independent operators
shown in Table~\ref{table_numops_N}. This is the number of operators at each blocking level and 
we typically use 4 or 5 blocking levels in our calculations. This makes for quite a large 
basis of operators for all $R^{PC}$ quantum numbers, and indeed a very large number for most
of them. However this large number is slightly deceptive as at any given $\beta$ only two or perhaps
three blocking levels make an important contribution to the low-lying spectrum. In addition
the states in the $E$ representation are doubly degenerate and in the $T_1$ and $T_2$ are
triply degenerate, so one should divide the corresponding numbers in Table~\ref{table_numops_N}
by 2 or 3 in order to estimate the number of different energy levels accessible to
the basis.

For completeness we list our 12 basic loops in Table~\ref{table_loops}.
The links are labelled by $1,2,3$ to indicate their spatial directions, with
negative signs for backward going links, and the path ordering is from left to right.
So, for example, the plaquette in the $2-3$ plane would appear as $\{2,3,-2,-3\}$.
Our list includes the plaquette, three 6 link loops, four 8 link loops
and four 10 link loops.  We also show for each loop which representations
it contributes to once we include all the rotations and the inverses, doing so for the
real ($C=+$) and  imaginary ($C=-$) parts of the traces separately. This basis is
used for $N\geq 4$. For $SU(2)$ and $SU(3)$ we use a larger basis of 27 loops
that includes the 12 listed here. For some finite size studies we have used
a reduced set of 8 loops. 

We use these bases of operators in our variational calculations of the glueball spectra.
In each $R^{PC}$ sector we calculate a number of the lowest masses from the
correlation functions of the operators that our variational procedure selects as being
the best operators for those states within our basis. This relies on identifying an effective
mass plateau in the correlator, as described in Section~\ref{subsection_lattice_energies},
and performing an exponential fit to the data points on that plateau. So an important
question is: how reliably can we identify such a plateau? We will illustrate this
with our calculations in $SU(8)$ on the $20^330$ lattice at $\beta=47.75$. This
is at our smallest value of $a(\beta)$ which, being the closest to the continuum
limit, is one of the most important values in our subsequent extrapolations to the
continuum theory. Also it provides the best resolution in $t$ of our correlation
functions. We begin with the effective masses, $aE_{eff}(t=an_t)$, of the three
lightest $A_1^{++}$ states and of the two lightest $A_1^{-+}$ states, as shown in
Fig.\ref{fig_MeffA1_SU8}. As we will see below, the lightest two $A_1^{++}$ and
$A_1^{-+}$ states become the lightest two $J^{PC}=0^{++}$ and $0^{-+}$ glueballs
in the continuum limit. The third $A_1^{++}$ state is most likely part of the nonet of
spin states making up the $4^{++}$ glueball ground state (see below), although if not
then it would probably be the third scalar glueball. 
It is clear from Fig.\ref{fig_MeffA1_SU8}
that we have plausible effective mass plateaux for all the states, with the
weakest case being the excited $A_1^{-+}$ state. The solid lines show our
best mass estimates as obtained from fits to the correlation functions,
together with their error bands. The error band on the lightest  $A_1^{++}$
state, which is the mass gap of the gauge theory, is invisible on this plot,
so given the importance of this state we replot it in  Fig.\ref{fig_MeffA1b_SU8}
with an axis rescaling that exposes the errors. Here the solid line is the
best mass estimateand the dahsed lines bound the $\pm 1$ stadard deviation
error band. Fig.\ref{fig_MeffA1_SU8} illustrates
the obvious fact that as the masses get larger, the error to signal ratio
grows at any fixed value of $n_t$, so that at large enough masses the effective 
energies becomes too imprecise to unambiguously indicate the values of $n_t$
where the effective mass plateau begins. In this case we can turn to our
calculations of the excited $A_1^{-+}$ state at the smallest lattice spacings
in $SU(2)$ and $SU(3)$, where there is much better evidence for the effective
mass plateau beginning at $aM_{eff}(n_t=3a)$, and use that in estimating that
the effective mass plateau in this $SU(8)$ calculations begins at $n_t=3a$.
This is the type of argument we use for a number of the heavier states
at larger values of $N$ where our calculations do not extend to very small
values of $a(\beta)$.

As an aside, we also show in this
plot the effective mass plot for the ditorelon. It shows every sign of plateauing
to a value not far from $\sim 2aE_f\simeq 0.59$ as one would expect. It is clearly
important to make sure that one excludes such a state from the scalar glueball spectrum,
as we have done here, since it is located near the first excited glueball state
and would create a false level ordering if included.

We turn next to states which become $J=2$ glueballs in the continuum limit.
As we shall see below, the five components of a $J=2$ state are spread over
the two components of an $E$ energy level and the three of a $T_2$. As always
we average the masses of the `degenerate' doublets of $R=E$ and the triplets of
$R=T_1$ and $T_2$ to provide single effective masses in each case. What we show
in Fig.\ref{fig_MeffET2_SU8} are the resulting effective mass plots for a number
of such pairs of $E$ and $T_2$ states. The equality of the $E^{++}$ and $T_2^{++}$ masses
for the lightest two pairs is convincing, as it is for the lightest  $E^{-+}$
and $T_2^{-+}$. It is also plausible for the lightest $E^{--},T_2^{--}$
pair, and the lightest  $E^{+-},T_2^{+-}$ pair, and there is some convergence
for the second  $E^{-+},T_2^{-+}$ pair, even if these more massive states
do not show unambiguous plateaux. Again we show the ditorelon, this time in
the $E^{++}$ representation. In this spectrum it is nearly degenerate with
the  $E^{++}$  ground state. The fact that it does not appear in a nearly
degenerate $T_2^{++}$ state confirms that it is a finite volume state,
reflecting the limited rotational symmetries of the spatial volume.

If we examine the effective mass plots for the $T_1^{PC}$ states that we shall
later argue approach $J=1$ in the continuum limit, we see that
while the lightest $T_1^{+-}$ glueball has a well-defined effective mass plateau,
this is less evident for the more massive states.
Finally we remark that if we plot the  effective masses of our heaviest states,
such as those of the lightest five $A_1^{--}$ states,
the evidence for the effective mass plateaux is not strong, although one can
speculate on the presence of some the plateaux.
This will clearly result in substantial systematic errors on the corresponding mass
estimates, which we cannot readily quantify.

One lesson of the effective mass plots is that the heavier the state
the less reliable will be our error estimate. However these plots also tell us at what
$t=an_t$ the effective mass plateau begins for those states where this can be 
identified. Since the iterative blocking means that when  we vary $\beta$ our variationally
selected operators are of roughly constant physical size and shape, we can assume
that the overlap of a given state onto our basis will be roughly independent of $\beta$
and hence that the effective mass plateau will begin at a value of $t=a(\beta)n_t$ that
is roughly constant in physical units, i.e. at smaller $n_t$ as $a(\beta)$ grows.
So at larger values of $a(\beta)$ where $aE_{eff}(t)$ may be too large for us to identify 
a plateau, we can nonetheless use the value of $aE_{eff}(t_0)$, where $t_0$ is the
value where our above calculations tell us that a plateau begins, as an estimate of the mass.
This is something we do in our calculations, where appropriate.

We turn now to the glueball spectra that we obtain by the methods described above.
We obtain the spectra for the lightest glueballs in each $R^{PC}$ sector for
each of our gauge groups. As an example we display in \ref{table_Mlat_RPC_SU8}
our results for $SU(8)$.
Here we have removed the finite volume
ditorelon states in the $A_1^{++}$ and $E^{++}$ sectors whenever they are present, 
as discussed in Section~\ref{subsection_massV}, so that what
we present in the Table is our best estimate of the infinite volume glueball spectrum.
A similar Table for $SU(3)$ has been published separately in
\cite{AAMT-2020}.
%
All this assumes of course that none of these states
is a multiglueball state. Since we use single trace operators, their overlap onto
multiglueball states should decrease with increasing $N$ and in any case they would be
quite heavy. Some explicit calculations that provide evidence that such mutliglueball
states do not appear in our spectra are described in Section~\ref{subsection_scattstates}.

The reader will note that in some cases what we list as a higher excited state has a lower mass
than what is listed as a lower excited state. This typically involves states that are
nearly degenerate. Our procedure is to order the states according to the values of their
effective masses at $t=a$ which is the value of $t$ at which our variational calculation
typically operates. This nearly always corresponds to final mass estimates (from the plateaux)
that are in the same order, but very occasionally this not the case.
In the very few cases where the difference is well outside the statistical errors
we invert the ordering, but otherwise we do not. In addition it may be that
the continuum extrapolation leads to a level inversion; again we only take that on board
if the difference is well outside the errors. We have a good number of states that are nearly
degenerate with a neighbouring state and in these cases the variational procedure itself is
likely to mix them, and if that is done differently at different values of $\beta$ it can
lead to poor extrapolations to the continuum limit.

Finally we remark that
the ensembles of lattice fields used at each $N$ and $\beta$ have a reasonable distribution
of topological charge, either because the tunnelling between topological charge sectors is
sufficiently frequent, or because we have imposed such a distribution on the initial lattice
fields of the collection of sequences that make up the total ensemble, as discussed in
Section~\ref{subsection_Qfreezing}.

\subsection{strong-to-weak coupling transition}
\label{subsection_bulk}

Before moving on to the continuum extrapolation of the lattice masses that we have calculated
above, we briefly remark on the `bulk' transition that interpolates in $\beta\propto 1/g^2$
between the strong and weak coupling regimes. From eqns(\ref{eqn_Z},\ref{eqn_S}) we see that the
naive expectation is that at strong coupling, $\beta\to 0$, the natural expansion of a
lattice quantity is in positive powers of $\beta$ while at weak coupling, $\beta\to \infty$, 
the natural expansion is in positive powers of $1/\beta$. On the weak coupling side
asymptotic freedom promotes the dependence of physical quantities to exponentials,
$\propto \exp\{-c\beta\}$, and it is important to make sure that our calculations
are indeed in the weak coupling regime.

As remarked in Section~\ref{subsection_lattice_setup}, for $SU(N\leq 4)$ the transition
is known to be a smooth crossover, while for $SU(5)$ it is weakly first order,
and for $SU(N\geq 6)$ it is strongly first order. For $SU(N\leq 4)$ the location
of the crossover coincides with a dip in the mass gap which then drives a peak in the
specific heat. (The specific heat is proportional to the sum of the plaquette-plaquette correlator
and so peaks where the correlation length has a peak, i.e. where the mass gap has a dip.)
For $SU(2)$ this peak is around $\beta\sim 2.15$ as we see, for example, in Fig.8 of
\cite{KIGSMT-SU2-1983},
while for $SU(3)$ the peak is around $\beta\sim 5.4$ as we see, for example, in Fig.4a of
\cite{KIGSMT-SU3-1983}.
For $SU(4)$ the dip in the mass gap can be seen in Fig.1 of
\cite{BLMT_N}
and is located  around $\beta\sim 10.45$. For larger $N$ the transition has been shown in
\cite{BLMTUW_Tc}
to be first order, and strongly first order for $N\geq 6$.
In that case we have a strong hysteresis, i.e. if we lower $\beta$ slowly from large values 
then the transition occurs at $\beta=\beta_b^{\downarrow}$, while if we increase  $\beta$ slowly
from small values then the transition occurs at $\beta=\beta_b^{\uparrow}$, with 
$\beta_b^{\uparrow}$ significantly larger than $\beta_b^{\downarrow}$.
Values for these transitions can be found listed in Table 16 of 
\cite{BLMTUW_Tc}.
As described in that paper, calculations performed in the weak coupling false vacuum just
above $\beta=\beta_b^{\downarrow}$, and well below $\beta=\beta_b^{\uparrow}$, show no sign of
being affected by the simultaneous presence of the deeper true vacuum elsewhere
in field space. This fact has been exploited in previous large $N$ calculations.
For example all but one of the $\beta$ values in the $SU(8)$ calculations in
\cite{BLMTUW_N}
lie within this hysteresis window, as does the $SU(8)$ calculation in
\cite{BLARER_N}.
As is clear from Tables~\ref{table_param_SU2}-\ref{table_param_SU12}
all our calculations have been performed on the weak coupling side
of the cross-over or transition. In addition we note that
none of our $SU(5)$ or  $SU(6)$
calculations fall within the hysteresis window, only one of our $SU(8)$
calculations lies within this window, while some of the $SU(10)$ and about
half of the $SU(12)$ calculations lie within the window.

As remarked above, the main effect of the crossover at smaller $N$ is an anomalous
dip in the value of the mass gap, i.e. that of the lightest scalar glueball $am_{0^{++}}$.
This dip is caused by a nearby critical point which lies at the end of a first-order
transition line in an extended fundamental and adjoint coupling plane,
at which critical point the scalar glueball mass vanishes. For a detailed
calculation in $SU(3)$ see for example
\cite{Heller_bulk}.
(The fundamental coupling is the one that which appears in our path integral.)
As $N$ increases the first-order line intersects the fundamental axis and provides the
first-order bulk transition separating weak and strong coupling.
However until the critical point has moved far away from the fundamental
axis there may still be a dip in the mass gap near the first-order transition.
Indeed one can see some plausible evidence for this occuring in the behaviour of
the $SU(6)$ and $SU(8)$ mass gaps listed in 
\cite{BLMTUW_N}.
Since there is no theoretical reason to expect that this dip can be encoded in
the weak coupling expansion of the lattice action in powers of $a^2$, we minimise
the risk to the continuum extrapolation of the scalar glueball mass by simply excluding
from the continuum fit the value obtained at the largest value of $\beta$ when
including this value would require adding an additional $a^4$ correction to the fit.
As it turns out, we need to do this for all our $SU(N)$ groups except for $SU(2)$.

\subsection{continuum masses}
\label{subsection_massratios} 

For each of our $SU(N)$ lattice gauge theories  we now have the low-lying glueball spectra
for a range of values of $a(\beta)$. These are all given in lattice units as $aM$, and to
transform that to physical units we can take the ratio to the string tension, $a\surd\sigma$,
that we have simultaneously calculated. We can then extrapolate this ratio to the
continuum limit using the standard Symanzik effective action analysis
\cite{Symanzik_cont}
that tells us that for our lattice action the leading correction at tree-level is $O(a^2)$:
\be
\frac{aM(a)}{a\surd\sigma(a)} = \frac{M(a)}{\surd\sigma(a)}
=
\frac{M(0)}{\surd\sigma(0)} + a^2\sigma(a) + O(a^4).
\label{eqn_MKcont}
\ee
Here we have used the calculated string tension, $a^2\sigma(a)$, as the $O(a^2)$
correction. Clearly we could use any other calculated energy, and this would
differ at $O(a^4)$ in eqn(\ref{eqn_MKcont}). It is convenient to use $a^2\sigma(a)$
since its calculated value has very small errors.

The results of these continuum extrapolations are listed in
Tables~\ref{table_MK_R_SU2}-\ref{table_MK_R_SU12} for the gauge groups ranging
from $SU2)$ to $SU(12)$. (As always, each $E$ doublet, and each $T_2$ or $T_1$
triplet appears as a single state in our tables and discussions.)
A few entries are accompanied by a star denoting a
poor fit, $2.5 < \chi^2/n_{df} < 3.5$, or a double star denoting a very poor fit,
$\chi^2/n_{df} \geq 3.5$. For $SU(10)$ and $SU(12)$ we have only 5 values of $\beta$
and since the masses at the coarsest value of $\beta$ often have to be discarded
from the fit (not surprisingly) and since we are fitting 2 parameters,
we often have only 2 degrees of freedom. For lower $N$ we usually
have 3 and sometimes 4 degrees of freedom, which gives much more confidence in
the extrapolations. All of which is to say that while the fits  for larger $N$ are
certainly not trivial, it would be good to be able to do better in future calculations.

We now illustrate the quality of these linear continuum extrapolations for our most interesting
glueball states. We do so for $SU(4)$. In Fig.\ref{fig_MJ02ppK_cont_SU4} we show our
extrapolations of the lightest two $A_1^{++}$, $E^{++}$ and  $T_2^{++}$ states.
These states are of particular importance because, as we shall see latter on, they correspond
to the lightest two $J^{PC}=0^{++}$ and $2^{++}$ states. We see that all the linear fits
are convincing, even if in some cases we have to exclude from the fit the value at the
largest $a(\beta)$. In Fig.\ref{fig_MJ02mpK_cont_SU4} we show the corresponding plot for $P=-$
which, as we shall see latter on, correspond to the lightest two $J^{PC}=0^{-+}$ and $2^{-+}$
states. The lightest states have very plausible continuum extrapolations, although
the excited states, which are heavier than those for $P=+$, begin to show a
large scatter indicating a poor fit. In Fig.\ref{fig_MJ1K_cont_SU4} we show the
extrapolations of various $T_1^{PC}$ states that we shall later argue correspond to $J=1$,
and again we see fits that appear convincing for the lighter states and quite
plausible for the heavier states. 
Even for the heaviest states to which we can plausibly assign a continuum spin, such as the
states that pair up to give the  $J^{PC}=2^{--}$ and  $2^{+-}$ ground states and the states
that provide the seven components of the $3^{+-}$ ground  state, where the
larger errors lead to a greater scatter of points, continuum fits linear in $a^2\sigma$
are still plausible.

\subsection{$N\to\infty$ extrapolation}
\label{subsection_masslargeN} 

Amongst the various $SU(N)$ calculations, the one that is most interesting
from a phenomenological point of view is the $SU(3)$ one, and that is why we
devoted a separate paper to that case
\cite{AAMT-2020}.
From a theoretical point of view however the most interesting glueball spectra
are those of the $SU(N\to\infty)$ theory since the theoretical simplifications
in that limit make it the most likely case to be accessible to analytic
solution, whether complete or partial.

To obtain the $N=\infty$ spectrum from our results so far, we use the fact that
in the pure gauge theory the leading correction is $O(1/N^2)$. So we can extrapolate
the continuum mass ratios in Tables~\ref{table_MK_R_SU2}-\ref{table_MK_R_SU12}
using
\be
\left.\frac{M_i}{\surd\sigma}\right|_{N}
=
\left.\frac{M_i}{\surd\sigma}\right|_{\infty}
+ \frac{c_i}{N^2} + O\left(\frac{1}{N^4}\right).
\label{eqn_MKN}
\ee
The results of these extrapolations are presented in Table~\ref{table_MK_R_SUN}.
In this table the stars point to poor fits exactly as described earlier for the continuum fits.
The very poor fit for the first excited $A_1^{++}$ is due to a large mismatch between
the $SU(5)$ and $SU(6)$ mass estimates which may well be due to an inadequate treatment
of the ditorelon influence. (We can obtain a good fit to $N\geq 6$ and this
gives a mass $\sim 5.85(9)$ which  is the same within errors.)
Most of the fits are to $N\geq 2$ or $N\geq 3$ but some are over a more restricted range of
$N$ and this is indicated by a dagger.These states are the  $A_2^{++}$ ground state, fitted to $N\geq 4$,
the $T_2^{-+}$ second excited  state, also fitted to $N\geq 4$, the $T_2^{--}$ ground  state,
again fitted to $N\geq 4$, and finally the $A_2^{+-}$ ground state which is more arguable
since it was fitted to $N\leq 8$. From the practical point of view the most important
extrapolations are for those states to which we are able to assign a continuum spin
(in the next section). We therefore show these extrapolations in
Figs.\ref{fig_M0pp0mpK_N},\ref{fig_MJ2PCK_N},\ref{fig_MJ1K_N}
for states with $J=0,2,1$ respectively.

\subsection{continuum spins}
\label{subsection_spins} 

So far we have used the representations of the rotational symmetry of our cubic spatial
lattice to label our glueball states. However as we approach the continuum limit these
states will approach the continuum glueball states and these belong to representations of
the continuum rotational symmetry, i.e. they fall into degenerate multiplets of $2J+1$ states
where $J$ is the spin. In determining the continuum limit of the low-lying glueball spectrum,
it is clearly more useful to be able to assign the states to a given spin $J$, rather
than to the representations of the cubic subgroup which have a much less fine `resolution'
since all of $J=0,1,2,...,\infty$ are mapped to just 5 cubic representations. 
How the $2J+1$ states for a given $J$ are distributed amongst the representations of the
cubic symmetry subgroup is given, for the relevant low values of $J$, in Table~\ref{table_J_R}.
So, for example, we see that the seven states
corresponding to a $J=3$ glueball will be distributed over a singlet $A_2$, a degenerate
triplet $T_1$ and a degenerate triplet $T_2$, so seven states in total. These $A_2$, $T_1$ and $T_2$
states will be split by $O(a^2)$ lattice spacing corrections, generated by irrelevant operators.
So once $a$ is small enough these states will be nearly degenerate and one can use this
near-degeneracy to identify the continuum spin. 

This strategy works for the lightest states but becomes rapidly unrealistic for heavier
states. The latter will have larger statistical errors and will fall amongst other states
that become more densely packed as the energy increases, so that identifying apparent
near-degeneracies between states in different representations becomes highly ambiguous.
There exist more sytematic ways of assigning continuum spin to lattice states, such as
\cite{HMMT-2004,HM_Thesis,PCSDMT-2019},
but these are beyond the scope of this work. Accordingly we shall limit ourselves to the
lightest states where any ambiguity in identifying near-degeneracies is either small
or non-existent.

The states whose spin $J$  we feel confident in identifying are listed in
Table~\ref{table_M_J_R} and the resulting continuum masses are listed in
Table~\ref{table_MKJ_N2-5} and Table~\ref{table_MKJ_N6-12}.
We will now briefly illustrate the argument for the assignments,
taking $SU(8)$ as a typical example. Consider the masses in the $SU(8)$ column of
Table~\ref{table_MKJ_N6-12}. We have obtained these by interpreting the masses listed in
Table~\ref{table_MK_R_SU8} as follows. The ground state of $R^{PC}=A_1^{++}$ is much lighter than
any other mass and so must be the singlet $J^{PC}=0^{++}$ ground state. The first excited
$A_1^{++}$ state certainly has no `nearly-degenerate' partners amongst all the other $P,C=+,+$
states and so it must be the first excited $0^{++}$ glueball. The ground states of the
doubly degenerate $E^{++}$ and the triply degenerate $T_2^{++}$ states are nearly degenerate
and there are no other states with similar masses, so together they must provide the
five components of the $J^{PC}=2^{++}$ ground state. Similarly the first excited
$E^{++}$ doublet and $T_2^{++}$ triplet have no nearly degenerate companion states elsewhere,
and so they provide the first excited $J^{PC}=2^{++}$ state. The argument for the
ground and first excited $0^{-+}$ states is equally straightforward, as it is for
the $E^{-+}$ and the $T_2^{-+}$ ground states forming the $2^{-+}$ ground state.
The first excited $E^{-+}$ and the $T_2^{-+}$ states are consistent with forming the
first excited  $2^{-+}$ state, but here there is some ambiguity: it is also possible that
these states pair with the second excited $A_1^{-+}$ state and the $T_1^{-+}$ ground state
to make up the  $4^{-+}$ ground state. This ambiguity arises because the errors are
sufficiently large that both possibilities can be entertained. Here we can resolve the
ambiguity by observing that for neighbouring values of $N$, $SU(6)$ and  $SU(10)$,
the ambiguity is not present and the first excited $E^{-+}$ and the $T_2^{-+}$ states
form the first excited  $2^{-+}$ state, strongly suggesting that this is also the
case for $SU(8)$. This choice is not as solid as the earlier choices and fortunately
it is not an argument we need to make very often in coming to our choices of $J$.
With this choice we can plausibly assign the lightest $T_1^{-+}$ state
to the $1^{-+}$ ground state. (The second excited $A_1^{-+}$ state has a similar mass
but there are no further partners allowing us to assign it elsewhere according
to Table~\ref{table_J_R}.) Similar arguments to the above lead to the $J^{+-}$
and $J^{--}$ choices in  Table~\ref{table_M_J_R} and the corresponding masses
in Table~\ref{table_MKJ_N6-12}.
Determining the spin $J$ in the above way, clearly requires us not to miss any
intermediate states in the mass range of interest. For this one needs a
good enough overlap onto all the low-lying states and for that one needs a large
basis of operators, which has been our goal in these calculations. 

Using the arguments sketched above we infer the $J^{PC}$ glueball masses
listed in Tables~\ref{table_MKJ_N2-5},\ref{table_MKJ_N6-12}, from the
various masses listed in Tables~\ref{table_MK_R_SU2}-\ref{table_MK_R_SU12}.
In addition by inspecting the remaining states in
Tables~\ref{table_MK_R_SU2}-\ref{table_MK_R_SU12} we can infer some
lower bounds on the ground states in some other channels, and these
are listed in Table~\ref{table_MKJ_lowbound}. These estimates are
fairly rough and at that level apply to all our $SU(N)$ theories.

The most interesting glueball spectra are those for $SU(3)$ and for $SU(\infty)$:
the former for its potential phenomenological relevance and the latter for
its potential theoretical accessibility. 
The $SU(3)$ theory is sufficiently close to the real world of $QCD$ that it makes
sense to attempt to present the masses in physical units, using scales that
one believes to be relatively insensitive to the presence of quarks. We refer the
reader to
\cite{AAMT-2020}
for a more detailed discussion. Here we simply recall from our earlier discussion
that the Sommer scale $r_0$,
which is defined by the heavy quark potential at intermediate distances, is
believed to vary weakly with quark masses, with a physical value of
$r_0 \simeq 0.472(5)\mathrm{fm}$
\cite{Sommer-r0b}.
One can extract from the published data the value $r_0\surd\sigma=1.160(6)$
\cite{AAMT-2020}
This allows us to translate masses in units of $\surd\sigma$ to units of $r_0$
and finally to units of $\mathrm{GeV}$. Doing so we obtain the masses listed in
Table~\ref{table_MJ_N3}. This Table is similar to that in
\cite{AAMT-2020}
except that we have included lower bounds for most of the masses where
we previously lacked entries. 

We now turn to our main focus: the $N\to\infty$ limit.
We extrapolate the masses in Tables~\ref{table_MKJ_N2-5},\ref{table_MKJ_N6-12}
with $O(1/N^2)$ corrections, as in eqn(\ref{eqn_MKN}).
The extrapolations for most of these states are displayed
in Figs.\ref{fig_M0pp0mpK_N},\ref{fig_MJ2PCK_N},\ref{fig_MJ1K_N}.
This leads to the masses (in units of the string tension) listed in
Table~\ref{table_MK_J_SUN}. In making these extrapolations we have sometimes 
excluded the mass for the smallest value of $N$ ($N=2$ for $C=+$ and $N=3$ for $C=-$),
in order to achieve a significantly better fit. This we have done for the $2^{++}$,
$1^{-+}$ and $1^{--}$ ground states. This presumably reflects the need for at
least a further $1/N^4$ correction, but our number of data points is too small
for this. In some cases we obtain better fits by excluding the $N=12$ values,
and although this could be argued for on the basis that our range of $a$ for the
continuum extrapolation is relatively limited in that case, the danger of
`cherry-picking' has led us to avoid doing so.

The main features of these continuum spectra are that the lightest glueball
is the $J^{PC}=0^{++}$ scalar, with the $J^{PC}=2^{++}$ tensor about $50\%$ heavier,
and with the  $J^{PC}=0^{-+}$ pseudoscalar just slightly above that. The next
state is the $J^{PC}=1^{+-}$ vector, which is special in that it is the only
relatively light $C=-$ glueball, with most $C=-$ states being very heavy.
Just a little heavier than this vector is the first excited $0^{++}$ and then the
$2^{-+}$ ground state. The $N$ dependence of nearly all these states is weak and readily
absorbed into a $O(1/N^2)$ correction down to at least $SU(3)$; all this confirming
that the $SU(3)$ gauge theory is indeed `close to' $SU(\infty)$.

\subsection{scattering states}
\label{subsection_scattstates}

A given operator will project onto all states with the quantum numbers of that operator.
In particular our single trace glueball operators will project onto multi-glueball states
in addition to the glueballs that we are interested in. This means that some of the states
in the `glueball' spectra that we have calculated may in fact be multi-glueball
states. In the ideal case of a very large spatial volume these will be scattering states
where at sufficiently large times the states are far apart

There is some ambiguity here of course. Sufficiently heavy glueballs
will be unstable, and will decay into multiglueball states but to the extent
that their decay width is not too large, we still regard them as glueball states.
Indeed as $N$ increases the decay widths and the overlap of multiglueball
states onto our single trace basis of glueball operators
should decrease. So although we expect
this to be an issue primarily at smaller $N$, it is clearly important to address.
We do so here in some detail because it has been addressed only briefly and
occasionally in past glueball calculations.

A full analysis of scattering states and glueball decays would involve different
techniques to those used in this paper and so are outside the scope of this work.
Here we will, instead, perform some exploratory calculations to obtain an indication
of the impact of multiglueball states on our calculated glueball spectra. We limit
ourselves to considering states consisting of two glueballs since these are the lightest
ones and our calculated glueball spectra do not extend to masses that are much higher.
If there is weak mixing between double trace and single trace operators then we can
introduce two glueball scattering states by using double trace operators, such as
\be
\phi_{ab}(t) = (\phi_a(t)-\langle \phi_a\rangle)(\phi_b(t)-\langle \phi_b\rangle)
-\langle(\phi_a-\langle \phi_a\rangle)(\phi_b-\langle \phi_b\rangle)\rangle.
\label{eqn_phiab}
\ee
Here $\phi_a$ and $\phi_b$ are single trace operators. The vacuum subtractions ensure
that $\phi_{ab}$ does not have a projection onto the vacuum or a `trivial' projection
onto single glueball states through terms such as $\langle \phi_a\rangle \phi_b(t)$.
One can calculate the mass spectrum obtained when one adds such double trace operators
to the basis of single trace operators and compare it to that obtained using just the
single trace operators. If the resulting spectra are the same then this strongly
suggests that the spectrum obtained using single trace operators already includes
two glueball states. If, on the other hand, the spectrum obtained using the expanded basis
produces extra states that can be plausibly interpreted as two glueball scattering states,
then this suggests that the states obtained with our original basis of single trace
operators does not contain such states and that categorising them as single glueball
states is probably correct.

The potential number of operators $\phi_{ab}$ is clearly too huge to be practical
here and so we severely limit the number for our study as follows. Firstly, we take
both $\phi_a$ and $\phi_b$ to have zero momentum, so both the total and relative
momenta are zero. Of course this does not mean that the relative momentum of the
two glueballs has to be zero, since the relative momentum is not a conserved
quantum number in an interacting system, but one naively expects that the main overlap
will be onto zero relative momentum. Secondly we only keep the 2 or 3 blocking
levels that are most important at the $\beta$ we calculate. (We systematically avoid using
the largest blocking levels so as to exclude unwanted ditorelon states.) Thirdly, we take the
same blocking levels for $\phi_a$ and $\phi_b$. For $\phi_a$ we use only the rotationally
invariant sum of (blocked) plaquettes. That is to say it is in the $A_1^{++}$
representation. So $\phi_{ab}$  will be in the same representation as $\phi_b$. For
$\phi_b$ we take 3 different loops chosen so that we can have projections onto all
representations including $P=\pm$ and $C=\pm$. The lightest energy of the
corresponding asymptotic two glueball state would be the sum of the lightest
$A_1^{++}$ mass plus the lightest mass in the representation of $\phi_b$
if we were in an infinite spatial volume. Since our spatial volume is finite
the glueballs are interacting at all times and there will be a shift in the total
energy. However, to the extent that our spatial volume is not very small, this
shift should be small and we will use the naive sum as the quantity against
which to compare our supposed scattering states.

We perform calculations in $SU(3)$, which is the most physically interesting case
amongst our lower $N$ calculations, and in $SU(8)$ which is representative of our
large $N$ calculations. In $SU(3)$ we work at $\beta=6.235$ on the same $26^4$ lattice
used in our above glueball calculations. For completeness we have also carried out
some calculations on $18^326$ and $34^326$ lattices at the same value of $\beta$.
In $SU(8)$ we work at $\beta=46.70$ on a $16^324$ lattice which, again, is the same
as that used in our glueball calculations. The spatial volumes at larger $N$ are
smaller than those at smaller $N$, taking advantage of the expected suppression of
finite volume corrections with increasing $N$, and this is the reason for this
additional calculation in  $SU(8)$.

We calculate correlators of the double trace operators with each other as
well as with the single trace operators, since all of these are needed in the
variational calculation using the basis that combines single and double trace
operators. Because we need high statistics for glueball calculations, we calculate
the correlators at the same time as we generate the lattice fields. At that stage
we do not know the values of the vacuum subtractions in eqn(\ref{eqn_phiab})
and so we need to calculate a number of other correlators so that at the later
analysis stage we can reconstruct the desired subtracted correlators. The relevant
expressions are given in Appendix~\ref{section_appendix_scattstates}.

We begin with our $SU(3)$ calculation. In Fig.~\ref{fig_MeffG+GGA1++l26n_SU3} we
show the effective masses of the lightest few states in the $A_1^{++}$ representations
obtained using the basis of single trace operators (open circles) and the same basis
extended with our double trace operators (filled points). We see quite clearly
that the two sets of states match each other well except for one state, whose effective
masses are represented by a $\blacklozenge$, and which clearly has no partner amongst
the states from the single trace basis. This state appears to asymptote, at larger
$t$, to an energy that is close to twice the lightest $A_1^{++}$ mass, i.e. that of the
lightest free two glueball state, and so we infer that this is indeed the
lightest two glueball state. Apart from this state, the lightest 4 states in the
two bases match each other quite precisely. That is to say, we have good evidence
that the $A_1^{++}$ continuum masses listed in Table~\ref{table_MK_R_SU3} are indeed
those of single glueball states.

It is interesting to ask what kind of $A_1^{++}$ energy spectrum one obtains if one
uses a basis consisting only of the double trace operators. The result is shown
in Fig.~\ref{fig_MeffGGA1++l26n_SU3}. We see that the state that is lightest at small $t$
is consistent with being the lightest 2 glueball state, and that our variational
wave-functional has a good overlap onto that state. What is equally interesting is
the state that is the first excited state at small $t$ and whose effective mass
drops monotonically below that of our two glueball state as $t$ increases, consistent
with heading towards the mass of the lightest glueball. This presumably represents
the overlap of our double trace basis onto the lightest glueball. A rough estimate suggests
that this variationally selected operator has an overlap (squared) of $\sim 10\%$
which again points to little mixing of scattering states into our low-lying
glueball spectrum.

Clearly we would like to repeat our above analysis of the  $A_1^{++}$ spectrum for all 
the other representations. However once the lightest state in some $R^{PC}$ representation
is much heavier than in the above case the energy of the predicted scattering state
lies in a dense part of the spectrum, all the energies have much larger statistical errors
and we cannot follow the effective energies beyond the smallest values of $t$. That is to
say, the analysis becomes hopelessly ambiguous. We will therefore restrict ourselves to
a few cases where the energies are still manageably small. That is to say, we will
look at the $T_2^{++}$ which contains 3 components of the lightest $J=2^{++}$ state,
the  $A_1^{-+}$ where the lightest state is the interesting $0^{-+}$ pseudoscalar
(the lightest $P=-$ glueball),
the  $E^{-+}$ where the lightest state is the $2^{-+}$ pseudotensor, and the $T_1^{+-}$
which is the lightest $C=-$ state and where the lightest 3 component state is the
$C=-$ vector. We do not analyse any $R^{--}$ states since they are all too heavy.

We begin with the $T_2^{++}$ representation. Here our double trace operator
would project onto a two glueball state consisting of the lightest $A_1^{++}$
and $T_2^{++}$ glueballs if there were no mixing. Neglecting interactions
its energy is just the sum of these masses. This case differs from the $A_1^{++}$
representation discussed above because this is not the lightest scattering state
one can construct. Two $A_1^{++}$ glueballs with two units of angular momentum
are lighter, although one can expect an angular momentum threshold suppression
factor that would need to be quantified. Such a state requires the use of operators
with non-zero momentum which would mean extending our basis of operators well beyond
our choice for this exploratory study. It is plausible that what we learn using
our heavier two glueball states would also apply to these lighter ones. So
we carry out for the $T_2^{++}$ representation the analogue of the analysis
in Fig.~\ref{fig_MeffG+GGA1++l26n_SU3} for the $A_1^{++}$. Once again we find
that the inclusion of our double trace operators leads to an extra state
although because of the denser packing of the states it is less
prominent than in the  $A_1^{++}$ case. However the effective energy of this state is
consistent with decreasing towards that of a scattering state composed of the lightest
$A_1^{++}$ and $T_2^{++}$  glueball, albeit with substantial statistical uncertainty.
So it is very plausible that it is a scattering state composed of the lightest  $A_1^{++}$
and $T_2^{++}$ glueballs.
The spectrum from the double trace basis
shows quite clearly that the state that is lightest at small $t$ has an energy
roughly equal to the sum of the lightest $A_1^{++}$ and $T_2^{++}$ glueballs, showing
it to be a scattering state. It is also clear that the projection of these
double trace operators onto the lightest $T_2^{++}$ glueball is very small.
All this makes it plausible that the five or six lightest states in the $T_2^{++}$ channel
are single glueballs. However we recall our above caveat that there are lighter scattering
states that could, in principle, behave differently.

With the $T_1^{+-}$ representation our double trace operators project onto the
lightest two glueball state since the lightest  $T_1^{+-}$ is the lightest $C=-$ state.
Here we again find 
that the addition of the double trace operators to the single trace basis does
produce an extra state and that the effective energy of that state appears
to be decreasing towards the energy of the lightest scattering state, up to the
point where the errors grow too large to allow a statement. From this
we infer
that the lightest five $T_1^{+-}$ states obtained using the single trace basis are
not scattering states. In addition we find that in the
spectrum obtained using just the double trace operators the
lightest state appears to approach the expected scattering state and that
there apears to be no significant component of the lightest single glueballs
present in this basis, providing further evidence for little overlap
between our single and double trace operators.

In the case of the  $A_1^{-+}$ representation our double trace operators naturally
project onto a state with both an $A_1^{++}$ and an $A_1^{-+}$ glueball. Just as
for the $A_1^{++}$ this should be the lightest possible scattering state: although
one can obtain $P=-$ through unit angular momentum one would need a glueball
with non-zero spin to take us back to an overall $A_1$ state.
We show in Fig.~\ref{fig_MeffG+GGA1-+l26n_SU3}
how the inclusion of our double trace operators affects the spectrum. While the
ground and first excited states are unaffected, there is clearly an extra state
just above these, and a significant shift in the mass of the next state above this.
The effective energy of the (probable) extra state descends rapidly towards
the energy of the free scattering state, but the rapidly growing errors prevent
us from telling if it asymptotes to this value or continues to decrease.
For further evidence we plot in Fig.~\ref{fig_MeffGGA1-+l26n_SU3} the spectrum
obtained in the basis with only the double trace operators. Here we have a quite
dramatic contrast to what we have seen above in the $A_1^{++}$, $T_2^{++}$
and $T_1^{+-}$ representations: the lightest state is clearly the ground state
of the single trace spectrum. This tells us that here we have a large overlap
between single trace and double trace operators. So here we can have legitimate concerns
about the presence of multiglueball states in the low-lying spectrum obtained
using single trace operators.

The $E^{-+}$ representation contains the second lightest state in the $P,C=-,+$
sector, after the $A_1^{-+}$  ground state. Neglecting interactions, our double trace
operators would project onto the two glueball states composed of an $A_1^{++}$ and
an $E^{-+}$ glueball. There are scattering states that are slightly lighter
but they should have angular momentum threshold factors that effectively
cancel that advantage. (For example, the lightest  $A_1^{++}$ and $A_1^{-+}$
glueballs with two units of angular momentum.)
We find, when plotting the effective masses,
that the addition of the double trace operators to the single trace basis does
produce an extra state. However this state is nearly degenerate with another state
so it is unclear which is the extra one. (It is quite possible that our variational
procedure mixes the two.) In either case the effective energy of that state appears
to be decreasing towards the energy of the lightest scattering state, up to the
point where the errors grow too large to allow a statement. The spectrum of
states with just the double trace basis
shows quite clearly that the lightest state is the scattering state and that there
is no significant overlap of our double trace operators onto the low-lying spectrum
obtained with the single trace operators.

In summary, the above study in $SU(3)$ shows that two glueball states have very little
overlap onto most of the spectrum one obtains from single trace operators for the
lighter and hence most interesting states in our calculated glueball spectra. This is
something of a surprise in $SU(3)$ since one would not expect to be able to
invoke a large-$N$ suppression for $N=3$. The one apparent exception concerns
the $A_1^{-+}$ representation which contains the $0^{-+}$ pseudoscalar which
is of particular theoretical interest. This clearly merits a much more detailed
investigation.

We now turn to a similar study in $SU(8)$. This we should do since for $N\geq 8$
we use considerably smaller spatial volumes than for $SU(3)$, taking advantage of the
expected large-$N$ suppression of finite volume corrections. We perform our study on
a $16^324$ lattice at $\beta=46.70$ which corresponds to a lattice spacing that
is slightly larger than that of the above $SU(3)$ study if measured in units of the
string tension, $\sigma$, although equal when measured in units of the mass gap.
Our results are very much the same as for  $SU(3)$:
in the $A_1^{++},T_2^{++},T_1^{+-}$ and $E^{-+}$ representations there appear to
be very small overlaps between our would-be two glueball operators and the
single trace operators in the low-lying part of the spectrum that is of interest
to us in this paper. The striking exception, as in the case of $SU(3)$, is the 
$A_1^{-+}$ representation. This is intriguing both because this representation
contains the interesting $J^{PC}=0^{-+}$ pseudoscalar glueball and because one
would have naively expected the large-$N$ suppression of such overlaps between
single and double trace operators to have taken effect by $N=8$. There is
perhaps some hint of this in comparing the approaches to the respective $A_1^{-+}$ masses
of the ground states.

These results for $SU(3)$ and $SU(8)$ provide some reassurance that our spectra
in the glueball calculations of this paper are indeed those of single glueballs,
at least for the lower lying spectrum that is of the greatest physical and
theoretical interest.

\subsection{some comparisons}
\label{subsection_comparisons}

It is useful to compare our glueball results against those of earlier calculations both as
a check on the reliability of our errors, particularly the systematic errors that are the most
difficult to control. To reduce the comparison to manageable proportions we limit
ourselves to the continuum limit of the $J^{PC}=0^{++},2^{++},0^{-+},2^{-+},1^{+-},3^{+-},2^{--}$
ground states and to calculations that have appeared after the year 2000. The first five states
are the lightest ground states and are easy to identify, while the sixth involves a more
challenging identification of near-degeneracies amongst the cubic irreducible representations,
and the last is an example of the much heavier states over which we have less control.
We will also limit ourselves to $SU(3)$, where there are several calculations, and to $SU(8)$
as a representative of our larger $N$ calculations.

We list the $SU(3)$ and $SU(8)$ comparisons in Tables~\ref{table_MK_J_SU3_comp} and
\ref{table_MK_J_SU8_comp}. The calculations in
\cite{BLMT_N,BLMTUW_N,HM_Thesis}
are broadly of the same type as those in the present paper, making a comparison straightforward.
The calculations of
\cite{BLMT_N,BLMTUW_N}
were designed to explore the approach to the large $N$ limit and focussed upon the masses of
the $J^{PC}=0^{++},2^{++}$ ground states and the string tension. The primary goal of
\cite{HM_Thesis}
was to calculate the masses of higher spin glueballs which cannot be identified using the
naive approach of searching for near-degeneracies, as used in the present paper, and this required
the application of novel dedicated techniques. The calculations in 
\cite{MP-2005}
are significantly different to all of these, in that they use an alternative `improved' lattice
action and anisotropic lattices with a temporal lattice
spacing, $a_t$, smaller  than the spatial lattice spacing, $a_s$, by a factor of (roughly) 3 or 5.
In addition the continuum glueball masses are expressed in units of the Sommer scale $r_0$
rather than the string tension $\sigma$. We have used the value $r_0\surd\sigma = 1.160(6)$,
as fitted in
\cite{AAMT-2020},
to translate the values given for $r_oM_G$ in 
\cite{MP-2005}
to the values $M_G/\surd\sigma$ listed in Table~\ref{table_MK_J_SU3_comp}.

We begin with the $SU(3)$ comparison in Table~\ref{table_MK_J_SU3_comp}. Overall we see
that the results of the different calculations agree within about 2 standard deviations for
all the states. In particular the values in 
\cite{HM_Thesis}
are within one standard deviation of our values. On the lighter ground states our errors
are far smaller than those of the earlier calculations. However on the two heaviest states
the errors in
\cite{MP-2005}
are similar to ours, demonstrating the expected advantage of highly anisotropic lattices
for calculating the masses of heavy glueballs. Focussing now on possible differences,
we note that all the mass estimates from
\cite{MP-2005}
are slightly above ours, and indeed above those of
\cite{HM_Thesis}
It is possible that this has to do with the scale $r_0$. This scale is determined as the
spatial distance at which the heavy quark potential has a certain value of its derivative.
Thus its value is expressed in units of the spatial lattice spacing, i.e. as $r_0/a_s$,
and this has to be expressed as  $r_0/a_t$ in order to serve as a scale for a lattice
glueball mass, $a_tM_G$. This rescaling is clearly sensitive to even small shifts of the
anisotropy from the tree level values of 3 or 5. Moreover the values of $a_s$ used in
\cite{MP-2005}
are large which means that $r_0/a_s$ is not large, and this makes the accurate determination
of its value more delicate. So it is possible that all this is enough to lead to a $1-2\%$
shift in the estimates of $M/\surd\sigma$ in this calculation. Turning now to
\cite{BLMT_N,BLMTUW_N}
we see that the $0^{++}$ mass is about two standard deviations higher than our value
(or that of 
\cite{HM_Thesis}
).
A possible source of this is the fact that these older calculations have a more limited
range of higher $\beta$ values and include in their continuum extrapolations the smaller values
of $\beta$ where there is some effect of the dip in the mass gap associated with the
strong-to-weak coupling bulk transition, as discussed in Section~\ref{subsection_bulk}.
This dip in the mass at larger lattice spacings may then bias the straight line fit
in $a^2\sigma$ so that its $a=0$ intercept is a little too high. In the present
calculations the statistical errors are small enough that a good straight line fit
is no longer possible if one includes such large $a$ values and by excluding them any
such bias becomes smaller and the continuum intercept is a little lower. This discussion
is only relevant to the mass gap, since the other glueballs appear to be insensitive
to the bulk transition and have, for the most part, a small dependence on the
size of the lattice spacing.

Turning now to the $SU(8)$ comparison in Table~\ref{table_MK_J_SU8_comp}, we see that
for the states other than the $0^{++}$ the masses from the different calculations agree
within one standard deviation, although the errors on the earlier calculations are typically
about ten times larger than ours, leaving room for potential discrepancies.
The mass gap shows a signficant discrepancy but both of the earlier calculations include
masses calculated close to the bulk transition at $\beta=\beta_b^\downarrow$ where
the mass gap appears to have a significant dip, as discussed in Section~\ref{subsection_bulk},
which may well explain this minor difference.

The above comparisons naturally raise the question of why we chose to use the traditional Wilson
plaquette action instead of some alternative `improved' action, such as the one used in 
\cite{MP-2005}.
In glueball calculations the main improvement observed, for $SU(3)$, is to
largely remove the dip in the mass gap as one transitions from strong to weak coupling.
This means that calculations at coarser lattice spacings, and so on smaller lattices, may be
used in the continuum extrapolation of the mass gap. However a larger lattice spacing
means that the correlator signals tends to disappear into the statistical noise before one
has a well defined effective mass plateau. To remedy this one can use an anisotropic
lattice action with the temporal spacing $a_t$ much less than the spatial $a_s$.
However using contemporary resources there is no need to worry about the dip in the mass gap:
one can easily do calculations far from the dip, at much weaker coupling, as we have seen
in this paper. Moreover if one works at coarse lattice spacings using such an improved action 
one does not know in advance whether the behaviour of other glueballs will be improved
or worsened. And in any case, as we have seen in
Figs~\ref{fig_MJ02ppK_cont_SU4},\ref{fig_MJ02mpK_cont_SU4},\ref{fig_MJ1K_cont_SU4},
the lattice corrections to the glueball masses are already quite modest with
the standard Wilson plaquette  action. An advantage of the simple plaquette action
is that it is known to have exact reflection positivity and hence the transfer matrix has
real positive eigenvalues between zero and one, just as if it was the exponential of
a Hamiltonian. This means that one does not need to be
cautious in applying a variational calculation of the glueball spectrum. Of course
one can use an anisotropic plaquette action, which will also possess reflection positivity.
Such a calculation would produce a finer resolution in time of the correlator and so
could be very useful for calculating  heavier glueball masses. However it has a disadvantage
in setting the scale as described just above. One needs to know the relation between
$a_s$ and $a_t$ and since the chosen tree level anisotropy will receive corrections, this need
to be estimated, and the error on this estimate may be significant. Thus our suggestion for a
better calculation than the present one is that one does two calculations in parallel. One 
calculation should be with an isotropic action, and this will provide very accurate continuum
values for the light glueballs, including the mass gap $M_G$, in units of, for example,
the string tension, $\sigma$, or the Sommer scale $r_0$. The second calculation should be
with a strongly anisotropic action and provided that one has a large enough basis of operators,
this should provide accurate calculations of the heavier glueball masses in units of, say,
the mass gap. Here it is important that the spatial lattice spacings should extend to small
values, just as in the isotropic calculations, in order to guarantee the credibility of the
continuum extrapolations. Putting together both calculations allows all the masses to be expressed
in units of $\sigma$ or $r_0$. The action used could be either the plaquette action or some
improved action, although maintaining reflection positivity is desirable from a practical
point of view.

\section{Topological fluctuations}
\label{section_topology}

Euclidean $D=4$ $SU(N)$ gauge fields possess non-trivial
topological properties, characterised by a topological charge $Q$ which is integer-valued 
in a space-time volume with periodic boundary conditions. This charge can be expressed
as the integral over Euclidean space-time of a topological charge density, $Q(x)$, where
\begin{equation}
Q(x)= \frac{1}{32\pi^2} \epsilon_{\mu\nu\rho\sigma}
\mathrm{Tr}\{F_{\mu\nu}(x)F_{\rho\sigma}(x)\}.
\label{eqn_Q_cont}
\end{equation}
Since the plaquette matrix $U_{\mu\nu}(x) = 1 + a^2 F_{\mu\nu}(x) + ....$ on sufficiently
smooth fields, one can write a lattice topological charge density $Q_L(x)$ on
such fields as
\begin{equation}
Q_L(x) \equiv {\frac{1}{32\pi^2}} \epsilon_{\mu\nu\rho\sigma}
\mathrm{Tr}\{U_{\mu\nu}(x)U_{\rho\sigma}(x)\}
= a^4 Q(x) +O(a^6).        
\label{eqn_Q_lat}
\end{equation}
However this definition lacks the reflection antisymmetry of the continuum operator
in eqn(\ref{eqn_Q_cont}), since all the plaquettes $U_{\mu\nu}(x)$ are defined as forward
going in terms of our coordinate basis, so to recover these symmetry properties
we use the version of this operator that is antisymmetrised with respect to forward
and backward directions
\cite{DiVecchia-FFD}
(although this is unnecessary on smooth fields).

The fluctuations of $Q_L(x)$ are related to the expectation value of the composite operator
$Q_L^2(x)$ whose operator product expansion contains the unit operator
\cite{DiVecchia-FFD},
so these fluctuations are powerlike in $1/\beta$. On the other hand, the average of
$Q_L(x)$ is $O(a^4)$ and
hence exponentially suppressed in $\beta$. Thus as $\beta$ increases the fluctuations
around $Q_L(x)$ and $Q_L=\sum_x Q_L(x)$ diverge compared to the physically interesting
mean values. In addition the composite operator $Q_L(x)$ also receives a multiplicative
renormalisation $Z(\beta)$ such that $Z(\beta) \ll 1$  at accessible values of $\beta$
\cite{Pisa_ZQ}.

In practice all this means that one cannot extract the topological charge of a typical lattice
gauge field by directly calculating $Q_L=\sum_x Q_L(x)$ on that gauge field. However we note that
the fluctuations obscuring the value of $Q$ are ultraviolet, while the physically relevant
topological charge is on physical length scales. Thus if we perform a very limited local smoothening
of the fields to suppress the ultraviolet fluctuations, this should not affect physics
on long distance scales, and the value of $Q_L=\sum_x Q_L(x)$ calculated on these smoothened
fields should provide a reliable estimate of $Q$. Moreover, recalling that the total
topological charge of a gauge field is unchanged under smooth deformations, we can expect
that even under a moderately large amount of continued smoothening the value of  $Q_L$ will not
change, even though $Q_L(x)$ itself does gradually change. One convenient way to smoothen the
gauge fields is to locally minimise the action. Such a `cooling' of the original `hot'
lattice gauge field
\cite{MT-cool}
involves sweeping through the lattice one link at a time, precisely like the Monte Carlo
except that one chooses the new link matrix to be the one that minimises the total
action of the plaquettes containing that link matrix. This is a standard technique
that one can find described in more detail in, for example,
\cite{DSMT-Q}.
An alternative and attractive smoothing method with perturbatively proven
renormalisation properties is the gradient flow
\cite{Luscher:2010iy,Luscher:2011bx,Luscher:2013vga}
which has been shown to be numerically equivalent to cooling
\cite{Bonati:2014tqa,Alexandrou:2015yba,Alexandrou:2017hqw}.
Cooling typically performs nearly two orders of magnitude faster than the gradient flow and
since we are aiming for large statistics we adopt the simplest cooling method.
After the first couple of cooling sweeps the fields are already quite smooth, as we shall see below.
Since we are minimising the action and since in the continuum the minimum action field with a given
$Q$ is a multi-instanton field, we expect that under systematic cooling the lattice field will be driven
to become some multi-instanton field, which one can see by calculating the distribution $ Q_L(x)$ on such
a field. Of course, because of the discretisation of space-time the topological properties of lattice fields
are only approximately like those of a continuum field. One can deform a large instanton by gradually
shrinking its non-trivial core and on a lattice this core can shrink to within a hypercube. At this
point what was an instanton has been transformed into a gauge singularity and the value of $Q$
will now differ from its original value by one unit. (Equally, one can gradually grow an instanton
out of a hypercube.) This process can occur during cooling but it can equally occur during
the course of our Monte Carlo. In the latter case, it is these changes in $Q$ that allow us
to sample all possible values of $Q$ and hence maintain the ergodicity in $Q$ of our Markov process.
As $\beta\uparrow$ the distance between physical and ultraviolet scales grows and these
changes in $Q$ become increasingly suppressed -- in fact more strongly than the usual critical
slowing down (see below). This is as it should be: after all, the changes in topological charge
are no more than a (useful) lattice artifact. 

We have outlined in Section~\ref{subsection_Qfreezing} the specific reasons for the suppression
of changes in $Q_L$ when $a(\beta)\to 0$ at fixed $N$ and when $N\to \infty$ at fixed $a(\beta)$,
and how we deal with this problem in our glueball and string tension calculations.
Below, in Section~\ref{subsection_Qcooling}, we will give some evidence that the value of $Q_L$
on a cooled lattice field does indeed reflect the topology of the original lattice field.
Then, in Section~\ref{subsection_Qtunneling}, we will give some results on the rate of critical
slowing down, both as $a(\beta) \to 0$ and as $N\to\infty$. In Section~\ref{subsection_Qsusc}
we present our results for the continuum topological susceptibility in those cases, $2\leq N\leq 6$,
where the topological freezing is not too serious an obstacle. Finally in Section~\ref{subsection_QZ}
we present our numerical results for the factor $Z(\beta)$ that relates the value of $Q_L$
before cooling to its value, $Q_I$, after cooling. Our study of lattice topology has some
limitations of course: it is not a dedicated study but is constrained by being a byproduct of our
glueball and string tension calculations. Finally we remark that it is only in the calculations
of the topological susceptibility that the topological freezing is an obstacle; in our other calculations
in this Section it does not matter whether the topological charge we study is introduced by hand
or appears spontaneously.

\subsection{topology and cooling}
\label{subsection_Qcooling} 

As we cool our lattice fields the fluctuations in the measured value of $Q_L$ decrease and
it rapidly settles down to a value that is close to an integer. As a typical example we plot in
Fig.\ref{fig_Qcool20_su5} the number of fields with a given topological charge $Q_L$ after
2 coolings sweeps and the number after 20 cooling sweeps, taken from sequences of $SU(5)$ fields
on a $16^320$ lattice at $\beta=17.63$. With respect to the calculations in this paper, this 
corresponds to an intermediate value of $N$, and of $a(\beta)$, and of the lattice volume. As we
see, after 20 cooling sweeps the values of $Q_L$ are very strongly peaked close to 
integer values. The reason that the peak is not at an integer, even when the cooling has
erased the high frequency fluctuations, is that for an instanton of size $\rho$ on a lattice,  
we only obtain $Q_L \to 1$ in the limit $\rho/a \to \infty$, with deviations from unity
that are powers of $a/\rho$. However, as is clear from Fig.\ref{fig_Qcool20_su5}, there is 
no significant ambiguity in assigning to each field after 20 cooling sweeps an integer valued 
topological charge which we label $Q_I$. We can expect $Q_I$ to provide our most reliable
estimate of the topological charge $Q$ of the original lattice gauge field. For the calculations
with the largest values of $a(\beta)$ identifying the value $Q_I$ can be
less  clear-cut, but such lattices are of little importance in determining continuum physics.
We also see from Fig.\ref{fig_Qcool20_su5} that even after only 2 cooling sweeps the fields
fall into groups that only overlap slightly. Since 2 sweeps cannot affect anything other 
than the most local fluctuations, we can assume that this segregation into differing topological 
sectors directly reflects the topology of the gauge fields prior to any cooling. 

As a second example we show in Fig.\ref{fig_Qcool20_su8} the same type of plot for several sequences 
of $SU(8)$ fields generated at $\beta=47.75$ on a $20^330$ lattice. This corresponds to our
smallest lattice spacing in $SU(8)$. Here we see an even sharper peaking after 20 cooling sweeps
and, more interestingly, a very clear separation between the various sectors even after only 2 cooling 
sweeps. In contrast to the $SU(5)$ example, in this case there is essentially no tunnelling between
topological sectors, and the observed distribution of $Q$ has been imposed on the starting
configurations for the various sequences.  

The loss of ergodicity with respect to the topological charge is illustrated for $SU(8)$
in Fig.\ref{fig_Qseq_su8} where we show the values of $Q_L$ after 2 and 20 cooling sweeps
taken every 100 Monte Carlo sweeps for two sequences of 50000 sweeps generated at $\beta=47.75$.
In one sequence we have $Q_L\sim 1$ and in the other we have $Q_L\sim 2$. It is interesting
that even after only 2 cooling sweeps the separation in the measured values of $Q_L$ is
unambiguous. A similar plot for $SU(5)$ fields generated at $\beta=17.63$ is shown in 
Fig.\ref{fig_Qseq_su5}. Here the value of $Q_L$ remains unchanged over subsequences that
are typically a few thousand sweeps long, but the changes are sufficiently frequent that
in our overall ensemble of $\sim 2\times 10^6$ fields we may regard $Q_L$ as ergodic.
But just as for $SU(8)$ the values of $Q_L$ after only 2 cooling sweeps track the
values after 20 cooling sweeps.

The close relationship that we observe in Fig.\ref{fig_Qseq_su8} and Fig.\ref{fig_Qseq_su5}
between the values of $Q_L$ after 2 and 20 cooling sweeps is confirmed explicitly in
Fig.\ref{fig_Qcool20c2_su8}. Here we have taken all the $\beta=47.75$ $SU(8)$ fields on which
we have calculated $Q_L$ and we have extracted the three subsets that correspond to
$Q_L \simeq 1,2,3$ after 20 cooling sweeps. For each subset we plot the values that
$Q_L$ takes on fields after 2 cooling sweeps. As we see in Fig.\ref{fig_Qcool20c2_su8}
the three distributions do not overlap. So we can assign to each field a value of $Q$ that
is completely unambiguous (at least for our statistics) on the basis of the value of $Q_L$
measured after only 2 cooling sweeps, where the long-distance physics should be essentially
unchanged from that of the original uncooled lattice fields. As an aside, we show
in Fig.\ref{fig_Qcool20c1_su8} the same plot as in Fig.\ref{fig_Qcool20c2_su8}, but
with values of $Q_L$ calculated after only 1 cooling sweep. Here the distributions are
broader, so that one can no longer use the value of $Q_L$ to unambiguously assign the
field its value of $Q$, but the fact that even  after 1 cooling sweep the values
of $Q_L$ strongly reflect the value of $Q$ after 20 cooling sweeps is evident. 

It is instructive to see  how the measured values of $Q_L$ vary with the number of cooling
sweeps $n_c$. As an example, we take an ensemble of fields that have a topological charge
$Q=2$, as determined by the value of $Q_L$ after 20 cooling sweeps. For this same ensemble
we calculate how the average value of $Q_L$ varies with $n_c$ and also how the standard
deviation of its fluctuations vary with $n_c$. These quantities, labelled $\overline{Q}_L$
and $\sigma_{Q_L}$ respectively, are shown in Table~\ref{table_Q_nc_SU8} for three different
values of $\beta$ in $SU(8)$. As we see, in the uncooled ($n_c=0$) fields the
fluctuations $\sigma_{Q_{L}}$ are so large compared to the average charge $\bar{Q}_L(n_c=0)$
of those fields, that one cannot hope to estimate for individual fields the true value of $Q$ from
the value of $Q_L$ at $n_c=0$. However we also see that even after only 1 cooling sweep the value of
$\bar{Q}_L/\sigma_{Q_{L}}$ increases by a factor $\sim 20$ and after 2 cooling sweeps by $\sim 80$,
so that one can, with rapidly increasing reliability, assign a value of $Q$ to an individual
field on the basis of the value of $Q_L(n_c)$ at the first few cooling sweeps. What we also see
in Table~\ref{table_Q_nc_SU8} is that beyond the lowest few values of $n_c$ there is little
difference between the 3 ensembles, despite the fact that the lattice spacing changes
by a factor $\sim 1.7$. For larger $n_c$ the value of $Q_L(n_c)$ decreases slightly with
decreasing $\beta$ as one expects because the typical instanton size will decrease in lattice
units and the discretisation corrections are negative. For $n_c=0$ the decrease in $Q_L$
is larger and reflects the $\beta$-dependence of the renormalisation factor $Z(\beta)$ which
is driven by high frequency fluctuations, as analysed more quantitatively in
Section~\ref{subsection_QZ}. The rapid increase of $\sigma_{Q_L}$ with increasing $\beta$
is driven by two competing factors: the high frequency fluctuations per lattice site
decrease as an inverse power of $\beta$, but since these fluctuations are roughly
uncorrelated across lattice sites, there is a factor proportional to the square root
of the lattice volume in lattice units which increases exponentially with $\beta$ if our
volumes are roughly constant in physical units. At larger $\beta$ the latter factor
will dominate and it therefore becomes rapidly harder to relate $Q$ from $Q_L(n_c=0)$
as we approach the continuum limit. On the other hand, after the first few cooling sweeps
the high frequency fluctuations have been largely erased and the fluctuations of $Q_L$ no
longer increase with increasing $\beta$, so that it becomes easier to identify the
value of $Q$ from the value of $Q_L$.

As a footnote to the above we have calculated the same quantities as in
Table~\ref{table_Q_nc_SU8} but now for different $SU(N)$ groups. The calculations are at
roughly the same value of $a\surd\sigma$ as for the $SU(8)$ fields at $\beta=46.70$ in
Table~\ref{table_Q_nc_SU8}, i.e. at roughly the same value of the 't Hooft coupling, so that
the contribution of the high frequency fluctuations is roughly the same. We find that the
variation in  $\sigma_{Q_L}(n_c=0)$ with $N$ is consistent with being due to the difference
in the square root of the lattice volumes. At larger $n_c$ the value of $\sigma_{Q_L}(n_c)$
decreases with increasing $N$, which tells us that larger-scale non-perturbative fluctuations,
such as in the instanton size, are decreasing. Finally, the main practical point is that there
is very little variation with $N$ of $\overline{Q}_L$ at any value of $n_c$.

\subsection{tunneling between topological sectors}
\label{subsection_Qtunneling} 

In the continuum theory $Q$ cannot change under continuous deformations of the fields
unlike most other quantities, such as glueball correlation functions, so one expects to lose
ergodicity in $Q$ much faster than in such other quantities as one approaches the continuum
limit in a lattice calculation using a local Monte Carlo algorithm such as the heat bath.
In this section we shall provide our data on this `freezing' of $Q$ as a function
of $a$ and $N$ and then compare this to some theoretical expectations.

As described earlier, the value of $Q$ changes if the core of an instanton shrinks and
disappears within a hypercube (or the reverse of this process). When $a$ is small enough
such instantons can be described by dilute gas calculations since these very small
instantons are very rare and the effect on them of other background field fluctuations that exist on
physical length scales will be negligible. So the basic process is for $Q$ to change by one unit.
Therefore we use as our measure of topological freezing the average number of sweeps  between
changes of $Q$ by one unit. We call this $\tau_Q$. Since the probability of such a process
is clearly proportional to the space-time volume, which is not something we have tried to keep
constant in our calculations, we rescale our measured values of $\tau_Q$ to a standard physical
volume, which we choose to be $V_0=l^4$ with $l\surd\sigma=3.0$,
so that we can compare the results of different calculations.
Our calculations of $Q$ have not been performed after every sweep, but typically with
gaps of  25 or 50 or 100 sweeps depending on the calculation. Our estimates of $\tau_Q$ can
only be reliable if they are much larger than this gap, since otherwise there could be multiple
changes of $Q$ within the gaps that we are missing, and so we do not includes those ensembles
where this issue arises -- typically at the coarser lattice spacings. However even if $\tau_Q$
appears to be much larger than the gap, we will occasionally see that $\Delta Q$, the change in
$Q$ across a gap, is greater than unity. In this case we assume that there have been $|\Delta Q|$
jumps within the gap and we make that part of our final estimate of  $\tau_Q$. In practice
this makes a very small difference to our results. By counting the number of changes of $Q$ in our
total sequence of lattice fields we can obtain the average distance between such changes, $\tau_Q$,
once we have renormalised to our standard volume $V_0$.

As $\tau_Q \to \infty$ the above definition is adequate. However when $\tau_Q$ is not very large
one can worry about the unwanted contribution of near-dislocations that occur across
a measurement of $Q$. We have in mind 
a small instanton that appears out of a hypercube shortly before a measurement, survives the
cooling (because of its environment)  and so contributes to the value of $Q$, but then quickly
disappears without becoming a larger physical instanton long before the next measurement.
In a sequence of measurements of $Q$ such an event would be characterised by a jump  $\Delta Q = \pm 1$
at one measurement followed by  the opposite jump,  $\Delta Q = \mp 1$, at the next measurement.
Of course it could be that when this happens we are seeing two independent events, with an
instanton appearing from one hypercube and after the measurement an anti-instanton appearing
out of a quite different hypercube. The characteristic of the latter events is that the signs of
the changes in $Q$ at neighbouring measurements are uncorrelated. Correcting for these we obtain
the measure  $\Tilde{\tau}_Q$ (normalised to our standard volume) which excludes this
estimate of near-dislocations. Whether this estimate is entirely reliable is arguable, so it is
reassuring that the differences between $\Tilde{\tau}_Q$ and $\tau_Q$ are not large,
and which one we use does not alter our conclusions below.

We present in Table~\ref{table_tauQ_SUN} our results for $\tau_Q$  and  $\Tilde{\tau}_Q$
from the sequences of fields generated in our glueball and string tension calculations.
The excluded values of $\beta$ either correspond to cases where $\tau_Q$ is not much
larger than the gap between measurements, and this includes all of our $SU(2)$ calculations,
or where $\tau_Q$ has become so large that we see no changes in $Q$ at all, which includes
almost all of our $SU(10)$ and $SU(12)$ calculations. The prominent qualitative features of our
results are that for any given gauge group both  $\tau_Q$  and  $\Tilde{\tau}_Q$ increase rapidly
as $a(\beta)$ decreases and, at roughly equal values of $a(\beta)$, they increase rapidly
as $N$ increases.

One can formulate some theoretical expectations for the behaviour of $\tau_Q$ as one decreases
$a$ and increases $N$. Just before shrinking through a hypercube an instanton will be very small
with size $\rho \sim a$ where the number density, $D(\rho)d\rho$  can be estimated using the
standard semiclassical formula
\cite{Coleman_Q}
\begin{equation}
D(\rho) \propto \frac{1}{\rho^5}
  \frac{1}{g^{4N}}
  \exp\left\{-\frac{8\pi^2}{g^2(\rho)}\right\}
    \stackrel{N\to\infty}{\propto}
 \frac{1}{\rho^5}\left\{ 
  \exp\left\{-\frac{8\pi^2}{g^2(\rho)N}
    \right\}\right\}^N
    \stackrel{\rho = a}{\propto}
 \left(a\Lambda\right)^{\frac{11N}{3}-5},  
\label{eqn_DI}
\end{equation}
where $\Lambda$ is the dynamical length scale of the theory. This is of
course a very asymptotic expression: we have neglected the powers of $g$ because they only
contribute powers of $\log(a\Lambda)$, and we have used the 1-loop expression for $g^2(a)$ which
is, as we have seen in Section~\ref{section_coupling}, inadequate for our range of
bare couplings. We also note that this expression only tells us what is the probability
of a very small instanton. In addition there will be a factor for the small instanton with
$\rho \sim O(a)$ to finally shrink completely within a hypercube: this `tunneling event' may
contribute an important factor depending on the lattice action being used.

The first qualitative feature of eqn(\ref{eqn_DI}) is that if we increase $N$ at
fixed $a$, we should find 
\begin{equation}
  \tau_Q \propto \frac{1}{D(\rho)} \propto \left\{\frac{1}{a\Lambda}\right\}^{\frac{11N}{3}-5}
\longrightarrow
  \ln\{\tau_Q\} = b + c N,
\label{eqn_IN}
\end{equation}
where $c$ depends on the value of $a$ and $b$ is some undetermined constant. In our
simulations we have some that correspond to almost equal values of $a\surd\sigma$, and hence of
$a\Lambda$, across several values of $N$. These are for $N=3,4,5,6$ at $a\surd\sigma \simeq 0.15$
and for $N=8,10,12$ at  $a\surd\sigma \simeq 0.33$. In Fig.~\ref{fig_tauQ_suN}
we plot the values of $\ln\{\tilde{\tau}_Q\}$ against $N$ and we see that the
behaviour is roughly linear as predicted from eqns(\ref{eqn_DI},\ref{eqn_IN}).

The second qualitative feature of eqn(\ref{eqn_IN}) is that if we vary $a$ at
fixed $N$, we should find 
\begin{equation}
  \tau_Q \propto \frac{1}{D(\rho)} \propto \left\{\frac{1}{a\Lambda}\right\}^{\frac{11N}{3}-5}
\longrightarrow
  \ln\{\tau_Q\} = b + \left\{\frac{11N}{3}-5\right\} \ln \left\{\frac{1}{a\Lambda}\right\}.
\label{eqn_Ia}
\end{equation}
In  Fig.~\ref{fig_tauQ_KsuN} we show plots of $\ln\{\tau_Q\}$ versus  $\ln\{1/a\surd\sigma\}$
for $N\in[3,8]$ and we see that the plots are roughly linear as predicted by
eqns(\ref{eqn_DI},\ref{eqn_Ia}). (As an aside, the fact that the $SU(2)$ values of $\tau_Q$ never
become large enough to be useful is no surprise given that the asymptotic dependence predicted
by eqn(\ref{eqn_Ia}) is quite weak, $\tau_Q\propto a^{7/3}$.)
The fitted coefficients of the $\ln\{1/a\surd\sigma\}$
term are listed in Table~\ref{table_tauQ_a} and compared to the value in eqn(\ref{eqn_Ia}).
We do not expect a good agreement since we know that the one loop expression for $g^2(a)$ is a poor
approximation in our range of bare couplings but it is interesting that our calculated values
listed in Table~\ref{table_tauQ_a}  do reflect the trend of the asymptotic theoretical values.

\subsection{topological susceptibility}
\label{subsection_Qsusc}

The simplest topological quantity that one can calculate is the topological susceptibility,
$\chi_t = \langle Q^2 \rangle / V$, where $V$ is the space-time volume. This is a
particularly interesting quantity because of the way it enters into estimates of the physical
$\eta^{\prime}$ mass through the Witten-Veneziano sum rule
\cite{Veneziano_eta,Witten_eta}.
There have been many calculations of this quantity and here we will add to these our calculations
for the gauge groups $SU(2)$, $SU(3)$, $SU(4)$, $SU(5)$ and for those of our $SU(6)$
ensembles where there are enough fluctuations in the value of $Q$ to make an estimate plausible.

We typically calculate the topological charge $Q$ after every 25 or 50 or 100 Monte Carlo
sweeps on most of the lattice ensembles that we use for our glueball and/or string
tension calculations. We estimate the value of $Q$ from the value of the lattice $Q_L$ after 20
cooling sweeps. The assignment of an integer value, $Q_I$, after 20 cooling sweeps is, as we
have seen, largely unambiguous. In Tables~\ref{table_QQ_SU2},~\ref{table_QQ_SU3_SU4} and
~\ref{table_QQ_SU5_SU6} we list our values for $Q^2$ on the lattices and at the couplings shown.

To obtain the continuum limit we perform a conventional extrapolation
\be
\left.\frac{\chi_t^{\frac{1}{4}}}{\surd\sigma}\right|_a
=
\left.\frac{\chi_t^{\frac{1}{4}}}{\surd\sigma}\right|_0
+ c a^2\sigma,
\label{eqn_chi_cont}
\ee
where we systematically remove from the fit the values corresponding to the largest $a$ until we
get an acceptable fit. We show the resulting continuum extrapolations for $N=2,3,4,5$ in
Fig.\ref{fig_khiIK_cont} and list the resulting continuum values in Table~\ref{table_khiK_SUN_cont}.
For $SU(2)$ and $SU(3)$ we do not include the value at the smallest $a$ since it is clearly too low
and it is plausible that it is due to a gradual loss of ergodicity in $Q$ accompanied by an
increasingly unreliable estimate of the errors. We see from Fig.\ref{fig_khiIK_cont} that
As we increase $N$ our fits are able to include values from increasingly coarse $a(\beta)$.
This is presumably related to the fact that the `bulk' cross-over/transition between
strong and weak coupling becomes rapidly sharper as $N$ increases
\cite{BLMTUW05}.
(It is particularly smooth for $SU(2)$.)
As shown in Table~\ref{table_khiK_SUN_cont} all the fits are acceptable.
We have performed separate continuum extrapolations of the susceptibilities obtained from
the non-integer lattice charges, $Q_L$, and the integer charges, $Q_I$, labelling these
$\chi_L$ and $\chi_I$ respectively. In the continuum limit these should be the same and we see
from Table~\ref{table_khiK_SUN_cont} that this is indeed so, within errors, for all except
the case of $SU(2)$, where the difference between the two values can be taken as a
systematic error that is additional to the statistical errors. 

Once we have the continuum susceptibilites we can extrapolate them to $N=\infty$  as shown
in Fig.\ref{fig_khiK_N}:
\be
\left.\frac{\chi_I^{\frac{1}{4}}}{\surd\sigma}\right|_N
=
0.3681(28) + \frac{0.471(15)}{N^2},
\label{eqn_chiIN}
\ee
and 
\be
\left.\frac{\chi_L^{\frac{1}{4}}}{\surd\sigma}\right|_N
=
0.3655(27) + \frac{0.448(15)}{N^2}.
\label{eqn_chiN}
\ee
The two are within errors as one would expect. We note that this value is consistent within errors
with the $N=\infty$ extrapolation in
\cite{bonanno_Q}
which uses a novel technique
\cite{hasenbusch_Q}
for ameliorating the problem of topological freezing.

Finally we return to the case of $SU(3)$ since it also has some phenomenological interest.
Here our analysis differs slightly from our earlier analysis in
\cite{AAMT-2020}
and our final result
\be
\left.\frac{\chi_I^{\frac{1}{4}}}{\surd\sigma}\right|_{SU(3)}
=
0.4246(36)
\label{eqn_chiIN3}
\ee
is about one standard deviation higher. To transform this into physical units
we translate from units in terms of $\surd\sigma$ to the standard scale $r_0$
and then to $\mathrm{MeV}$ units just as we did in eqn(\ref{eqn_LamMS_SU3}) of
Section~\ref{subsection_Lambda}, giving
\be
r_0\surd\sigma = 1.160(6) \Longrightarrow   \chi_I^{\frac{1}{4}} = 206(4)\mathrm{MeV}.
\label{eqn_chiIN3_MeV}
\ee
This is within errors of the value we presented in
\cite{AAMT-2020}
and the value $\chi_I^{\frac{1}{4}} \stackrel{SU3}{=} 208(6)\mathrm{MeV}$ obtained in
\cite{MLFP_SU3_10}
using the gradient flow technique. As an aside we note that the value of  
$\chi_I^{\frac{1}{4}}/\surd\sigma$ at $N=\infty$, as given in eqn(\ref{eqn_chiIN}),
is $\sim 13\%$ lower than the $SU(3)$ value, i.e. $\sim 179$`MeV' if
we simply rescale the value in eqn(\ref{eqn_chiIN3_MeV}).

\subsection{${Z_Q(\beta)}$ and lattice $\theta$ parameter}
\label{subsection_QZ} 

Consider the ensemble of Monte Carlo lattice gauge fields that correspond to an integer valued 
topological charge $Q$. If we calculate the charge $Q_L$ of each of these fields, prior to
any cooling, the average value will be related to $Q$ by
\cite{Pisa_ZQ}
\be
<Q_L> = Z_Q(\beta) Q,
\label{eqn_ZQ}
\ee
where  $Z_Q(\beta)$ will depend (weakly) on $N$ and negligibly on the lattice volume (as long
as it is not very small). Since the deviation of  $Z_Q(\beta)$ from unity is driven by high
frequency lattice fluctuations, it is of little physical interest in itself. However its
value is important if, for example, one wishes to study the $\theta$ dependence of the theory by
adding a term $i\theta Q$ to the continuum action and, correspondingly, a term $i\theta_L Q_L$
to the lattice action, since one sees that
\cite{Pisa_theta}
\be
\theta_L \simeq  Z^{-1}_Q(\beta) \theta.
\label{eqn_theta}
\ee
Primarily for this reason we have calculated  $Z_Q(\beta)$ in our lattice calculations
and will also provide interpolating functions that will give  $Z_Q(\beta)$ at values
of $\beta$ that lie between our measured values.

A first estimate for $Z_Q(\beta)$ can be obtained in perturbation theory, giving at one loop
\cite{Pisa_ZQ}
\be
Z_Q(\beta) \stackrel{\beta\to\infty}{=} 1 - (0.6612 N^2 - 0.5)\frac{1}{\beta} + O(\beta^{-2})
\label{eqn_Zpert}
\ee
which already tells us that for our range of $\beta$ we will have $Z_Q(\beta) \ll 1$. The
fact that the one loop correction is so large tells us that the one loop estimate is likely
to be not very accurate, and indeed that proves to be the case. Moreover as we increase the
lattice spacing the typical instanton becomes smaller and the value of $Q_L$ acquires
significant corrections that are powers of $a(\beta)$ and which are additional to any
perturbative corrections. So in constructing our interpolating function for $Z_Q$ we
simply use the form
\be
Z^{int}_Q(\beta) = 1 -  z_0 g^2N - z_1 (g^2N)^2 \quad ; \quad g^2N = \frac{2N^2}{\beta}, 
\label{eqn_Zint}
\ee
where $g^2N$ is the 't Hooft coupling, and we make no attempt 
to constrain the parameters $z_0$ and $z_1$ to perturbative values.
This turns out to be an adequate fitting function to our values of  $Z_Q(\beta)$. It
is however important to note that while this works as an interpolating function, it is
likely to fail increasingly badly the further one uses it away from the fitted range of $\beta$
as an extrapolating function.

Our values for $Z_Q(\beta)$ are obtained from fits such as those shown in Fig.\ref{fig_ZQ_su8},
and are listed in Tables~\ref{table_ZQA} and \ref{table_ZQB}.
Interpolating these values with eqn(\ref{eqn_Zint}) gives the values for $z_0$ and $z_1$ listed
in Table~\ref{table_ZQint}. Physically the most relevant interpolating function is the one
for $SU(3)$:
\be
Z^{int}_Q(\beta) \stackrel{su3}{=} 1 - 0.162(10)g^2N - 0.0425(31) (g^2N)^2 \quad , \quad \chi^2/n_{df} = 0.62
\label{eqn_Zintsu3}
\ee
In the case of $SU(2)$ we have only a few entries because most of our earliest calculations
did not include calculating $Q_L$ on the fields prior to cooling. In addition in the case
of $SU(2)$ the identification of an integer $Q$ after 20 cooling sweeps possesses small
but significant ambiguities, which rapidly disappear as $N$ increases. We can also
fit our interpolating functions in $N$, thus obtaining
\be
z_0 = 0.179(12) - \frac{0.08(15)}{N^2}   \quad , \quad \chi^2/n_{df} = 1.00
\label{eqn_Z0suN}
\ee
and
\be
z_1 = 0.0482(46) - \frac{0.072(50)}{N^2}   \quad , \quad \chi^2/n_{df} = 1.13
\label{eqn_Z1suN}
\ee
which should be reliable over a wide range of $N$ as long as we are not too far outside
the range of the t'Hooft coupling $\lambda=g^2N$ of our calculations in
Tables~\ref{table_ZQA} and \ref{table_ZQB}. Finally we remind the reader that all these
results for $Z_Q(\beta)$ only apply to calculations with the standard Wilson plaquette action
and with the definition of the lattice topological charge $Q_L$ used here.

\section{Conclusions}
\label{section_conclusion} 

Our primary goal in this paper has been to calculate the glueball spectra of a range
of $SU(N)$ gauge theories, in the continuum limit, with enough precision to obtain plausible
extrapolations to the theoretically interesting $N=\infty$ limit. This provides
the first calculation of the masses of the ground states 
in all the $R^{PC}$ channels, as well as some excited states in most channels,
in the continuum limit of the $N\to\infty$ gauge theory.

Our results, for $N=2,3,4,5,6,8,10,12$, were obtained using standard lattice gauge theory 
methods. Although the issue of topological freezing at larger $N$ in $SU(N)$ gauge theories 
is not expected to be important for glueball spectra
\cite{Witten_98,LDDGMEV_Q},
we confirmed this explicitly in some extensive $SU(8)$ calculations, and in addition
we chose to minimise any remnant bias at larger $N$ by modifying the usual update algorithm, 
explicitly imposing the expected distribution of topological charge on the starting 
lattice fields of our ensemble of Monte Carlo sequences.
We employed a large basis of single-trace glueball operators, which allowed us
to calculate the ground state and some excited states for each of the $R^{PC}$ channels,
where $R$ labels the representation of the rotation symmetry group appropriate
to our cubic lattice, and $P,C$ label the parity and charge conjugation. The large basis
gives us confidence that we are not missing any low-lying states and this in turn
allows us to match near-degeneracies between states with different $R$ so as to
assign continuum spin quantum numbers to a significant number of glueball states.

Our results have greatly extended existing calculations while largely confirming existing
results; in particular the important conclusion that $SU(3)$ is `close to' $SU(\infty)$.
As before, one finds that the lightest glueball is the $J^{PC}=0^{++}$ scalar ground state,
with a mass that ranges from $M_{0^{++}}\sim 3.41\surd\sigma$ for $SU(3)$ to
$M_{0^{++}}\sim 3.07\surd\sigma$ for $SU(\infty)$, where $\sigma$ is the string tension,
and that the next heavier glueballs are the tensor with a mass $M_{2^{++}} \simeq 1.5 M_{0^{++}}$,
and the pseudoscalar, $0^{-+}$, which is nearly degenerate with the tensor. One then
has the $1^{+-}$ with $M_{1^{+-}} \sim 1.85 M_{0^{++}}$, and this is the only relatively light
$C=-$ state. At roughly the same mass is the first excited $0^{++}$ and then the lightest
pesudotensor with $M_{2^{-+}} \sim 1.95 M_{0^{++}}$. All other states are heavier than twice
the lightest scalar, with most of the $C=-$ ground states being very much heavier than that.
One sees a number of near-degerenacies which may or may not be significant.
The continuum glueball masses (in units of the string tension) for the various $R^{PC}$
channels are listed in Tables~\ref{table_MK_R_SU2}-\ref{table_MK_R_SUN} and for the
$J^{PC}$ channels in Tables~\ref{table_MKJ_N2-5}-\ref{table_MK_J_SUN}.

Since our calculations are on a finite spatial volume we have had to identify and exclude
the `ditorelon' states composed of a pair of mutually conjugate flux tubes that wind
around our periodic spatial torus. These are states whose projection onto our single
trace operators will vanish as $N\to\infty$. We also need to exclude any multiglueball
scattering states. Our (albeit partial) explicit calculations using the corresponding
double trace operators strongly suggest that these states do not appear in the glueball
spectra that we claim to obtain using our single trace basis.

In calculating the glueball spectra we have also calculated a number of other quantities
that could be calculated simultaneously. Our calculations of the string tension were
primarily intended to provide a scale in which to express our glueball masses.
However they also provided a scale for the lattice spacing $a$ at each value of the
bare lattice coupling, $g_L(a)$, which we were able to use to obtain values of the
dynamical scale $\Lambda_{\overline{MS}}$ for all our values of $N$. These improved
upon earlier calculations of this kind
and for $SU(3)$ provided a value in physical units of
$\Lambda_{\overline{MS}} \simeq 263(4)[9]{\mathrm{MeV}}$ which
is consistent with values obtained using more dedicated methods
At the same time we were able to confirm that keeping fixed the running 't Hooft coupling 
$g^2(a)N$, with $a$ being kept fixed in units of the string tension, is the way to
approach the $N=\infty$ limit along lines of constant physics.

In addition to the fundamental string tension we calculate the tension of $k=2$ flux tubes,
in order to analyse the way that $\sigma_{k=2}/\sigma$ approaches $N=\infty$. We find that
a leading $O(1/N)$ correction works better than the $O(1/N^2)$ expected from standard
large-$N$ counting, but that the latter cannot be completely excluded. We speculate
that the  $O(1/N)$ behaviour is sub-asymptotic, with the $O(1/N^2)$ correction 
settling in for flux tubes of length $l\leq l_c$ once $N$ is large enough that the $k=2$ 
flux tube becomes a pair of weakly interacting fundamental flux tubes for $l\leq l_c$
with  $l_c\uparrow$ as  $N\uparrow$. That is to say, the $l,N\to\infty$ limit is
not uniform.

Since for our glueball calculations  we need to monitor the onset of the freezing
of the topology of our lattice fields, we have performed extensive calculations of the
topological charge along with the glueball calculations. Using these calculations 
we obtain values for the continuum limit of the topological susceptibility for the
$SU(N\leq 6)$ gauge theories. The freezing of topology means that we have no values
for $N > 6$ or for our smaller $SU(6)$ lattice spacings. Nonetheless this does not
prevent us achieving a usefully precise value of the $N=\infty$ topological suseptibility,
$\chi_I^{\frac{1}{4}}/\surd\sigma = 0.3681(28)$. We have also calculated the renormalisation
factor $Z_Q(\beta)$ that relates the value of our particular (but standard) lattice topological
charge, as obtained on the Monte carlo generated lattice gauge fields, to the true integer
valued topological charge of those fields. This is useful if one wishes to include
a topological $\theta$-term in the action, and so we also include functions that
interpolate between our values of $\beta$. We can do so for all our $SU(N)$ groups
because these calculations can be equally well determined using ensembles of fields
where the topological charge has been inserted through the initial fields of the
Markovian chains.

The present study could be improved in several ways. A definitive study of ditorelon states
and the lightest multiglueball states for all $R^{PC}$ sectors would be useful.
The heaviest states need a finer resolution in the correlation functions for the
mass identification to become completely unambiguous: this could be achieved by using
an anisotropic lattice such that the timelike lattice spacing is much smaller than the
spacelike one, a technique that has occasionally been put to good purpose in the past
\cite{KIGSMT-SU2-1983,KIGSMT-SU3-1983,MP-1999,MP-2005}.
Perhaps most important would be the incorporation of more effective techniques for
determining the continuum spins of the glueball states, as for example in
\cite{PCSDMT-2019}
for 2 space dimensions and in
\cite{HM_Thesis}
for our case here of 3 spatial dimensions.

\section*{Acknowledgements}

AA has been financially supported by the European Union's Horizon 2020 research and innovation
programme ``Tips in SCQFT'' under the Marie Sk\l odowska-Curie grant agreement No. 791122.
MT acknowledges support by Oxford Theoretical Physics and All Souls College. The numerical
computations were carried out on the computing cluster in Oxford Theoretical Physics.

\clearpage

\begin{appendix}
\setcounter{table}{0}
\renewcommand{\thetable}{A\arabic{table}}

\section{Lattice running couplings}
\label{section_appendix_couplings}

For pure gauge theories the perturbative $\beta$ function for the running coupling
in a coupling scheme $s$ is given by
\be
\beta(g_s)
=
-a\frac{\partial g_s}{\partial a}
=
-b_0  g^3_s - b_1  g^5_s - b^s_2  g^7_s + ...,
\label{eqn_bfunction2}
\ee
where $a$ is the length scale on which the coupling is calculated and on which it
depends.  The first two coefficients $b_0$ and $b_1$ are scheme independent while the
coefficients $b^s_{j\geq 2}$ depend on the scheme. Integrating between scales $a_0$
and $a$, we obtain
\be
\frac{a}{a_0} 
=
\exp{\left(-\int^{g(a)}_{g(a_0)} \frac{dg}{\beta(g)}\right)}.
\label{eqn_aa0}
\ee
(The label $s$ on $g$ is to be understood.)
The integrand is singular as $g\to 0$ and for any calculations it is convenient
to separate out the singular pieces. Since the issue arise for  $g\to 0$ we
can expand $1/\beta(g)$ in powers of $g$ and we then readily see that we can
separate out the singular terms as follows,
\beq
\frac{a}{a_0} 
& = &
e^{+\int^{g(a)}_{g(a_0)} dg \left(
 - \frac{b_1}{b^2_0 g} + \frac{1}{b_0 g^3}
 \right) }
e^{-\int^{g(a)}_{g(a_0)} dg \left(\frac{1}{\beta(g)}
 - \frac{b_1}{b^2_0 g} + \frac{1}{b_0 g^3}
 \right) }       \nonumber \\
& = &
\left(\frac{g^2(a)}{g^2(a_0)}\right)^{-\frac{b_1}{2b^2_0}}
e^{-\frac{1}{2b_0}\left(\frac{1}{g^2(a)}-\frac{1}{g^2(a_0)}\right)}
e^{-\int^{g(a)}_{g(a_0)} dg \left(\frac{1}{\beta(g)}
 - \frac{b_1}{b^2_0 g} + \frac{1}{b_0 g^3}
 \right) }.
\label{eqn_aa0b}
\eeq
In the second line we have integrated the `singular' terms, and the remaining
integral will now be finite as $g(a_0) \to 0$. So we can break up the integral as
\be
e^{-\int^{g(a)}_{g(a_0)} dg \left(\frac{1}{\beta(g)}
 - \frac{b_1}{b^2_0 g} + \frac{1}{b_0 g^3}
 \right) }
=
e^{+\int^{g(a_0)}_0 dg \left(\frac{1}{\beta(g)}
 - \frac{b_1}{b^2_0 g} + \frac{1}{b_0 g^3}
 \right) }
e^{-\int^{g(a)}_0 dg \left(\frac{1}{\beta(g)}
 - \frac{b_1}{b^2_0 g} + \frac{1}{b_0 g^3}
 \right) }
\label{eqn_aa0c}
\ee
where each integral will be well-defined since the singularities at $g=0$
have been removed. Separating the terms in $a$ and $a_0$ in
eqns(\ref{eqn_aa0b},\ref{eqn_aa0c}) we can write
\be
\frac{a}{a_0}  = \frac{F(g(a))}{F(g(a_0))},
\label{eqn_aa0d}
\ee
where we define
\be
F(g)
\equiv
\left(b_0g^2\right)^{-\frac{b_1}{2b^2_0}}
e^{-\frac{1}{2b_0g^2}}
e^{-\int^{g}_0 dg \left(\frac{1}{\beta(g)}
 - \frac{b_1}{b^2_0 g} + \frac{1}{b_0 g^3}
 \right) }.
\label{eqn_aa0e}
\ee
Note that the factor of $b_0$ that we have inserted in the first term on the right
of eqn(\ref{eqn_aa0e}) will cancel in eqn(\ref{eqn_aa0d}) so we are free to insert
it if we wish. We see from eqn(\ref{eqn_aa0d}) that $a/F(g^2(a))$ is
independent of the scale $a$ on which the coupling is defined and is a constant.
So we can now define a dynamical energy scale $\Lambda$ by
\beq
\Lambda \equiv \frac{a_0}{F(g(a_0))}
\Longrightarrow
a & = & \frac{1}{\Lambda} F(g(a)) \nonumber \\
& = &
\frac{1}{\Lambda} 
\left(b_0g^2\right)^{-\frac{b_1}{2b^2_0}}
e^{-\frac{1}{2b_0g^2}}
e^{-\int^{g(a)}_0 dg \left(\frac{1}{\beta(g)}
 - \frac{b_1}{b^2_0 g} + \frac{1}{b_0 g^3}
 \right) }.
\label{eqn_aa0f}
\eeq
The scale $\Lambda$ defined here coincides with the conventional $\Lambda$ parameter
that appears in the standard 2-loop expression for the running coupling.
(It is to ensure this equality that we inserted the factor of $b_0$ above.)
This scale will clearly depend on the coupling scheme and within a given scheme
the value we obtain for it will depend on our approximation to $\beta(g)$. 

For our lattice action  only the first 3 coefficients in the
$\beta$-function are known. In that case, collecting terms, it is convenient
to rewrite eqn(\ref{eqn_aa0f}) as
\be
a
\stackrel{3 loop}{=}
\frac{1}{\Lambda} 
\left(b_0g^2(a)\right)^{-\frac{b_1}{2b^2_0}}
e^{-\frac{1}{2b_0g^2(a)}}
e^{-\frac{1}{2} \int^{g^2(a)}_0 dg^2
\left(\frac{b_0b_2-b^2_1-b_1b_2g^2}{b^3_0+b^2_0b_1g^2+b^2_0b_2g^4}
\right) }
\equiv
\frac{1}{\Lambda} F_{3l}(g(a)),
\label{eqn_ag3loop}
\ee
where we denote by $F_{3l}(g)$ the 3-loop approximation to $F(g)$.
The integrand is a smooth function of $g^2$ and so the integral can be calculated
accurately for any given $g(a)$ using any elementary numerical integration
method. If we retain only the first 2 coefficients of $\beta(g)$ then we can
do the integral analytically to obtain
\be
a
\stackrel{2 loop}{=}
\frac{1}{\Lambda} 
e^{-\frac{1}{2b_0g(a)^2}}
\left(\frac{b_1}{b^2_0}+\frac{1}{b_0g^2(a)}\right)^{\frac{b_1}{2b^2_0}}
\equiv
\frac{1}{\Lambda} F_{2l}(g(a))
\label{eqn_ag2loop}
\ee
as the exact 2-loop result, where we denote by $F_{2l}(g)$ the 2-loop approximation to
$F(g)$.

The above perturbative expressions for $a\Lambda$ can be turned into expressions
for $a\mu$ where $\mu$ is some physical mass or energy:
\be
a\mu
=
\left.a\Lambda \frac{\mu}{\Lambda}\right|_{a}
=
\left.\frac{\mu}{\Lambda}\right|_{a=0} (1 + c_{\mu}a^2\mu^2 + O(a^4)) F(g(a)),
\label{eqn_amu}
\ee
where $F(g(a))$ is defined in eqn(\ref{eqn_aa0e}) and $c_{\mu}$ is an unknown constant.
Here we use the standard tree-level expansion for a dimensionless ratio of physical
energy scales, which here is $\mu/\Lambda$. This expression marries perturbative
(logarithmic)
and power-like dependences on $a$ in a plausible way. It is of course arguable: for
example the perturbative expansion is at best asymptotic and this can introduce other
power-like corrections. In practice we shall use this for the string tension,
$\mu=\surd\sigma$, and we will drop $O(a^4)$ and higher order terms.
That is to say we will attempt to fit the calculated string tensions with
\be
a\surd\sigma
=
\left.\frac{\surd\sigma}{\Lambda}\right|_{a=0} (1 + c_{\sigma}a^2\sigma) F_{3l}(g(a))
\label{eqn_aK}
\ee
and we shall be doing so in the mean-field coupling scheme $s=I$. To calculate
$F_{3l}(g(a))$ in that scheme we only need to know the coefficient $b_2^{s=I}$,
since $b_0$ and $b_1$ are scheme independent with values
\be
b_0={\frac{1}{(4\pi)^2}} {\frac{11}{3}} N, \quad 
b_1={\frac{1}{(4\pi)^4}} {\frac{34}{3}} N^2.
\label{eqn_b0b1}
\ee
We can begin with the well-known value of $b_2^{s=\overline{MS}}$
\cite{MSbar_4loop}
\be
b^{\overline{MS}}_2={\frac{1}{(4\pi)^6}} {\frac{2857}{54}} N^3.
\label{eqn_b2MS}
\ee
To obtain $b^s_2$ in the improved lattice coupling scheme $I$ we first transform
to the plaquette action lattice coupling scheme, $s=L$, using
\cite{betaL_3loop}
\be
b^L_2=2b^{\overline{MS}}_2 - b_1l_0 +b_0l_1,
\label{eqn_b2La}
\ee
where
\be
l_0=\frac{1}{8N} - 0.16995599 N, \qquad
l_1=-\frac{3}{128N^2}+0.018127763-0.0079101185N^2.
\label{eqn_b2Lb}
\ee
The lattice coupling, $g^2_L$, satisfies the $\beta$-function in eqn(\ref{eqn_bfunction2}) with
$s=L$, and the mean-field coupling will satisfy a $\beta$-function with the same $b_0$ and $b_1$
coefficients but with a different coefficient, $b^I_2$, of the $g^7$ power. To
determine $b^I_2$ one can use the expression for  $\langle \mathrm{Tr} U_p \rangle$
as a power series in $g^2_L$ to write $g^2_I$ as
\be
g^2_I = g^2_L \langle \frac{1}{N} \mathrm{Tr} U_p \rangle
=  g^2_L \left(1-w_1g^2_L-w_2g^4_L- ...\right),
\label{eqn_gIgL}
\ee
where
\cite{plaq_pert}
\be
w_1=\frac{(N^2-1)}{8N}, \qquad
w_2=\left(N^2-1\right)\left(0.0051069297 -\frac{1}{128N^2}\right).
\label{eqn_plaq_pert}
\ee
We now insert the expression for $g^2_I$ in eqn(\ref{eqn_gIgL}) into the $\beta$-function
for $g^2_L$ in eqn(\ref{eqn_bfunction2}) giving us, after some elementary manipulation,
\be
b^I_2=b^L_2+w_2b_0-w_1b_1.
\label{eqn_b2I}
\ee
So: using eqns(\ref{eqn_b2MS},\ref{eqn_b2Lb}) in eqn(\ref{eqn_b2La}) we obtain the
explicit expression for $b^L_2$ and then inserting that together with the functions in
eqn(\ref{eqn_plaq_pert}) and eqn(\ref{eqn_b0b1}) into eqn(\ref{eqn_b2I}) we
obtain the explicit expression for $b^I_2$ for any $N$. We now have the explicit
expressions for $b_0$, $b_1$ and $b^I_2$ which allow us to calculate the
value of $F_{3l}(g_I(a))$ in eqn(\ref{eqn_aK}) for any value of $N$ and $g_I(a)$.

\clearpage

\setcounter{table}{0}
\renewcommand{\thetable}{B\arabic{table}}

\section{Scattering states}
\label{section_appendix_scattstates}

We will restrict ourselves to states of two glueballs. We probe such states with product
operators $\phi_a(t)\phi_b(t)$ where $\phi_a$ and $\phi_b$ are typical single loop zero momentum operators
that are expected to project primarily onto single glueballs with chosen quantum numbers. Both the
individual and product operators will in general need vacuum subtraction for this to be the case. As usual
we will have some basis of single loop operators so the generic correlator will be of the form
\be
C_4(t) = \langle \phi'_a(t)\phi'_b(t) \phi'_c(0)\phi'_d(0) \rangle
-\langle \phi'_a\phi'_b\rangle \langle \phi'_c\phi'_d\rangle,
\label{eqn_corGGGGa}
\ee
where $\phi'_a(t) = \phi_a(t) - \langle \phi_a\rangle$ etc. This equation subtracts any vacuum
contribution to the individual operators as well as to their products. We have taken the operators
to be real. If, for example, we were to include operators  $\phi_i(t),\phi_j(t)$ with opposite
non-zero momenta then these would be complex. In that case we should change
$\phi_a(t),\phi_b(t) \to \phi^{\dagger}_a(t),\phi^{\dagger}_b(t)$ in eqn(\ref{eqn_corGGGGa}) and below.

Our glueball calculations require high statistics so we calculate our correlators during the
generation of the sequence of lattice fields. At this stage we can only calculate the correlators
of the fields $\phi$ without any vacuum subtraction -- we will only be able to calculate the
vacuum expectation values at the end of the computer simulation.
A short calculation tells us that the correlator of the $\phi'_i$ fields in eqn(\ref{eqn_corGGGGa})
can be written in terms of the correlators of the unsubtracted $\phi_i$ fields  as follows:
\begin{multline}
C_4(t) = \langle \phi_a(t)\phi_b(t) \phi_c(0)\phi_d(0) \rangle \\
-\langle\phi_a\rangle \langle\phi_b(t)\phi_c(0)\phi_d(0) \rangle
-\langle\phi_b\rangle \langle\phi_a(t)\phi_c(0)\phi_d(0) \rangle\\
-\langle\phi_c\rangle \langle\phi_a(t)\phi_b(t)\phi_d(0) \rangle
-\langle\phi_d\rangle \langle\phi_a(t)\phi_b(t)\phi_c(0) \rangle \\
+\langle\phi_a\rangle \langle\phi_c\rangle \langle \phi_b(t)\phi_d(0) \rangle
+\langle\phi_b\rangle \langle\phi_c\rangle \langle \phi_a(t)\phi_d(0) \rangle \\
+\langle\phi_a\rangle \langle\phi_d\rangle \langle \phi_b(t)\phi_c(0) \rangle
+\langle\phi_b\rangle \langle\phi_d\rangle \langle \phi_a(t)\phi_c(0) \rangle \\
-\langle\phi_a\phi_b\rangle \langle\phi_c\phi_d\rangle
+2\langle\phi_a\phi_b\rangle \langle\phi_c\rangle \langle\phi_d\rangle
+2\langle\phi_c\phi_d\rangle \langle\phi_a\rangle \langle\phi_b\rangle
-4\langle\phi_a\rangle \langle\phi_b\rangle \langle\phi_c\rangle \langle\phi_d\rangle.
\label{eqn_corGGGGb}
\end{multline}
Of course when some of the operators have non-vacuum quantum numbers, then the corresponding
vacuum expectation values will vanish and the above expression will simplify in obvious ways.

In addition to the above we may also be interested in the overlap of single loop operators,
which mainly project onto single glueballs, with the above product operators, which one
expects to mainly project onto two glueballs. That is to say correlators such as
\be
C_3(t) = \langle \phi'_a(t)\phi'_b(t) \phi'_c(0) \rangle
-\langle \phi'_a\phi'_b\rangle \langle \phi'_c\rangle
= \langle \phi'_a(t)\phi'_b(t) \phi'_c(0) \rangle
\label{eqn_corGGGa}
\ee
since $\langle \phi'_c\rangle=0$ by definition. In terms of the unsubtracted operators we find
\begin{multline}
C_3(t) = \langle \phi_a(t)\phi_b(t) \phi_c(0) \rangle 
-\langle\phi_a\rangle \langle\phi_b(t)\phi_c(0) \rangle
-\langle\phi_b\rangle \langle\phi_a(t)\phi_c(0) \rangle \\
-\langle\phi_c\rangle \langle\phi_a\phi_b\rangle
+2\langle\phi_a\rangle \langle\phi_b\rangle \langle\phi_c\rangle. 
\label{eqn_corGGGb}
\end{multline}

In this paper we can only calculate light masses with any reliability so we restrict ourselves
to double loop operators where one loop is in the $A_1^{++}$ representation and therefore has
some projection onto the lightest glueball. The second operator will then be in the representation
$R^{PC}$ in which we happen to be interested. If $R^{PC} \neq A_1^{++}$ then eqn(\ref{eqn_corGGGGb})
simplifies drastically 
\begin{multline}
C_4(t) = \langle \phi_a(t)\phi_b(t) \phi_c(0)\phi_d(0) \rangle \\
-\langle\phi_a\rangle \langle\phi_b(t)\phi_c(0)\phi_d(0) \rangle
-\langle\phi_c\rangle \langle\phi_a(t)\phi_b(t)\phi_d(0) \rangle
+\langle\phi_a\rangle \langle\phi_c\rangle \langle \phi_b(t)\phi_d(0) \rangle
-\langle\phi_a\phi_b\rangle \langle\phi_c\phi_d\rangle
\label{eqn_corGGGGd}
\end{multline}
since with $\phi_a,\phi_c$ being $A_1^{++}$ and $\phi_b,\phi_d$ not being $A_1^{++}$ means
that not only $\langle \phi_b\rangle = \langle \phi_d\rangle = 0$ but also that
products like  $\langle \phi_a \phi_b\rangle$ are zero. Similarly if  $\phi_a$ in
eqn(\ref{eqn_corGGGa}) is $A_1^{++}$ and  $\phi_b$ is in some $R^{PC} \neq A_1^{++}$ then
$\phi_c$ must be in the same $R^{PC} \neq A_1^{++}$ for $C_3(t)$ not to be zero.
In that case we will have
\be
C_3(t) = \langle \phi_a(t)\phi_b(t) \phi_c(0) \rangle 
-\langle\phi_a\rangle \langle\phi_b(t)\phi_c(0) \rangle.
\label{eqn_corGGGc}
\ee
If however $\phi_b$ and $\phi_c$ are in $A_1^{++}$ then we need the full expression.


\end{appendix}

\clearpage

\clearpage

\begin{table}[htb]
\centering
\begin{tabular}{|cc|ccc|} \hline
\multicolumn{5}{|c|}{SU(2) ; $l\surd\sigma\sim 4$ } \\ \hline
$\beta$ & $l^3l_t$ & $\tfrac{1}{2}\text{ReTr}\langle U_p\rangle$ & 
$a\surd\sigma$ & $am_G$  \\ \hline
 2.2986 & $12^316$  & 0.6018259(46)  & 0.36778(69)  & 1.224(16)  \\
 2.3714 & $14^316$  & 0.6226998(36)  & 0.29023(50)  & 1.025(12)  \\
 2.427  & $20^316$  & 0.6364293(15)  & 0.24013(41)  & 0.8469(76)  \\
 2.509  & $22^320$  & 0.6537214(10)  & 0.18011(22)  & 0.6563(56)  \\
 2.60   & $30^4$    & 0.6700089(5)   & 0.13283(30)  & 0.5001(41)  \\
 2.70   & $40^4$    & 0.6855713(3)   & 0.09737(23)  & 0.3652(35)  \\ \hline
\end{tabular}
\caption{Parameters of the main $SU(2)$ calculations: the inverse coupling, $\beta$, the lattice size, the
  average plaquette, the string tension, $\sigma$, and the  mass gap, $m_G$.}
\label{table_param_SU2}
\end{table}

\begin{table}[htb]
\centering
\begin{tabular}{|cc|ccc|} \hline
\multicolumn{5}{|c|}{SU(3) ; $l\surd\sigma\sim 4$}  \\ \hline
$\beta$ & $l^3l_t$ & $\tfrac{1}{3}\text{ReTr}\langle U_p\rangle$ & 
$a\surd\sigma$ & $am_G$  \\ \hline
5.6924 & $10^316$ & 0.5475112(71) & 0.3999(58)  & 0.987(9)    \\
5.80   & $12^316$ & 0.5676412(36) & 0.31666(66) & 0.908(12)   \\
5.8941 & $14^316$ & 0.5810697(18) & 0.26118(37) & 0.7991(92) \\
5.99   & $18^4$ &  0.5925636(11)  & 0.21982(77) & 0.7045(65)  \\
6.0625 & $20^4$ &  0.6003336(10)  & 0.19472(54) & 0.6365(43)  \\
6.235  & $26^4$ &  0.6167723(6)   & 0.15003(30) & 0.4969(29)  \\
6.3380 & $30^4$ &  0.6255952(4)   & 0.12928(27) & 0.4276(37)  \\
6.50   & $38^4$ &  0.6383531(3)   & 0.10383(24) & 0.3474(22)  \\ \hline
\end{tabular}
\caption{Parameters of the main $SU(3)$ calculations: the inverse coupling, $\beta$, the lattice size, the
  average plaquette, the string tension, $\sigma$, and the  mass gap, $m_G$.}
\label{table_param_SU3}
\end{table}

\begin{table}[htb]
\centering
\begin{tabular}{|cc|ccc|} \hline
\multicolumn{5}{|c|}{SU(4) ; $l\surd\sigma\sim 4$}  \\ \hline
$\beta$ & $l^3l_t$ & $\tfrac{1}{4}\text{ReTr}\langle U_p\rangle$ & 
$a\surd\sigma$ & $am_G$  \\ \hline
10.70  & $12^316$ & 0.5540665(24)  & 0.3021(5)   & 0.8406(48)  \\
10.85  & $14^320$ & 0.5664268(15)  & 0.25426(38)  & 0.7611(54)  \\
11.02  & $18^320$ & 0.5782610(11)  & 0.21434(28)  & 0.6605(33)  \\
11.20  & $22^4$   & 0.5893298(6)  & 0.18149(49)  & 0.5709(34)  \\
11.40  & $26^4$   & 0.6004374(4)  & 0.15305(34)  & 0.4864(30)  \\
11.60  & $30^4$   & 0.6106057(3)  & 0.13065(21)  & 0.4132(44)  \\ \hline
\end{tabular}
\caption{Parameters of the main $SU(4)$ calculations: the inverse coupling, $\beta$, the lattice size, the
  average plaquette, the string tension, $\sigma$, and the  mass gap, $m_G$.}
\label{table_param_SU4}
\end{table}

\begin{table}[htb]
\centering
\begin{tabular}{|cc|ccc|} \hline
\multicolumn{5}{|c|}{SU(5) ; $l\surd\sigma\sim 3.1$} \\ \hline
 $\beta$ & $l^3l_t$ & $\tfrac{1}{5}\text{ReTr}\langle U_p\rangle$ & 
$a\surd\sigma$ & $am_G$  \\ \hline
16.98   & $10^316$  & 0.5454873(28)  & 0.3033(8)    & 0.8241(68)  \\
17.22   & $12^316$  & 0.5587002(18)  & 0.2546(6)    & 0.7517(51)  \\
17.43   & $14^320$  & 0.5685281(10)  & 0.22217(37)  & 0.6751(44)  \\
17.63   & $16^320$  & 0.5769707(9)   & 0.19636(35)  & 0.5961(79)  \\
18.04   & $20^324$  & 0.5924012(6)   & 0.15622(38)  & 0.4783(44)  \\
18.375  & $24^330$  & 0.6036547(4)   & 0.13106(30)  & 0.4078(38)  \\ \hline
\end{tabular}
\caption{Parameters of the main $SU(5)$ calculations: the inverse coupling, $\beta$, the lattice size, the
  average plaquette, the string tension, $\sigma$, and the  mass gap, $m_G$.}
\label{table_param_SU5}
\end{table}

\begin{table}[htb]
\centering
\begin{tabular}{|cc|ccc|} \hline
\multicolumn{5}{|c|}{SU(6) ; $l\surd\sigma\sim 3.1$} \\ \hline
$\beta$ & $l^3l_t$ & $\tfrac{1}{6}\text{ReTr}\langle U_p\rangle$ & 
$a\surd\sigma$ & $am_G$  \\ \hline
24.67    & $10^316$  & 0.5409011(28)  & 0.30658(34)  & 0.8240(41)  \\
25.05    & $12^316$  & 0.5557062(13)  & 0.25177(23)  & 0.7395(50)  \\
25.32    & $14^320$  & 0.5646185(10)  & 0.22208(35)  & 0.6673(32)  \\
25.55    & $16^320$  & 0.5715585(9)   & 0.20153(34)  & 0.6112(41)  \\
26.22    & $20^324$  & 0.5894540(4)   & 0.15480(36)  & 0.4751(53)  \\
26.71    & $24^330$  & 0.6009861(3)   & 0.12867(27)  & 0.3886(37)  \\ \hline
\end{tabular}
\caption{Parameters of the main $SU(6)$ calculations: the inverse coupling, $\beta$, the lattice size, the
  average plaquette, the string tension, $\sigma$, and the  mass gap, $m_G$.}
\label{table_param_SU6}
\end{table}

\begin{table}[htb]
\centering
\begin{tabular}{|cc|ccc|} \hline
\multicolumn{5}{|c|}{SU(8) ; $l\surd\sigma\sim 2.6$} \\ \hline
 $\beta$ & $l^3l_t$ & $\tfrac{1}{8}\text{ReTr}\langle U_p\rangle$ & 
$a\surd\sigma$ & $am_G$  \\ \hline
44.10  & $8^316$   & 0.5318034(31) & 0.32589(62)  & 0.8246(66)  \\
44.85  & $10^316$  & 0.5497960(15) & 0.25791(40)  & 0.7461(53)  \\
45.50  & $12^320$  & 0.5622253(9)  & 0.21851(45)  & 0.6409(38)  \\
46.10  & $14^320$  & 0.5723242(8)  & 0.18932(38)  & 0.5617(43)  \\
46.70  & $16^324$  & 0.5815072(5)  & 0.16557(38)  & 0.4909(43)  \\
47.75  & $20^330$  & 0.5959878(3)  & 0.13253(26)  & 0.4075(28)  \\ \hline
\end{tabular}
\caption{Parameters of the main $SU(8)$ calculations: the inverse coupling, $\beta$, the lattice size, the
  average plaquette, the string tension, $\sigma$, and the  mass gap, $m_G$.}
\label{table_param_SU8}
\end{table}

\begin{table}[htb]
\centering
\begin{tabular}{|cc|ccc|} \hline
\multicolumn{5}{|c|}{SU(10) ; $l\surd\sigma\sim 2.6$} \\ \hline
 $\beta$ & $l^3l_t$ & $\tfrac{1}{10}\text{ReTr}\langle U_p\rangle$ & 
$a\surd\sigma$ & $am_G$  \\ \hline
69.20    & $8^316$   & 0.5292925(29) & 0.33024(35)  & 0.8282(60)  \\
70.38    & $10^316$  & 0.5478565(13) & 0.25987(30)  & 0.7435(42)  \\
71.38    & $12^320$  & 0.5602903(9)  & 0.21988(32)  & 0.6451(44)  \\
72.40    & $14^320$  & 0.5713707(6)  & 0.18845(20)  & 0.5549(59)  \\
73.35    & $16^324$  & 0.5807004(4)  & 0.16399(19)  & 0.4952(44)  \\ \hline
\end{tabular}
\caption{Parameters of the main $SU(10)$ calculations: the inverse coupling, $\beta$, the lattice size, the
  average plaquette, the string tension, $\sigma$, and the  mass gap, $m_G$.}
\label{table_param_SU10}
\end{table}

\begin{table}[htb]
\centering
\begin{tabular}{|cc|ccc|} \hline
\multicolumn{5}{|c|}{SU(12) ; $l\surd\sigma\sim 2.6$} \\ \hline
$\beta$ & $l^3l_t$ & $\tfrac{1}{12}\text{ReTr}\langle U_p\rangle$ & 
$a\surd\sigma$ & $am_G$  \\ \hline
 99.86   & $8^316$   & 0.5275951(27)   & 0.33341(40)  & 0.8243(52)  \\
 101.55  & $10^316$  & 0.5464461(12)   & 0.26162(32)  & 0.7384(51)  \\
 103.03  & $12^320$  & 0.55936304(61)  & 0.21915(25)  & 0.6432(32)  \\
 104.55  & $14^320$  & 0.57087665(49)  & 0.18663(38)  & 0.5521(42)  \\
 105.95  & $16^324$  & 0.58043063(32)  & 0.16197(27)  & 0.4920(37)  \\ \hline
\end{tabular}
\caption{Parameters of the main $SU(12)$ calculations: the inverse coupling, $\beta$, the lattice size, the
  average plaquette, the string tension, $\sigma$, and the  mass gap, $m_G$.}
\label{table_param_SU12}
\end{table}

\begin{table}[htb]
\centering
\begin{tabular}{|c|c|c||c|c|c|} \hline
\multicolumn{6}{|c|}{$aE_{eff}(t=2a)$ : $SU(8)$ at $\beta=45.50$ on $14^320$} \\ \hline
  $R^{PC}$   & all $Q$ & $Q=0$ &  $R^{PC}$   & all $Q$ & $Q=0$  \\ \hline \hline
$A_1^{++}$  & 0.6512(21) & 0.6497(25) & $A_1^{-+}$ & 1.119(8)  & 1.098(9)  \\
          &  1.199(9)  & 1.209(9)  &          & 1.581(23)  & 1.587(21)  \\
          &  1.243(12) & 1.236(11) &          & 1.908(43)  & 1.937(51)  \\
          &  1.565(21) & 1.551(23) &          &            &           \\ \hline
$A_2^{++}$  &  1.668(27) & 1.642(25) & $A_2^{-+}$ & 2.21(8)  & 2.51(11)  \\
          &  1.983(48) & 2.037(48) &          &  2.30(9)  & 2.23(11)  \\ \hline
$E^{++}$   & 1.020(4)  & 1.020(5)   & $E^{-+}$  & 1.354(9)  & 1.356(11)  \\
          & 1.230(8)  & 1.233(8)   &          & 1.757(24)  & 1.765(25)  \\
          & 1.435(10) & 1.424(9)   &          & 2.060(51)  & 2.157(50)  \\ \hline
$T_1^{++}$  & 1.670(14) & 1.673(18) & $T_1^{-+}$ & 1.841(25)   & 1.861(24)  \\
          & 1.728(20) & 1.708(19) &          & 1.976(27)   & 1.950(26)  \\
          & 2.011(27) & 2.041(35) &          & 1.925(28)   & 1.928(28)  \\
          & 2.053(35) & 2.037(31) &          &             &           \\ \hline
$T_2^{++}$  & 1.032(4)  & 1.033(3) & $T_2^{-+}$ & 1.360(8)    &  1.356(9) \\
          & 1.460(9)  & 1.447(8)  &          & 1.727(15)   & 1.714(17)  \\
          & 1.670(15) & 1.648(18) &          & 1.881(20)   & 1.900(24)  \\
          & 1.692(14) & 1.691(17) &          &             &          \\ \hline
$A_1^{+-}$  & 2.10(7)  & 2.10(8)   & $A_1^{--}$ & 2.10(7)   & 2.10(8)  \\
          & 2.33(9)   & 2.41(12)   &         & 2.41(12)  & 2.40(13)  \\ \hline
$A_2^{+-}$  & 1.557(18) & 1.577(21) & $A_2^{--}$ & 2.00(5)  & 2.06(7)  \\
          & 1.852(29)  & 1.754(26) &          & 2.15(8)  & 2.25(9)  \\
          & 2.13(6)    & 2.27(10)  &          &          &           \\ \hline
$E^{+-}$   & 1.980(35)  & 1.981(33) & $E^{--}$ & 1.690(18) & 1.686(21)  \\
          & 2.155(48)  & 2.119(41) &          & 2.072(39) & 1.998(42)  \\
          & 2.198(54)  & 2.087(50) &          & 2.198(57) & 2.204(66)  \\ \hline
$T_1^{+-}$  & 1.266(5)   & 1.271(6)  & $T_1^{--}$ & 1.738(18) & 1.734(18)  \\
          & 1.542(9)   & 1.557(10) &          & 1.973(29) &  1.920(26) \\
          & 1.656(13)  & 1.648(11) &          & 2.088(32) &  2.001(31) \\
          & 1.859(18)  & 1.860(19) &          &           &    \\ \hline
$T_2^{+-}$  & 1.571(11)  & 1.543(11) & $T_2^{--}$ & 1.721(15) & 1.715(21)  \\
          & 1.880(19)  & 1.881(20) &          & 1.888(23) & 1.889(28)  \\
          & 1.900(20)  & 1.915(27) &          & 2.043(27) & 2.006(26)  \\
          & 1.996(34)  & 1.994(30) &          &           &          \\ \hline \hline
$l_{k=1}$  & 0.5963(11) & 0.5932(14) & $l_{k=2}$ & 1.1204(33) & 1.1235(28)  \\
          & 1.2954(33) & 1.2918(42) &           & 1.3876(57)  & 1.3810(64)  \\ \hline
\end{tabular}
\caption{Comparison of glueball and flux tube energies obtained on fields with
  topological charge $Q=0$ against fields with a `normal' distribution of $Q$.
  In $SU(8)$ on a $14^320$ lattice at $\beta=45.50$. Energies extracted
  from best (variationally selected)  correlators between $t=a$ and $t=2a$.
  Glueballs labelled by representation of cubic rotation symmetry $R$, parity $P$ and charge
  conjugation $C$. Flux tubes are fundamental, $l_{k=1}$, and $k=2$, $l_{k=2}$.}
\label{table_GKvsQ_SU8_l14}
\end{table}


\clearpage

\begin{table}[htb]
\centering
\begin{tabular}{|cc|cc|cc|} \hline
  \multicolumn{2}{|c|}{SU(2) $\beta=2.427$} & \multicolumn{2}{|c|}{SU(3) $\beta=6.235$}
   & \multicolumn{2}{|c|}{SU(6) $\beta=25.35$} \\ \hline
  $l$ & $a\surd\sigma$ & $l$ & $a\surd\sigma$ & $l$ & $a\surd\sigma$ \\ \hline
12   & 0.2380(3)  & 18  & 0.1491(5)  & 12  & 0.2196(5)    \\
14   & 0.2387(4)  & 26  & 0.1499(4)  & 14  & 0.2200(7)    \\
16   & 0.2390(8)  & 34  & 0.1506(6)  & 18  & 0.2181(17)    \\
20   & 0.2399(10) &     &   &     &     \\
24   & 0.2396(16) &     &   &     &     \\ \hline
  \multicolumn{2}{|c|}{SU(4) $\beta=11.02$} & \multicolumn{2}{|c|}{SU(4) $\beta=11.60$}
   & \multicolumn{2}{|c|}{SU(8) $\beta=45.50$} \\ \hline
  $l$ & $a\surd\sigma$ & $l$ & $a\surd\sigma$ & $l$ & $a\surd\sigma$ \\ \hline
 18  & 0.2143(3)  & 24  & 0.1301(5)  & 12  &  0.2187(4)   \\
 22  & 0.2142(6)  & 30  & 0.1307(3)  & 14  &  0.2189(4)   \\ \hline
  \multicolumn{2}{|c|}{SU(10) $\beta=71.38$} & \multicolumn{2}{|c|}{SU(12) $\beta=103.03$}
   &  \multicolumn{2}{|c|}{}  \\ \hline
   $l$ & $a\surd\sigma$ & $l$ & $a\surd\sigma$ &  & \\ \hline
 12  & 0.2199(4)  & 12  & 0.2198(2)  &   &     \\
 14  & 0.2195(5)  & 14  & 0.2202(7)  &   &     \\ \hline
\end{tabular}
\caption{String tensions obtained using eqn(\ref{eqn_NG}), for  (fundamental) flux tubes
  of length $l$ for various groups. A test of finite volume corrections.}
\label{table_V_k1_SUN}
\end{table}

\begin{table}[htb]
\centering
\begin{tabular}{|cc|cc|cc|} \hline
  \multicolumn{2}{|c|}{SU(4) $\beta=11.02$} & \multicolumn{2}{|c|}{SU(4) $\beta=11.60$}
   & \multicolumn{2}{|c|}{SU(6) $\beta=25.35$} \\ \hline
  $l$ & $a\surd\sigma_{k=2}$ & $l$ & $a\surd\sigma_{k=2}$ & $l$ & $a\surd\sigma_{k=2}$ \\ \hline
 18  & 0.2490(12)  & 24  & 0.1523(5)  & 12  & 0.2779(9) \\
 22  & 0.2480(24)  & 30  & 0.1534(5)  & 14  & 0.2846(6) \\
     &             &     &            & 18  & 0.2841(8) \\  \hline
 \multicolumn{2}{|c|}{SU(8) $\beta=45.50$} & \multicolumn{2}{|c|}{SU(10) $\beta=71.38$} &
 \multicolumn{2}{|c|}{SU(12) $\beta=103.03$} \\ \hline
  $l$ & $a\surd\sigma_{k=2}$ & $l$ & $a\surd\sigma_{k=2}$ & $l$ & $a\surd\sigma_{k=2}$ \\ \hline
 12  & 0.2833(9)  & 12  & 0.2901(10)  & 12  & 0.2925(11)    \\
 14  & 0.2885(24) & 14  & 0.2946(22)  & 14  & 0.2987(23)    \\ \hline
\end{tabular}
\caption{String tensions obtained using eqn(\ref{eqn_NG})  for  $k=2$  flux tubes
  of length $l$ for various groups. Testing finite volume corrections.}
\label{table_V_k2_SUN}
\end{table}

\begin{table}[htb]
\centering
\begin{tabular}{|cc|ccc|} \hline
\multicolumn{5}{|c||}{SU(2)} \\ \hline
$\beta$ & lattice & $\tfrac{1}{2}\text{ReTr}\langle U_p\rangle$ & $aE_f$ & $a\surd\sigma$ \\ \hline
2.2986  & $8^316$   & 0.6018323(78)  & 0.9310(55) & 0.36590(93)   \\
2.3714  & $10^316$  & 0.6227129(41)  & 0.7159(41) & 0.28779(71)   \\
2.427   & $12^316$  & 0.6364295(26)  & 0.5831(32) & 0.23750(56)   \\
2.452   & $14^320$  & 0.6420346(28)  & 0.5852(51) & 0.21791(83)   \\
2.509   & $16^320$  & 0.6537206(12)  & 0.4399(15) & 0.17857(26)   \\
2.60    & $22^330$  & 0.6700085(8)   & 0.3370(17) & 0.13279(29)   \\
2.65    & $26^334$  & 0.6780431(5)   & 0.2914(15) & 0.11342(26)   \\
2.70    & $28^340$  & 0.6855710(4)   & 0.2228(13) & 0.09698(24)   \\
2.75    & $34^346$  & 0.6926656(2)   & 0.2048(16) & 0.08375(21)   \\
2.80    & $40^354$  & 0.6993804(2)   & 0.1817(16) & 0.07242(28)   \\ \hline
\end{tabular}
\caption{Energies of periodic flux tubes of length $l$ in $SU(2)$ and derived string tensions
  at the couplings $\beta=4/g^2$ on the lattices $l^3l_t$.}
\label{table_Ksmall_SU2}
\end{table}

\begin{table}[htb]
\centering
\begin{tabular}{|cc|ccc|} \hline
\multicolumn{5}{|c|}{SU(3)} \\ \hline
$\beta$ & lattice & $\tfrac{1}{3}\text{ReTr}\langle U_p\rangle$ & $aE_f$ & $a\surd\sigma$ \\ \hline
5.6924 & $8^316$  & 0.547503(7)   & 1.1588(37) & 0.4010(23)  \\
5.80   & $10^316$ & 0.567642(5)   & 0.8862(26) & 0.31603(84) \\
5.8941 & $12^316$ & 0.581069(4)   & 0.7298(47) & 0.2613(14) \\
5.99   & $14^320$ & 0.5925655(13) & 0.5984(26) & 0.21959(76) \\
6.0625 & $14^320$ & 0.6003331(21) & 0.4517(25) & 0.19509(65) \\
6.235  & $18^326$ & 0.6167715(13) & 0.3369(26) & 0.14899(59) \\
6.338  & $22^330$ & 0.6255948(8)  & 0.3182(19) & 0.12948(44) \\
6.50   & $26^338$ & 0.6383532(5)  & 0.2334(17) & 0.10319(41) \\
6.60   & $32^340$ & 0.64566194(21) & 0.2255(15) & 0.09024(26)  \\
6.70   & $36^344$ & 0.65260033(18) & 0.1933(16) & 0.07898(28)  \\ \hline
\end{tabular}
\caption{Energies of periodic flux tubes of length $l$ in $SU(3)$ and derived string tensions
  at the couplings $\beta=6/g^2$ on the lattices $l^3l_t$.}
\label{table_Ksmall_SU3}
\end{table}

\begin{table}[htb]
\centering
\begin{tabular}{|c||c|c||c|} \hline
\multicolumn{4}{|c|}{continuum $k=2$ string tensions} \\ \hline
 group   &  $\sigma_k/\sigma_f$:NG  &  $\sigma_k/\sigma_f$:$l\to\infty$ & Casimir scaling \\ \hline
 $SU(4)$   & 1.381(14)  & 1.381(14)  & 1.333  \\
 $SU(5)$   & 1.551(11)  & 1.551(11)  & 1.500  \\
 $SU(6)$   & 1.654(13)  & 1.654(13)  & 1.600  \\
 $SU(8)$   & 1.731(11)  & 1.794(28)  & 1.714  \\
 $SU(10)$  & 1.733(15)  & 1.796(29)  & 1.778   \\
 $SU(12)$  & 1.792(16)  & 1.857(29)  & 1.818   \\ \hline
\end{tabular}
\caption{Ratio of the $k=2$ string tension to the fundamental for various $SU(N)$ gauge theories.
  Values labelled NG are obtained using eqn(\ref{eqn_NG}). Values labelled $l\to\infty$ denote our
  best estimates in that limit. The values corresponding to `Casimir scaling'
  are shown for comparison.}
\label{table_sigmak2}
\end{table}


\begin{table}
\begin{center}
\begin{tabular}{|c|ccc|c|c|}\hline
  $N$ & $a\sqrt{\sigma}\in$ & $\sqrt{\sigma}/\Lambda^{3loop}_I$ & $\chi^2/n_{df}$
  & $\sqrt{\sigma}/\Lambda^{2loop}_I$ & $\Lambda^{3loop}_{\overline{MS}}/\sqrt{\sigma}$ \\ \hline  
2  & [0.133,0.072] & 4.535(16)  & 0.42 & 4.914(16) &  0.5806(21)[570]  \\
3  & [0.195,0.079] & 4.855(11)  & 1.40 & 5.210(20) &  0.5424(13)[185]  \\ 
4  & [0.254,0.131] & 5.043(10)  & 0.48 & 5.532(11) &  0.5222(11)[230]  \\ 
5  & [0.255,0.131] & 5.090(14)  & 0.31 & 5.622(15) &  0.5174(15)[245]  \\ 
6  & [0.252,0.129] & 5.105(11)  & 1.62 & 5.664(48)$^{\ast}$ & 0.5158(11)[250]   \\ 
8  & [0.258,0.133] & 5.148(17)  & 0.13 & 5.727(20) &  0.5115(17)[250]  \\ 
10 & [0.260,0.164] & 5.221(40)$^{\ast}$  & 3.30  & 5.845(44)$^{\ast}$ & 0.5044(20)[270]   \\ 
12 & [0.262,0.162] & 5.193(13)  & 0.45 & 5.823(15) &  0.5071(13)[270]  \\ \hline 
\end{tabular}
\caption{Results for $\Lambda_I$ in units of the string tension using the exact 3-loop
  $\beta$-function, with the 2-loop result for comparison. Poor fits denoted by $\ast$.
  Resulting $\Lambda_{\overline{MS}}$ is shown with statistical errors and an estimate
  of the (correlated) systematic error in square brackets.}
\label{table_Lambda_fitN}
\end{center}
\end{table}

\begin{table}
\begin{center}
\begin{tabular}{|c|c|c|c|c|c|c|}\hline 
$N$ &  $c_0$ & $c_{\sigma}$ & $a\sqrt{\sigma}\in$ & $\beta\in$ & $\chi^2/n_{df}$ & $n_{df}$ \\ \hline
2  & 4.510(15)  & 4.98(25)  & [0.133,0.072] & [2.60,2.80]   &  0.40 & 3   \\
3  & 4.827(12)  & 2.48(11)  & [0.195,0.079] & [5.99,6.70]   &  1.29 & 4 \\
4  & 5.017(8)   & 1.622(40) & [0.302,0.131] & [10.70,11.60] &  0.98 & 4  \\ 
5  & 5.068(11)  & 1.435(56) & [0.303,0.131] & [16.98,18.375] & 0.38 & 4  \\ 
6  & 5.064(12)  & 1.541(68) & [0.252,0.129] & [25.05,26.71]  & 1.45 & 3  \\ 
8  & 5.143(9)   & 1.265(34) & [0.326,0.133] & [44.10,47.75]  & 1.75 & 4  \\
10 & 5.176(10)  & 1.285(35) & [0.260,0.164] & [70.38,73.35]  & 3.06 & 2   \\
12 & 5.148(12)  & 1.387(43) & [0.262,0.162] & [101.55,105.95] & 0.41 & 2  \\  \hline
\end{tabular}
\caption{Fitted values of the coefficients $c_{0}$ and  $c_{\sigma}$ that
  determine the interpolation function in eqn(\ref{eqn_aKgI}) for each of our $SU(N)$
  lattice gauge theories. In each case the range in $a\surd\sigma$ and of $\beta$ of
  the fit is shown as is the $\chi^2$ per degree of freedom and the number
  of degrees of freedom.}
\label{table_interp_gI}
\end{center}
\end{table}

\begin{table}[htb]
\centering
\begin{tabular}{|c|c|c|c|} \hline
\multicolumn{4}{|c|}{$aE_{eff}(t=2a)$ : $SU(2)$ at $\beta=2.427$} \\ \hline
  $R^{P}$  &  $12^316$ & $14^316$  &  $20^316$  \\ \hline
$A_1^{+}$  & 0.840(11)   & 0.854(8) & 0.847(8)  \\
           & 1.225(26)   &           &            \\ 
           & 1.418(17)   & 1.367(15) & 1.408(19)  \\ 
           &             & 1.593(27) &            \\ 
           & 1.611(36)   & 1.721(51) & 1.780(45)  \\ 
           & 1.871(40)   & 1.839(46) & 1.872(46)  \\  \hline
$A_2^{+}$  & 1.680(40)   & 1.805(38) & 1.942(72)  \\  \hline
$E^{+}$    & 1.220(9)    & 1.254(14) & 1.284(13)  \\
           & 1.317(17)   &           &            \\ 
           &             & 1.549(20) &            \\ 
           & 1.632(30)   & 1.672(33) &  1.672(40) \\ 
           & 1.774(31)   & 1.820(39) &  1.832(42) \\   \hline
$T_1^{+}$  & 1.925(42)   & 1.875(27) & 1.855(37)  \\ 
           & 1.977(48)   & 1.972(34) & 1.916(50)  \\  \hline
$T_2^{+}$  & 1.297(12)   & 1.278(14) & 1.289(11)  \\ 
           & 1.739(20)   & 1.701(30) & 1.672(21)  \\ 
           & 1.774(34)   & 1.849(22) & 1.877(30)  \\  \hline
$A_1^{-}$  & 1.471(30)   & 1.540(30) & 1.526(36)  \\
           & 1.71(4)     & 1.87(9)   & 1.94(9)  \\  \hline
$A_2^{-}$  & 2.31(16)   & 2.21(17)   & 2.04(12)  \\  \hline
$E^{-}$    & 1.585(31)   & 1.680(28) & 1.652(34)  \\
           &  2.00(6)    & 2.04(7)   & 2.02(7)  \\ \hline
$T_1^{-}$  &  2.18(6)    & 2.15(6)   & 2.09(7)  \\ \hline
$T_2^{-}$  &  1.665(25)  & 1.632(33) & 1.620(21)  \\
           &  2.04(5)    & 2.00(4)   & 2.01(6)  \\ \hline
$l_{k=1}$   & 0.5804(38)   & 0.7140(41) & 1.0996(39)  \\
           & 1.367(23)    & 1.428(32)  & 1.677(9)  \\ \hline
\end{tabular}
\caption{Comparison of glueball  effective energies from $t=2a$ and flux tube energies ($l1$)
  obtained on $12^316$, $14^316$  and $20^316$ lattices at $\beta=2.427$ in $SU(2)$.
  Unmatched states are ditorelons.
  Glueballs labelled by representation of cubic rotation symmetry $R$ and parity $P$.
  Fundamental flux tube energies , $l_{k=1}$, are ground and first excited states.}
\label{table_GvsV_SU2}
\end{table}

\clearpage


\begin{table}[htb]
\centering
\begin{tabular}{|c|c|c||c|c|c|} \hline
\multicolumn{6}{|c|}{$aE_{eff}(t=2a)$ : $SU(12)$ at $\beta=103.03$} \\ \hline
  $R^{PC}$   & $12^320$ & $14^320$  &  $R^{PC}$   & $12^320$ & $14^320$  \\ \hline  \hline 
$A_1^{++}$  & 0.6434(32) & 0.6499(50)  &  $A_1^{-+}$ & 1.079(8) & 1.097(12) \\
          & 1.190(9)   & 1.200(14)   &            & 1.567(22)  & 1.591(33)  \\
          & 1.495(19)  & 1.579(25)   &            & 1.859(38)  & 2.037(91) \\
          & 1.604(26)  & 1.585(33)   &            &            &   \\ 
          & 1.660(26)  & 1.687(43)   &            &            &   \\ \hline
$A_2^{++}$  & 1.629(30) & 1.639(39)  & $A_2^{-+}$ & 2.14(10) & 2.21(11)  \\
          & 1.91(5)    & 2.11(10)    &          & 2.20(9)   & 2.30(13)  \\ \hline
$E^{++}$   & 1.021(5)   & 1.026(7)   & $E^{-+}$  & 1.346(10)  & 1.355(15)  \\
          & 1.429(14)  &  1.464(16)  &          & 1.723(21)  & 1.747(32)  \\
          & 1.610(19)  &  1.632(28)  &          & 2.105(33)  & 2.178(74)  \\ \hline
$T_1^{++}$  & 1.635(17) & 1.692(21) & $T_1^{-+}$ & 1.865(21)  & 1.885(39)  \\
          & 1.712(19) & 1.776(26)  &          & 1.945(30)  & 1.930(36)  \\
          & 2.090(32) & 2.062(47)  &          & 2.012(28)  & 2.026(49)  \\ \hline
$T_2^{++}$  & 1.040(4) & 1.031(7)  & $T_2^{-+}$ & 1.357(8) & 1.366(12)  \\
          & 1.465(9) & 1.473(16)  &          & 1.730(16)  & 1.748(23)  \\
          & 1.597(14) & 1.659(21)  &          & 1.891(26)  & 1.905(44)  \\
          & 1.652(14) & 1.676(25)  &          & 1.932(29)  & 1.975(40)  \\ \hline
$A_1^{+-}$  & 2.001(52) & 2.06(11)  & $A_1^{--}$ & 2.18(8)  & 2.09(11)  \\
          & 2.12(8)   & 2.23(13)   &            & 2.23(10) & 2.30(12)  \\ \hline
$A_2^{+-}$  & 1.564(22)  & 1.603(30)  & $A_2^{--}$ & 1.932(52) & 2.026(70)  \\
          & 1.823(36)  & 1.882(53)  &          & 2.209(85)  & 2.33(17)  \\
          & 1.97(8)    & 2.19(10)   &          &            &   \\ \hline
$E^{+-}$   & 1.903(29) & 2.011(43)  & $E^{--}$ & 1.682(17) & 1.677(30)  \\
          & 2.140(51) & 2.044(70)  &          & 2.053(44) & 2.087(54)  \\
          &           &            &          & 2.11(5)   & 2.26(9)   \\ \hline
$T_1^{+-}$  & 1.269(6)  & 1.268(8)  & $T_1^{--}$ & 1.696(18)  & 1.693(22)  \\ 
          & 1.543(12)  & 1.546(16)  &          & 1.921(25)  & 1.897(44)  \\
          & 1.678(16)  & 1.690(18)  &          & 1.976(33)  & 2.023(45)  \\
          & 1.850(21)  & 1.880(27)  &          & 2.21(6)    & 2.18(6) \\ \hline
$T_2^{+-}$  & 1.534(13)  & 1.541(15) & $T_2^{--}$ & 1.719(20) & 1.766(27)  \\
          & 1.878(24)  & 1.870(27)  &          & 1.893(29)  & 1.935(31)  \\
          & 1.850(22)  & 1.896(36)  &          & 2.014(32)  & 2.035(37)  \\
          & 1.988(32)  & 1.987(43)  &          & 2.19(5)    & 2.19(6)  \\ \hline \hline
$l_{k=1}$  & 0.4812(13) & 0.5993(40) &  $l_{k=2}$ & 0.935(8)   & 1.172(18)  \\ \hline
\end{tabular}
\caption{Comparison of glueball effective energies from $t=2a$  and flux tube energies ($l1$,$l2$)
  obtained on $12^320$ and $14^320$ lattices at $\beta=103.03$ in $SU(12)$. 
  Glueballs labelled by representation of cubic rotation symmetry $R$, parity $P$ and charge
  conjugation $C$. Flux tube energies are fundamental, $l_{k=1}$, and $k=2$, $l_{k=2}$.}
\label{table_GvsV_SU12}
\end{table}

\begin{table}[htb]
\centering
\begin{tabular}{|c|cc|cc|} \hline
\multicolumn{5}{|c|}{$aE_{eff}(t=2a)$ : $SU(12)$ at $\beta=103.03$} \\ \hline
$R^{PC}$   & \multicolumn{2}{|c|}{$12^320$} &  \multicolumn{2}{|c|}{$14^320$} \\ 
          & $bl=1-4$  &  $bl=1-5$   &  $bl=1-4$  &  $bl=1-5$ \\ \hline 
$A_1^{++}$ & 0.6434(32) & 0.6432(32) & 0.6499(50) & 0.6497(50)  \\
          &            & 1.172(10)  &            &           \\
          & 1.190(9)   & 1.179(14)  & 1.200(14)  & 1.200(14) \\
          &            &            &            & 1.424(23) \\ 
          & 1.462(20)  & 1.495(19)  & 1.588(23)  & 1.433(29) \\ 
          & 1.604(26)  & 1.601(29)  & 1.585(33)  & 1.571(35) \\
          & 1.660(26)  & 1.666(30)  & 1.687(43)  & 1.659(40) \\ 
          & 1.753(25)  & 1.780(31)  & 1.786(54)  & 1.780(52) \\ \hline 
$E^{++}$   & 1.021(5)    & 1.019(5)  & 1.026(7)   & 1.026(7) \\
          & 1.525(13)[1.01(14)] & 1.118(7)  &    &          \\
          &             &           &            & 1.273(20) \\
          & 1.429(14)   & 1,373(12)  &  1.464(16)  & 1.446(16) \\
          & 1.610(19)   & 1.610(18)  &  1.635(29)  & 1.632(28) \\
          & 1.657(19)   & 1.650(18)  &  1.663(24)  & 1.666(25) \\ \hline 
\end{tabular}
\caption{Comparison of $A_1^{++}$ and $E^{++}$ glueball effective energies from $t=2a$  
  obtained on $12^320$ and $14^320$ lattices at $\beta=103.03$ in $SU(12)$.
  Using operators up to blocking levels 4 and 5 respectively, as shown.}
\label{table_GvsV_SU12B}
\end{table}

\begin{table}[htb]
\centering
\begin{tabular}{|c|c|c|c|c|} \hline
\multicolumn{5}{|c|}{number of operators: $N\in [4,12]$} \\ \hline
  $R$   & P=+,C=+ & P=-,C=+ &  P=+,C=-   &  P=-,C=-   \\ \hline
$A_1$  & 12 & 7   & 6   & 8 \\
$A_2$  &  7 & 5   & 7   & 7 \\ 
$E$    & 36 & 24  & 24  & 30 \\
$T_1$  & 48 & 54  & 66  & 51 \\
$T_2$  & 60 & 60  & 60  & 54 \\ \hline
\multicolumn{5}{|c|}{number of operators: $N=2,3$} \\ \hline
  $R$   & P=+,C=+ & P=-,C=+ &  P=+,C=-   &  P=-,C=-   \\ \hline
$A_1$  & 27  & 9  & 8   & 11 \\ 
$A_2$  & 14  & 6  & 13  & 11 \\ 
$E$    & 80  & 30 & 40  & 44 \\ 
$T_1$  & 78  & 84(75) & 132 & 81 \\ 
$T_2$  & 108 & 96(87) & 108 & 84 \\ \hline
\end{tabular}
\caption{Number of operators in rotational representation , $R$, with parity, $P$, and charge conjugation, $C$,
  as used in our various $SU(N)$ calculations, at each blocking level. Some numbers for $SU(2)$
  differ from $SU(3)$ and are in brackets; also no $C=-$ for $SU(2)$} 
\label{table_numops_N}
\end{table}

\begin{table}[htb]
\centering
\begin{tabular}{|c|c|} \hline
\multicolumn{2}{|c|}{operator loops} \\  \hline
  $loop$   &   $R^{PC}$     \\ \hline
  \{2,3,-2,-3\}  &   $A_1^{++},E^{++}$ \\ 
                 &   $T_1^{+-}$ \\ \hline
  \{1,2,2,-1,-2,-2\} &   $A_1^{++},A_2^{++},E^{++}$   \\ 
                 &   $T_1^{+-},T_2^{+-}$  \\ \hline
  \{1,2,3,-1,-2,-3\} &   $A_1^{++},T_2^{++}$   \\ 
                 &  $A_2^{+-},T_1^{+-}$   \\ \hline
  \{1,3,2,-3,-1,-2\} &   $A_1^{++},E^{++},T_2^{++},T_1^{-+},T_2^{-+}$   \\ 
                 &  $T_1^{+-},T_2^{+-},A_1^{--},E^{--},T_2^{--}$   \\ \hline
  \{1,2,2,-1,3,-2,-3,-2\} &  $A_1^{++},A_2^{++},E^{++},T_1^{++},T_2^{++},A_1^{-+},A_2^{-+},E^{-+},T_1^{-+},T_2^{-+}$    \\ 
                 &    $A_1^{+-},A_2^{+-},E^{+-},T_1^{+-},T_2^{+-},A_1^{--},A_2^{--},E^{--},T_1^{--},T_2^{--}$  \\ \hline
  \{1,3,-1,-3,-1,-2,1,2\} &  $A_1^{++},E^{++},T_1^{++},T_2^{++},A_1^{-+},E^{-+},T_1^{-+},T_2^{-+}$    \\ 
                 &   $A_1^{+-},E^{+-},T_1^{+-},T_2^{+-},A_1^{--},E^{--},T_1^{--},T_2^{--}$  \\ \hline
  \{1,2,3,-1,-3,-3,-2,3\} &  $A_1^{++},E^{++},T_1^{++},T_2^{++},A_1^{-+},E^{-+},T_1^{-+},T_2^{-+}$    \\ 
                 &   $A_2^{+-},E^{+-},T_1^{+-},T_2^{+-},A_2^{--},E^{--},T_1^{--},T_2^{--}$  \\ \hline
  \{1,3,1,2,-3,-1,-1,-2\} &   $A_1^{++},A_2^{++},E^{++},T_1^{++},T_2^{++},A_1^{-+},A_2^{-+},E^{-+},T_1^{-+},T_2^{-+}$   \\ 
                 &   $A_1^{+-},A_2^{+-},E^{+-},T_1^{+-},T_2^{+-},A_1^{--},A_2^{--},E^{--},T_1^{--},T_2^{--}$  \\ \hline
  \{1,2,2,2,-1,3,-2,-3,-2,-2\} &  $A_1^{++},A_2^{++},E^{++},T_1^{++},T_2^{++},A_1^{-+},A_2^{-+},E^{-+},T_1^{-+},T_2^{-+}$    \\ 
                 &   $A_1^{+-},A_2^{+-},E^{+-},T_1^{+-},T_2^{+-},A_1^{--},A_2^{--},E^{--},T_1^{--},T_2^{--}$  \\ \hline
  \{1,2,2,2,-1,-2,3,-2,-3,-2\} &   $A_1^{++},A_2^{++},E^{++},T_1^{++},T_2^{++},T_1^{-+},T_2^{-+}$   \\ 
                 &  $T_1^{+-},T_2^{+-},A_1^{--},A_2^{--},E^{--},T_1^{--},T_2^{--}$    \\ \hline
  \{-3,1,3,1,2,-3,-1,3,-1,-2\} &   $A_1^{++},A_2^{++},E^{++},T_1^{++},T_2^{++},A_1^{-+},A_2^{-+},E^{-+},T_1^{-+},T_2^{-+}$    \\ 
                 &   $A_1^{+-},A_2^{+-},E^{+-},T_1^{+-},T_2^{+-},A_1^{--},A_2^{--},E^{--},T_1^{--},T_2^{--}$  \\ \hline
  \{-3,1,3,1,2,3,-1,-3,-2,-1\} &    $A_1^{++},A_2^{++},E^{++},T_1^{++},T_2^{++},A_1^{-+},A_2^{-+},E^{-+},T_1^{-+},T_2^{-+}$   \\  
                 &   $A_1^{+-},A_2^{+-},E^{+-},T_1^{+-},T_2^{+-},A_1^{--},A_2^{--},E^{--},T_1^{--},T_2^{--}$  \\ \hline
\end{tabular}
\caption{The 12 loops used as the basis of our glueball calculations for $N\geq 4$. These generate contributions
  to the representations as shown.}
\label{table_loops}
\end{table}






\begin{table}[htb]
\centering
\begin{tabular}{|c|c|c|c|c|c|c|} \hline
\multicolumn{7}{|c|}{$SU(8):aM_G$} \\ \hline
 $R^{PC}$ & $\beta=44.10$ & $\beta=44.85$ & $\beta=45.50$ & $\beta=46.10$ & $\beta=46.70$  & $\beta=47.75$    \\ 
          & $8^316$ & $10^316$ & $12^320$  & $14^320$ & $16^324$ & $20^330$    \\ \hline \hline
$A^{++}_1$ & 0.8246(66) & 0.7461(53) & 0.6409(38) & 0.5617(43) & 0.4909(43) & 0.4075(28)   \\
          & 1.615(26)  & 1.191(50)  & 1.112(37) & 1.081(22) & 0.930(13) & 0.755(10)   \\
          & 2.01(6)    & 1.764(39)  & 1.42(9)  & 1.252(41) & 1.146(27)  & 0.936(18)   \\ \hline
$A^{++}_2$ & 2.42(11)   & 1.828(46)  & 1.615(25) & 1.408(26) & 1.180(27) & 0.969(16)    \\ \hline
$E^{++}$   & 1.504(13)  & 1.184(8)  & 1.0063(51) & 0.8813(48) & 0.7676(67) & 0.6192(69)   \\
          &  2.145(41) & 1.689(18) & 1.442(11)  & 1.299(34)  & 1.070(16)  & 0.831(8)  \\ 
          &            & 1.855(28) & 1.579(15)  & 1.393(13) & 1.184(22)  &  0.954(11)  \\ \hline
$T^{++}_1$ & 2.50(8)   & 1.970(29)  & 1.517(27)  & 1.397(40) & 1.204(21)  & 0.982(7)   \\
          &           & 1.932(29)  & 1.536(90)  & 1,392(43) & 1.238(29)  & 0.996(10) \\ \hline
$T^{++}_2$ & 1.536(12)  & 1.191(22)  & 1.033(10) & 0.8795(58) & 0.7725(64) & 0.6195(22)   \\
          & 2.092(41)  & 1.689(15)  & 1.428(39) & 1.188(21) & 1.101(12)  & 0.870(16)   \\ 
          & 2.41(8)  & 1.875(28)  & 1.621(76)  & 1.357(34) & 1.162(14)  & 0.963(8)   \\ 
          &         & 1.889(30)  & 1.52(8)  & 1.344(35) & 1.201(18)  & 0.975(24)   \\ \hline
$A^{-+}_1$ & 1.618(33) & 1.256(13)  & 1.009(20)  & 0.902(17) & 0.8021(95) & 0.6365(81)   \\
          & 2.21(10)  & 1.75(5)   & 1.42(8)   & 1.23(5)  & 1.074(82)   &  0.963(35)  \\ \hline
$A^{-+}_2$ &           & 2.27(13)  & 2.07(6)   & 1.857(38) & 1.588(32)  &  1.257(33) \\ \hline
$E^{-+}$   & 1.994(56) & 1.573(17) & 1.295(41)  & 1.131(21) & 1.013(11)  & 0.8082(65)   \\
          & 2.72(14)  & 2.019(54)  & 1.734(20) & 1.498(14) & 1.320(9)   & 1.101(15)  \\ \hline
$T^{-+}_1$ &           & 2.29(15)  & 1.844(23)  & 1.571(65) & 1.324(32)  & 1.108(11)   \\ \hline
c          &           &           & 1.990(24)  & 1.639(12) & 1.447(39)  & 1.142(18)  \\ \hline
$T^{-+}_2$ & 1.959(26)  & 1.47(5)  & 1.271(24)  & 1.146(16) & 0.982(10)  & 0.807(10)  \\
          & 2.62(11)  & 2.043(43) & 1.741(18)  & 1.506(11) & 1.298(25)  & 1.065(11)  \\ \hline
$A^{+-}_1$ &          & 2.47(16)  & 2.103(73)  & 1.831(40) & 1.552(19)  & 1.209(32)   \\ \hline
$A^{+-}_2$ &  2.25(10) & 1.834(37)  & 1.529(18) & 1.275(44) & 1.176(26)  & 0.928(13)   \\ \hline
$E^{+-}$   &           & 2.196(52) & 1.921(33) & 1.659(26) & 1.463(38) & 1.117(18)    \\ \hline
$T^{+-}_1$ & 1.898(26)  & 1.510(12)  & 1.281(7)  & 1.102(6) & 0.9556(82)  & 0.7706(46)   \\
          & 2.23(6)   & 1.812(27)  & 1.539(13)  & 1.328(6) & 1.1724(41)  & 0.9269(78)   \\
          &          & 1.914(21)  & 1.648(14)  & 1.437(10) & 1.221(16)  & 0.980(8)   \\ \hline
$T^{+-}_2$ & 2.33(6)  & 1.781(21)  & 1.544(15)  & 1.294(30) & 1.133(14)  & 0.922(7)   \\
          & 2.53(12) & 2.237(46)  & 1.811(22)  & 1.50(6)  & 1.327(32)  &  1.093(15)  \\
          & 2.53(10) & 2.265(50)  & 1.815(20)  & 1,618(18) & 1.343(33) &  1.119(13)  \\ \hline
$A^{--}_1$ &         & 2.62(18)   & 2.05(7)   & 1.888(35) & 1.633(23)  &  1.254(36)  \\ \hline
$A^{--}_2$ &        & 2.22(10)  & 1.94(6)  & 1.74(17)  & 1.368(67)   &  1.114(24) \\ \hline
$E^{--}$   & 2.42(9) & 2.012(40)  & 1.663(19)  & 1.441(11) & 1.250(31)  & 1.009(14)   \\ \hline
$T^{--}_1$ & 2.55(9) & 1.906(30)  & 1.564(69)  & 1.414(42) & 1.186(22)  & 0.974(29)   \\ \hline
$T^{--}_2$ & 2.48(7) & 1.981(32)  & 1.709(16)  & 1.473(10) & 1.2920(56) & 1.037(9)  \\
          &         & 2.24(5)   & 1.896(22)  & 1.601(13) & 1.4108(81)  & 1.109(14)  \\ \hline \hline
$aE_f$    & 0.7067(33)  & 0.5506(21)  & 0.4778(24)  & 0.4204(20) & 0.3674(20)  & 0.2943(14)  \\ 
$aE_{k=2}$ & 1.268(20)  & 0.9948(85)  & 0.8716(61)  & 0.7923(69) & 0.6735(46)  & 0.5483(29)  \\ \hline
\end{tabular}
\caption{$SU(8)$ lattice glueball masses for all $R^{PC}$ representations, with
  the fundamental, $aE_f$, and $k=2$, $aE_{k=2}$, flux tube energies.} 
\label{table_Mlat_RPC_SU8}
\end{table}


\clearpage


\begin{table}[htb]
\centering
\begin{tabular}{|c|c|c|c|c|} \hline
\multicolumn{3}{|c|}{$SU(2)$: $M_G/\surd\sigma$  continuum limit} \\ \hline
  $R$   & P=+,C=+ & P=-,C=+     \\ \hline
$A_1$  &  3.781(23)  &  6.017(61)  \\
    &  6.126(38)  &  8.00(15)   \\ 
    &  7.54(10)   &             \\ \hline
$A_2$  &  7.77(18)   &  9.50(18)$^\star$ \\
    &  8.56(21)   &             \\ \hline
$E$   &  5.343(30)  &  7.037(67)  \\
    &  6.967(62)  &  8.574(83)  \\
    &  7.722(82)  &  9.58(16)   \\ \hline
$T_1$  &  8.14(10)   &  9.06(13)   \\
    &  8.46(12)   &  9.40(13)   \\
    &  9.67(9)    &  9.83(16)   \\ \hline
$T_2$  &  5.353(23)  &  6.997(65)  \\
    &  7.218(52)  &  8.468(86)  \\
    &  8.23(10)   &  9.47(10)   \\ \hline
\end{tabular}
\caption{$SU(2)$ continuum limit of glueball masses in units of the string tension,
  for all representations, $R$, of the rotation symmetry of a cube, for
  both values of parity, $P$.
  Ground states and some excited states. Stars
  indicate poor fits (see text).}
\label{table_MK_R_SU2}
\end{table}

\begin{table}[htb]
\centering
\begin{tabular}{|c|c|c|c|c|} \hline
\multicolumn{5}{|c|}{$SU(3)$: $M_G/\surd\sigma$  continuum limit} \\ \hline
  $R$   & P=+,C=+ & P=-,C=+ &  P=+,C=-   &  P=-,C=-   \\ \hline
$A_1$  & 3.405(21)  & 5.276(45) & 9.32(28)  &  9.93(49)  \\
    & 5.855(41)  & 7.29(13)  &           &  10.03(47) \\
    & 7.38(11)   & 9.18(26)  &           &           \\ 
    & 7.515(50)  & 9.37(22)  &           &            \\  \hline
$A_2$  & 7.705(85)  & 9.80(22)  & 7.384(90) &  8.96(15)  \\
    & 8.61(20)   & 11.17(30) & 8.94(10)  &  10.21(20) \\ 
    &            &           & 8.90(21)  &             \\  \hline
$E$   & 4.904(20)  & 6.211(56) & 8.77(12)  &  7.91(10)  \\
    & 6.728(47)  & 8.23(9)   & 9.03(23)  &  9.39(18)  \\ 
    & 7.49(9)    & 9.47(16)  & 10.39(21) &  10.40(22) \\ 
    & 7.531(60)  &           &           &            \\  \hline
$T_1$  & 7.698(80)  & 8.48(12)  & 6.065(40) &  8.31(10)  \\
    & 7.72(11)   & 8.57(13)  & 7.21(8)   &  9.30(14)$^{\star\star}$  \\ 
    & 9.31(11)$^\star$ & 8.66(15) & 7.824(56) & 9.72(15)  \\ 
    &            & 9.56(28)  & 8.92(10)  &            \\  \hline
$T_2$  & 4.884(19)  & 6.393(45) & 7.220(86) &  8.198(80) \\
    & 6.814(31)  & 8.15(7)   & 8.72(11)  &  8.99(11)$^{\star\star}$  \\ 
    & 7.716(70)$^{\star\star}$  & 9.23(12)$^\star$ & 9.060(80) &  9.69(13)  \\ 
    & 7.677(71)  &           &  &   \\  \hline
\end{tabular}
\caption{$SU(3)$ continuum limit of glueball masses in units of the string tension,
  for all representations, $R$, of the rotation symmetry of a cube, for
  both values of parity, $P$,
  and charge conjugation, $C$. Ground states and some excited states. Stars
  indicate poor fits (see text).}
\label{table_MK_R_SU3}
\end{table}

\begin{table}[htb]
\centering
\begin{tabular}{|c|c|c|c|c|} \hline
\multicolumn{5}{|c|}{$SU(4)$: $M_G/\surd\sigma$  continuum limit} \\ \hline
  $R$   & P=+,C=+ & P=-,C=+ &  P=+,C=-   &  P=-,C=-   \\ \hline
$A_1$  & 3.271(27)  & 5.020(46) & 9.22(22)$^\star$  & 10.27(28)  \\
    & 5.827(62)  & 7.33(11)  & 10.43(21) & 9.95(29)  \\
    & 7.50(11)   & 8.98(23)  &           &   \\
    & 7.73(8)    &           &           &   \\ \hline
$A_2$  & 7.32(12)   & 9.67(19)  & 6.87(26)  & 8.89(13)  \\
    & 8.42(22)   & 10.69(45)$^\star$ & 8.66(16) & 10.99(35)  \\
    &            &           & 9.57(19)  &   \\ \hline
$E$   & 4.721(27)  & 6.130(52) & 8.73(20)  & 7.80(11)  \\
    & 6.702(45)  & 7.91(13)  & 9.28(19)  & 9.55(13)$^\star$  \\
    & 7.271(86)  & 9.13(22)  & 9.58(20)  & 10.06(26)  \\
    & 7.586(84)  &           &           &   \\ \hline
$T_1$  & 7.42(12)   & 8.47(11)  & 5.956(42) & 7.603(84)  \\
    & 7.50(9)    & 8.59(16)  & 7.11(8)   & 8.92(12)$^{\star\star}$  \\
    & 9.10(18)   & 8.62(17)$^{\star\star}$ & 7.508(74) & 9.86(22) \\
    & 9.54(23)   &           & 8.91(13)  &            \\ \hline
$T_2$  & 4.750(16)  & 6.203(33) & 7.010(68) & 7.787(96)  \\
    & 6.687(51)  & 8.05(10)$^{\star\star}$ & 8.53(11) & 8.42(14)  \\
    & 7.411(81)  & 8.25(16)  & 8.86(8)   & 9.80(17)    \\
    & 7.492(68)  &           & 8.94(20)  &             \\ \hline
\end{tabular}
\caption{$SU(4)$ continuum limit of glueball masses in units of the string tension,
  for all representations, $R$, of the rotation symmetry of a cube, for
  both values of parity, $P$, and charge conjugation, $C$.
  Ground states and some excited states. Stars
  indicate poor fits (see text).}
\label{table_MK_R_SU4}
\end{table}

\begin{table}[htb]
\centering
\begin{tabular}{|c|c|c|c|c|} \hline
\multicolumn{5}{|c|}{$SU(5)$: $M_G/\surd\sigma$  continuum limit} \\ \hline
  $R$   & P=+,C=+ & P=-,C=+ &  P=+,C=-    &  P=-,C=-   \\ \hline
$A_1$  & 3.156(31)  & 4.832(40)  & 9.04(22)  & 10.36(18)  \\
    & 5.689(53)  & 7.38(11)   & 9.48(30)  &  9.70(36)  \\
    & 7.36(14)   & 8.03(30)$^\star$  &     &             \\
    & 7.53(14)   &            &           &             \\ \hline
$A_2$  & 7.31(14)   & 9.80(20)   & 7.333(59) & 8.05(23)  \\
    & 8.85(32)   & 9.93(57)   & 8.75(19)  & 11.03(37)  \\ \hline
$E$   & 4.692(22)  & 6.152(60)  & 8.54(19)  & 8.000(51)  \\
    & 6.590(62)  & 8.19(23)   & 9.45(18)  & 9.02(22)   \\
    & 7.31(10)   & 9.51(17)   & 9.53(20)  & 10.22(13)$^{\star\star}$  \\
    & 7.28(10)   &            &           &             \\ \hline
$T_1$  & 7.396(73)  & 8.29(13)   & 5.915(45) & 7.62(15)  \\
    & 7.19(12)   & 8.56(15)   & 7.018(62) & 8.53(17)$^{\star\star}$  \\
    & 8.99(22)   & 8.59(16)   & 7.624(74) & 9.91(26)    \\
    &            &            & 8.36(14)  &             \\ \hline
$T_2$  & 4.686(30)  & 6.208(47)  & 7.051(72) & 7.87(11)     \\
    & 6.621(55)  & 7.97(11)   & 8.28(15)$^\star$ & 8.52(9)  \\
    & 7.338(84)  & 8.45(16)   & 8.44(16)  & 10.17(9)$^{\star}$  \\
    & 7.53(10)   &            & 9.18(13)  &   \\ \hline
\end{tabular}
\caption{$SU(5)$ continuum limit of glueball masses in units of the string tension,
  for all representations, $R$, of the rotation symmetry of a cube, for
  both values of parity, $P$, and charge conjugation, $C$.
  Ground states and some excited states. Stars
  indicate poor fits (see text).}
\label{table_MK_R_SU5}
\end{table}

\begin{table}[htb]
\centering
\begin{tabular}{|c|c|c|c|c|} \hline
\multicolumn{5}{|c|}{$SU(6)$: $M_G/\surd\sigma$  continuum limit} \\ \hline
  $R$   & P=+,C=+ & P=-,C=+ &  P=+,C=-   &  P=-,C=-   \\ \hline
$A_1$  & 3.102(32)  & 4.967(43) & 9.37(22)  & 10.46(17)  \\
    & 6.020(57)  & 7.14(13)  & 10.58(21) & 10.53(20)$^{\star\star}$  \\
    & 7.51(12)   & 8.72(31)  &           &          \\
    & 7.59(13)$^\star$  &     &           &          \\ \hline
$A_2$  & 7.46(13)   & 9.37(25)  & 7.169(96) & 8.71(16)$^\star$  \\
    & 9.58(24)$^{\star\star}$ & 10.48(43) & 9.00(11) & 10.87(38)    \\
    &            &           & 9.12(20)  &           \\ \hline
$E$   & 4.706(30)  & 6.098(55) & 8.80(11)  & 8.106(50)  \\
    & 6.43(11)   & 7.93(15)  & 8.95(17)  & 9.48(33)   \\
    & 7.06(13)   & 9.58(19)  & 9.34(19)  & 9.88(11)   \\
    & 7.19(17)   &           &           &            \\ \hline
$T_1$  & 7.435(80)  & 8.405(88) & 5.847(43) & 7.41(13)   \\
    & 7.57(10)   & 8.75(11)  & 7.066(80) & 8.90(11)    \\
    & 9.17(19)   & 9.23(11)$^\star$  & 7.552(85) & 9.86(26)$^\star$  \\
    &            &           & 8.46(11)  &             \\ \hline
$T_2$  & 4.649(28)  & 6.157(50) & 6.997(80) & 7.64(14)   \\
    & 6.577(71)  & 7.92(14)  & 8.20(19)  & 8.51(9)   \\
    & 7.189(72)  & 8.25(11)$^\star$ & 8.864(61) & 10.20(9)  \\
    & 7.02(10)$^\star$   &           & 9.15(11)  &            \\ \hline
\end{tabular}
\caption{$SU(6)$ continuum limit of glueball masses in units of the string tension,
  for all representations, $R$, of the rotation symmetry of a cube, for
  both values of parity, $P$, and charge conjugation, $C$.
  Ground states and some excited states. Stars
  indicate poor fits (see text).}
\label{table_MK_R_SU6}
\end{table}

\clearpage

\begin{table}[htb]
\centering
\begin{tabular}{|c|c|c|c|c|} \hline
\multicolumn{5}{|c|}{$SU(8)$: $M_G/\surd\sigma$  continuum limit} \\ \hline
  $R$   & P=+,C=+ & P=-,C=+ &  P=+,C=-   &  P=-,C=-   \\ \hline
$A_1$  & 3.099(26)  & 4.755(58) & 8.97(31)  & 9.69(35)  \\
    & 5.87(7)    & 6.90(24)  & 10.06(24) & 9.50(39)  \\
    & 7.18(13)   & 8.25(26)  &           &           \\
    & 7.61(7)    &           &           &           \\ \hline
$A_2$  & 7.28(14)   & 9.80(29)  & 7.00(12)  & 8.26(28)  \\
    & 9.06(34)   & 11.28(25) & 8.60(18)  & 11.40(23)  \\
    &            &           & 8.87(48)  &            \\ \hline
$E$   & 4.658(32)  & 6.091(59) & 8.53(19)  & 7.60(11)    \\
    & 6.32(7)$^\star$  & 8.23(13) & 8.94(24)$^{\star\star}$ & 9.18(25)  \\
    & 7.26(11)   & 9.46(24)  & 8.81(35)  & 9.57(29)   \\ \hline
$T_1$  & 7.318(74)  & 8.24(15)  & 5.801(38) & 7.09(17)  \\
    & 7.51(11)   & 8.60(27)  & 7.089(55) & 8.35(18)  \\
    & 9.21(22)   & 8.46(22)  & 7.43(9)$^\star$ & 9.60(25) \\
    & 9.21(25)   &           & 8.43(12)  &   \\ \hline
$T_2$  & 4.661(21)  & 5.995(61) & 6.908(67) & 7.863(66)  \\
    & 6.599(84)  & 8.00(11)  & 8.01(16)  & 8.31(12)  \\
    & 7.16(8)    & 8.38(16)  & 8.51(12)$^\star$ & 9.40(24)  \\
    & 7.23(16)   &           & 8.92(15)  &   \\ \hline
\end{tabular}
\caption{$SU(8)$ continuum limit of glueball masses in units of the string tension,
  for all representations, $R$, of the rotation symmetry of a cube, for
  both values of parity, $P$, and charge conjugation, $C$.
  Ground states and some excited states. Stars
  indicate poor fits (see text).}
\label{table_MK_R_SU8}
\end{table}

\begin{table}[htb]
\centering
\begin{tabular}{|c|c|c|c|c|} \hline
\multicolumn{5}{|c|}{$SU(10)$: $M_G/\surd\sigma$  continuum limit} \\ \hline
  $R$   & P=+,C=+ & P=-,C=+ &  P=+,C=-   &  P=-,C=-   \\ \hline
$A_1$  & 3.102(37) & 4.835(60) & 8.92(37)  & 10.04(36)  \\
    & 5.99(13)  & 7.03(14)  & 8.73(60)  &  9.79(55)  \\
    & 7.15(17)  & 8.90(68)  &           &            \\
    & 7.87(13)  &           &           &            \\ \hline
$A_2$  & 7.20(18)  & 8.99(35)  & 6.78(15)  & 9.06(44)   \\
    & 8.35(50)  & 11.74(70) & 7.83(44)  & 10.70(70)  \\
    &           &           & 8.29(103)?&   \\ \hline
$E$   & 4.587(27) & 6.12(10)  & 8.64(20)  & 7.60(14)  \\
    & 6.49(9)   & 8.03(12)  & 9.04(30)  & 9.59(33)  \\
    & 7.34(12)  & 9.36(46)  & 9.05(30)  & 9.92(22) \\ \hline
$T_1$  & 7.14(17)  & 8.49(28)  & 5.776(41) & 7.27(17)  \\
    & 7.84(27)  & 8.11(62)  & 7.039(62) & 9.00(24)$^\star$  \\
    & 9.32(18)$^{\star\star}$ & 9.53(100)$^\star$ & 7.79(13) & 10.17(52)  \\
    & 9.69(22)$^{\star\star}$ &   & 8.75(21)$^{\star\star}$ &          \\ \hline
$T_2$  & 4.600(30) & 5.94(8)   & 7.071(55) & 7.89(16)  \\
    & 6.638(82) & 7.52(17)  & 7.60(24)  & 7.82(26)$^\star$  \\
    & 7.00(18)  & 7.88(30)$^{\star\star}$ & 8.60(13) &  9.26(36) \\
    & 7.19(22)  &           & 8.77(28)  &           \\ \hline
\end{tabular}
\caption{$SU(10)$ continuum limit of glueball masses in units of the string tension,
  for all representations, $R$, of the rotation symmetry of a cube, for
  both values of parity, $P$, and charge conjugation, $C$.
  Ground states and some excited states. Stars
  indicate poor fits (see text).}
\label{table_MK_R_SU10}
\end{table}

\begin{table}[htb]
\centering
\begin{tabular}{|c|c|c|c|c|} \hline
\multicolumn{5}{|c|}{$SU(12)$: $M_G/\surd\sigma$  continuum limit} \\ \hline
  $R$   & P=+,C=+ & P=-,C=+ &  P=+,C=-   &  P=-,C=-   \\ \hline
$A_1$  & 3.151(33)  & 4.696(65) & 9.63(25)   & 9.57(32)  \\
    & 5.914(82)  & 6.94(19)  & 10.50(51)  & 9.90(58)  \\
    & 7.07(18)   & 9.56(44)$^\star$ &      &           \\
    & 7.61(19)   &           &            &          \\ \hline
$A_2$  & 7.74(22)   & 10.16(26) & 6.49(17)   & 9.03(22)  \\
    & 9.64(26)   & 12.09(74) & 8.90(22)   & 10.66(60)  \\
    &            &           & 9.46(46)   &   \\ \hline
$E$   & 4.647(33)  & 6.22(10)  & 8.52(22)   & 7.80(14)  \\
    & 6.613(80)  & 8.28(12)$^\star$ & 9.56(43) & 9.88(20) \\
    & 7.67(13)   &           &            & 10.24(29)    \\ \hline
$T_1$  & 6.64(27)   & 8.59(12)  & 5.741(60)  & 7.50(17)$^\star$  \\
    & 7.59(21)   & 8.95(16)  & 7.005(78)  & 8.83(14)     \\
    & 9.34(18)$^{\star\star}$ & 9.30(30)$^\star$ & 7.48(14) & 9.51(48)  \\
    &            &           & 8.80(23)   &   \\ \hline
$T_2$  & 4.645(33)  & 5.97(7)   & 6.88(10) &  8.20(13)      \\
    & 6.764(57)  & 8.05(15)  & 8.16(25) &  8.27(31)$^\star$ \\
    & 7.35(15)$^\star$ & 8.89(22) & 8.48(25) & 10.52(21)$^{\star\star}$  \\
    & 7.20(19)   &           & 8.21(30) &                \\ \hline
\end{tabular}
\caption{$SU(12)$ continuum limit of glueball masses in units of the string tension,
  for all representations, $R$, of the rotation symmetry of a cube, for
  both values of parity, $P$, and charge conjugation, $C$.
  Ground states and some excited states. Stars
  indicate poor fits (see text).}
\label{table_MK_R_SU12}
\end{table}

\clearpage

\begin{table}[htb]
\centering
\begin{tabular}{|c|c|c|c|c|} \hline
\multicolumn{5}{|c|}{$SU(\infty)$:  $M_G/\surd\sigma$ } \\ \hline
  $R$   & P=+,C=+ & P=-,C=+ &  P=+,C=-   &  P=-,C=-   \\ \hline
$A_1$  & 3.072(14)         & 4.711(25)            & 9.26(16)        &  10.10(18) \\
    & 5.805(31)$^{\star\star}$ & 7.050(68)             &                &  10.14(23) \\
    & 7.294(63)         &                       &                &   \\ \hline
$A_2$  & 7.40(12)$^\dagger$ & 9.73(12)              & 7.142(75)$^{\dagger\star}$  & 8.61(13)$^\star$ \\
    & 9.14(14)$^\star$   & 11.12(24)             & 8.77(10)      &  11.38(21)   \\ \hline
$E$   & 4.582(14)         & 6.108(44)             & 8.63(10)      &  7.951(53)$^{\star\star}$ \\
    & 6.494(33)$^\star$  & 8.051(60)             & 9.14(15)      &  9.55(13)                   \\
    & 7.266(50)$^\star$  &                       &               &  9.84(12)                 \\ \hline
$T_1$  & 7.250(47)         & 8.412(76)             & 5.760(25)     & 7.134(86)  \\
    & 7.337(60)$^\star$  & 8.79(10)              & 7.020(39)     & 8.65(9)  \\
    & 9.142(82)         & 9.08(12)$^{\star\star}$ & 7.470(55)     & 9.81(17)  \\ 
    &                   &                       & 8.422(84)     &             \\ \hline
$T_2$  & 4.578(11)         & 5.965(28)             & 6.957(41)     &  7.96(8)$^\dagger$ \\
    & 6.579(30)         & 7.883(57)             & 7.93(11)      &  8.22(8)            \\
    & 7.121(45)         & 8.45(14)$^\dagger$     & 8.63(7)$\star$ & 10.26(10)$^{\star\star}$  \\
    & 7.122(76)         &     &     &  \\ \hline
\end{tabular}
\caption{Continuum glueball masses in units of the string tension,
  in the limit $N\to\infty$. Fits are to $N\geq 2$ or $N\geq 3$ except
  for values labelled with a $\dagger$, and $\star$ indicates a  poor fit,
  as explained in the text. 
  Labels are $R$ for the representations of the rotation symmetry of a cube,
  $P$ for parity and  $C$ for charge conjugation.}
\label{table_MK_R_SUN}
\end{table}

\begin{table}[htb]
\centering
\begin{tabular}{|ccc|} \hline
\multicolumn{3}{|c|}{continuum $J \sim$ cubic $R$} \\ \hline
$J$    &        &  cubic $R$  \\  \hline
 0   & $\sim$ & $A_1$   \\
 1   & $\sim$ & $T_1$      \\
 2   & $\sim$ & $E+T_2$     \\
 3   & $\sim$ & $A_2+T_1+T_2$     \\
 4   & $\sim$ & $A_1+E+T_1+T_2$     \\
 5   & $\sim$ & $E+2T_1+T_2$      \\
 6   & $\sim$ & $A_1+A_2+E+T_1+2T_2$     \\
 7   & $\sim$ & $A_2+E+2T_1+2T_2$     \\
 8   & $\sim$ & $A_1+2E+2T_1+2T_2$     \\ \hline
\end{tabular}
\caption{Projection of continuum spin $J$ states onto the cubic representations $R$.}
\label{table_J_R}
\end{table}

\begin{table}[htb]
\centering
\begin{tabular}{|llc|} \hline
\multicolumn{3}{|c|}{continuum $J^{PC}$ from cubic $R$} \\ \hline
$J^{PC}$    &        &  cubic $R^{PC}$  \\  \hline
$0^{++}$gs  & $\sim$ & $A_1^{++}$gs     \\ 
$0^{++}$ex1 & $\sim$ & $A_1^{++}$ex1    \\ 
$2^{++}$gs  & $\sim$ & $E^{++}$gs + $T_2^{++}$gs    \\  
$2^{++}$ex1 & $\sim$ & $E^{++}$ex1 + $T_2^{++}$ex1    \\ 
$3^{++}$gs  & $\sim$* & $A_2^{++}$gs + $T_1^{++}$gs(ex1) + $T_2^{++}$ex3(ex2)    \\ 
$4^{++}$gs  & $\sim$* & $A_1^{++}$ex2 + $E^{++}$ex2 +$T_1^{++}$ex1(gs) + $T_2^{++}$ex2(ex3)   \\ \hline
 $0^{-+}$gs  &  $\sim$  &  $A_1^{-+}$gs     \\ 
 $0^{-+}$ex1 &  $\sim$  &  $A_1^{-+}$ex1    \\
 $2^{-+}$gs  &  $\sim$  &  $E^{-+}$gs + $T_2^{-+}$gs   \\
 $2^{-+}$ex1 &  $\sim$  &  $E^{-+}$ex1 + $T_2^{-+}$ex1   \\ 
 $1^{-+}$gs  &  $\sim$  & $T_1^{-+}$gs    \\ \hline
 $2^{+-}$gs  &  $\sim$  &  $E^{+-}$gs + $T_2^{+-}$ex2   \\
 $1^{+-}$gs  &  $\sim$  & $T_1^{+-}$gs   \\ 
 $1^{+-}$ex1 &  $\sim$  & $T_1^{+-}$ex2   \\ 
 $3^{+-}$gs  &  $\sim$  & $A_2^{+-}$gs + $T_1^{+-}$ex1 + $T_2^{+-}$gs   \\  \hline
 $2^{--}$gs  &  $\sim$  & $E^{--}$gs + $T_2^{--}$gs \\
 $1^{--}$gs  &  $\sim$  & $T_1^{--}$gs   \\ \hline
\end{tabular}
\caption{Identification of continuum $J^{PC}$ states from the results for the cubic representations
  in  Tables~\ref{table_MK_R_SU2}-\ref{table_MK_R_SUN}.
  Ground state denoted by $gs$, $i$'th excited state by $exi$. Where
  there is some ambiguity, a single star denotes 'likely'
  while two stars indicate 'significant uncertainty'.}
\label{table_M_J_R}
\end{table}

\begin{table}[htb]
\centering
\begin{tabular}{|l|c|c|c|c|} \hline
\multicolumn{5}{|c|}{$M(J^{PC})/\surd\sigma$ continuum limit} \\ \hline
  $J^{PC}$      & $SU(2)$ & $SU(3)$ & $SU(4)$ & $SU(5)$ \\ \hline
 $0^{++}$ {gs}   & 3.781(23)  & 3.405(21)  & 3.271(27)  & 3.156(31)  \\
 $0^{++}$ {ex1}  & 6.126(38)  & 5.855(41)  & 5.827(62)  & 5.689(53)  \\
 $2^{++}$ {gs}   & 5.349(20)  & 4.894(22)  & 4.742(15)  & 4.690(20)  \\
 $2^{++}$ {ex1}  & 7.22(6)$^+$ & 6.788(40) & 6.694(40)  & 6.607(45)  \\
 $3^{++}$ {gs}   & 8.13(8)$^*$ & 7.71(9)$^*$ &          & 7.29(9)     \\ 
 $4^{++}$ {gs}   &            & 7.60(12)$^*$ & 7.36(9)$^*$ & 7.41(10) \\ \hline
 $0^{-+}$ {gs}   & 6.017(61)  & 5.276(45)  & 5.020(46)  & 4.832(40)  \\ 
 $0^{-+}$ {ex1}  & 8.00(15)   & 7.29(13)   & 7.33(11)   & 7.38(11)  \\
 $2^{-+}$ {gs}   & 7.017(50)  & 6.32(9)    & 6.182(33)  & 6.187(50)  \\
 $2^{-+}$ {ex1}  & 8.521(65)  & 8.18(8)    & 7.91(13)$^*$ & 8.02(11)  \\
 $1^{-+}$ {gs}   & 9.06(13)   & 8.48(12)   & 8.47(11)   & 8.29(13)  \\ \hline
 $2^{+-}$ {gs}   &            & 8.91(15)   & 8.84(11)   &  8.49(15)  \\
 $1^{+-}$ {gs}   &            & 6.065(40)  & 5.956(42)  & 5.915(45)  \\
 $1^{+-}$ {ex1}  &            & 7.82(6)    & 7.51(8)    & 7.62(8)  \\
 $3^{+-}$ {gs}   &            & 7.27(12)   & 7.03(7)    & 7.13(7)  \\  \hline
 $2^{--}$ {gs}   &            & 8.08(15)   & 7.79(9)   & 7.97(6)  \\
 $1^{--}$ {gs}   &            & 8.31(10)   & 7.60(9)   & 7.62(15)  \\ \hline
\end{tabular}
\caption{Continuum limit of glueball masses, in units of the string tension,
  for those $J^{PC}$ representations we can identify. Ground state denoted by $gs$,
  $i$'th excited state by $exi$. Stars explained in text.}
\label{table_MKJ_N2-5}
\end{table}

\begin{table}[htb]
\centering
\begin{tabular}{|l|c|c|c|c|} \hline
\multicolumn{5}{|c|}{$M(J^{PC})/\surd\sigma$ continuum limit} \\ \hline
  $J^{PC}$      & $SU(6)$ & $SU(8)$ & $SU(10)$ & $SU(12)$ \\ \hline
 $0^{++}$ {gs}   & 3.102(32)  & 3.099(26)  & 3.102(37)  & 3.151(33)  \\
 $0^{++}$ {ex1}  & 6.020(57)  & 5.87(7)    & 5.99(13)   & 5.914(82)  \\
 $2^{++}$ {gs}   & 4.678(30)  & 4.660(20)  & 4.594(25)  & 4.646(30)  \\
 $2^{++}$ {ex1}  & 6.54(7)    & 6.60(9)$^*$ & 6.57(9)   & 6.71(6)  \\
 $3^{++}$ {gs}   & 7.44(8)$^*$ & 7.34(11)$^*$ & 7.14(18) &   \\ 
 $4^{++}$ {gs}   &            & 7.20(7)$^*$ & 7.32(15)  &   \\ \hline
 $0^{-+}$ {gs}   & 4.967(43)  & 4.755(58)  & 4.835(60)  & 4.696(65)  \\ 
 $0^{-+}$ {ex1}  & 7.14(13)   & 6.90(24)   & 7.03(14)   & 6.94(19)  \\
 $2^{-+}$ {gs}   & 6.148(50)  & 6.043(55)  & 6.01(8)   &  6.05(7) \\
 $2^{-+}$ {ex1}  & 7.92(14)   & 8.10(11)   & 7.86(12)$^*$ & 8.16(15)  \\
 $1^{-+}$ {gs}   & 8.41(9)    & 8.24(15)   & 8.49(28)  & 8.59(12)  \\ \hline
 $2^{+-}$ {gs}   & 8.83(9)    & 8.52(12)   & 8.62(12)  & 8.50(17)    \\
 $1^{+-}$ {gs}   & 5.847(43)  & 5.801(38)  & 5.776(41) & 5.741(60)  \\
 $1^{+-}$ {ex1}  & 7.55(9)    & 7.43(9)$^*$ & 7.79(13) & 7.48(14)  \\
 $3^{+-}$ {gs}   & 7.067(70)  & 6.999(70)  & 7.035(50) & 6.89(10)  \\ \hline
 $2^{--}$ {gs}  & 8.05(5)$^*$ & 7.78(7)   & 7.74(15)  & 8.00(15)  \\
 $1^{--}$ {gs}   & 7.41(13)   & 7.09(17)   & 7.27(17)  & 7.50(17)  \\ \hline
\end{tabular}
\caption{Continuum limit of glueball masses, in units of the string tension,
  for those $J^{PC}$ representations we can identify. Ground state denoted by $gs$,
  $i$'th excited state by $exi$. Stars explained in text.}
\label{table_MKJ_N6-12}
\end{table}

\clearpage

\begin{table}[htb]
\centering
\begin{tabular}{|cccccc|} \hline
\multicolumn{6}{|c|}{$M(J^{PC})/\surd\sigma$ lower bounds} \\ \hline
  $1^{++}$ &  $3^{-+}$ & $0^{+-}$& $2^{+-}$ & $0^{--}$ & $3^{--}$ \\ \hline
 $\ge 9.0$  & $\ge 9.7$  &$\ge 9.2$   & $\ge 8.6$  &$\ge 10.0$   & $\ge 8.6$   \\ \hline
\end{tabular}
\caption{Lower bounds on masses of some of the low-$J$ ground states not
  appearing in Tables~\ref{table_MKJ_N2-5},\ref{table_MKJ_N6-12}.}
\label{table_MKJ_lowbound}
\end{table}

\begin{table}[htb]
\centering
\begin{tabular}{|l|l|l|l|l|} \hline
\multicolumn{5}{|c|}{$M_G \, GeV \quad SU(3)$} \\ \hline
  $J$   & P=+,C=+ & P=-,C=+ &  P=+,C=-   &  P=-,C=-   \\ \hline
 0 gs   & 1.653(26)   & 2.561(40) & $\ge 4.52(15)$ &  $\ge 4.81(24)$   \\
 0 ex1  & 2.842(40)   & 3.54(8)   &               &    \\
 2 gs   & 2.376(32)   & 3.07(6)   & 4.24(8)*      &  3.92(9)   \\
 2 ex1  & 3.30(5)     & 3.97(7)   & $\ge 4.38(13)$ & $\ge 4.55(11)$  \\
 1 gs   & $\ge 4.52(6)$ & 4.12(8)  & 2.944(42)    &  4.03(7)  \\ 
 1 ex1  &             & 4.16(8)*  & 3.80(6)       & $\ge 4.51(9)$    \\
 1 ex2  &             & 4.20(9)*  &               &    \\
 3 gs   & 3.74(7)*    &  $\ge 4.75(13)$ & 3.53(8)   & $\ge 4.35(9)$   \\
 4 gs   & 3.69(8)*    &  $\ge 4.45(14)$ & 4.38(8)** & $\ge 4.81(24)$   \\ \hline
\end{tabular}
\caption{Continuum limit of $SU(3)$ glueball masses, in physical $GeV$ units
  for those $J^{PC}$ representations we can identify, with lower bounds in those
  cases where this is not possible. Ground state denoted by $gs$,
  first excited state by $ex$. Stars denote ambiguity.}
\label{table_MJ_N3}
    \end{table}

\begin{table}[htb]
\centering
\begin{tabular}{|c|c|c|c|c|} \hline
\multicolumn{5}{|c|}{$M_J/\surd\sigma$ ; $SU(\infty)$} \\ \hline
  $J$      & P=+,C=+   & P=-,C=+   &  P=+,C=-   &  P=-,C=-   \\ \hline
 $0$ {gs}  & 3.072(14) & 4.711(26) &  $\ge 9.26(16)$     & $\ge 10.10(18)$    \\ 
 $0$ {ex}  & 5.845(50) & 7.050(68) &                     &    \\ 
 $2$ {gs}  & 4.599(14) & 6.031(38) & 8.566(76)           &  7.910(56)  \\ 
 $2$ {ex}  & 6.582(36) & 7.936(54) &  $\ge 9.14(15)$     & $\ge 9.55(13)$    \\ 
 $1$ {gs}  & $\ge 9.14(9)$ & 8.415(76)  & 5.760(25)      & 7.26(11)   \\ 
 $1$ {ex}  &           &                & 7.473(57)      & $\ge 8.65(9) $  \\ 
 $3$ {gs}  & 7.263(56) & $\ge 9.73(12)$ & 6.988(41)      & $\ge 8.61(13) $   \\ 
 $4$ {gs}  & 7.182(71) & $\ge 8.79(10)$ & $\ge 9.26(16)$ & $\ge 10.10(18)$   \\  \hline
\end{tabular}
\caption{Large $N$ extrapolation of continuum glueball masses, in units of the string tension,
  for those $J^{PC}$ representations we can identify, with lower bounds in those
  cases where this is not possible. Ground state denoted by $gs$,
  first excited state by $ex$.}
\label{table_MK_J_SUN}
\end{table}

\begin{table}[htb]
\centering
\begin{tabular}{|c|c|c|c|c|c|c|c|} \hline
\multicolumn{8}{|c|}{$M/\surd\sigma$ ; continuum $SU(3)$} \\ \hline
  paper     & $0^{++}$ & $2^{++}$ & $0^{-+}$ & $2^{-+}$ & $1^{+-}$ & $3^{+-}$ & $2^{--}$    \\ \hline
 this work                      & 3.405(21)  & 4.894(22) &  5.276(45) & 6.32(9)  & 6.065(40) & 7.27(12) & 8.08(15)  \\
 2005: ref\cite{HM_Thesis}      & 3.347(68)  & 4.891(65) &  5.11(14)  & 6.32(11) & 6.06(15)  & 7.43(20) & 8.32(29)  \\
 2005: ref\cite{MP-2005}$^\star$ & 3.59(15)   & 5.03(11)  &  5.39(11) &  6.40(14) & 6.27(11)  & 7.58(12) & 8.42(14)  \\
 2004: ref\cite{BLMTUW_N}       & 3.55(7)    & 4.78(9)   &  &  &  &   &   \\
 2001: ref\cite{BLMT_N}         & 3.61(9)    & 5.13(22)  &  &  &  &   &   \\  \hline
\end{tabular}
\caption{Some glueball masses in units of the string tension in the continuum limit of the $SU(3)$ gauge theory:
  a comparison between this work and some earlier work. The starred calculation includes our tansformation of
the units from the Sommer scale to the string tension.}
\label{table_MK_J_SU3_comp}
\end{table}

\begin{table}[htb]
\centering
\begin{tabular}{|c|c|c|c|c|c|c|c|} \hline
\multicolumn{8}{|c|}{$M/\surd\sigma$ ; continuum $SU(8)$} \\ \hline
  paper     & $0^{++}$ & $2^{++}$ & $0^{-+}$ & $2^{-+}$ & $1^{+-}$ & $3^{+-}$ & $2^{--}$    \\ \hline
 this work                 & 3.099(26)  & 4.660(20) & 4.755(58) & 6.043(55) & 5.801(38) & 7.00(7)  & 7.78(7)  \\
 2005: ref\cite{HM_Thesis} & 3.32(15)   & 4.65(19)  & 4.72(32)  & 5.67(40)  & 5.70(29)  & 7.74(79) & 7.3(1.4)  \\
 2004: ref\cite{BLMTUW_N}  & 3.55(12)   & 4.73(22) &  &  &  &   &   \\  \hline
\end{tabular}
\caption{Some glueball masses in units of the string tension in the continuum limit of the $SU(8)$ gauge theory:
  a comparison between this work and some earlier work.}
\label{table_MK_J_SU8_comp}
\end{table}

\begin{table}[htb]
\centering
\begin{tabular}{|c|lc|lc|lc|} \hline
\multicolumn{7}{|c|}{$Q_L(n_c)$ : $SU(8)$, $Q=2$} \\ \hline
       & \multicolumn{2}{|c|}{$\beta=45.50,12^320$} & \multicolumn{2}{|c|}{$\beta=46.70,16^324$} & \multicolumn{2}{|c|}{$\beta=47.75,20^330$}  \\
 $n_c$ &  $\bar{Q}_L$ &  $\sigma_{Q_{L}}$ & $\bar{Q}_L$ & $\sigma_{Q_{L}}$  & $\bar{Q}_L$ & $\sigma_{Q_{L}}$ \\ \hline
  0   & 0.275(21) & 0.836 & 0.333(23) & 1.282 &  0.410(35) & 1.956  \\
  1   & 1.199(6)  & 0.202 & 1.286(5)  & 0.265 &  1.345(7)  & 0.351  \\
  2   & 1.575(3)  & 0.070 & 1.631(2)  & 0.082 &  1.665(3)  & 0.103  \\
  3   & 1.699(2)  & 0.041 & 1.7465(7) & 0.044 &  1.7743(9) & 0.053  \\
  4   & 1.7576(7) & 0.030 & 1.8005(6) & 0.030 &  1.8254(6) & 0.035  \\
  8   & 1.8437(3) & 0.016 & 1.8781(3) & 0.014 &  1.8970(2) & 0.014  \\
  12  & 1.8744(3) & 0.013 & 1.9046(2) & 0.010 &  1.9212(1) & 0.009  \\ 
  16  & 1.8912(3) & 0.011 & 1.9186(2) & 0.008 &  1.9339(1) & 0.007  \\ 
  20  & 1.9021(3) & 0.010 & 1.9275(1) & 0.007 &  1.9418(1) & 0.006  \\   \hline
\end{tabular}
\caption{Lattice topological charge $Q_L$ as function of number of cooling sweeps, $n_c$, for fields
  which have $Q_L\simeq 2$ after 20 cooling sweeps: the average value, with error, and standard deviation.
  For small, intermediate and large $\beta$, in $SU(8)$, all with similar volumes in physical units.}
\label{table_Q_nc_SU8}
\end{table}


\begin{table}[htb]
\centering
\begin{tabular}{|ccc|ccc|} \hline
\multicolumn{6}{|c|}{topology tunnelling time} \\ \hline
$\beta$ &  $\tau_Q$ & $\Tilde{\tau}_Q$ & $\beta$ &  $\tau_Q$ & $\Tilde{\tau}_Q$ \\ \hline
\multicolumn{3}{|c|}{$SU(3)$} & \multicolumn{3}{|c|}{$SU(4)$} \\ \hline
 6.235 & 100(2)   & 112(2)   & 11.20 & 499(9)   & 520(9)    \\    
 6.338 & 247(6)   & 283(6)   & 11.40 & 2173(42) & 2275(45)    \\  
 6.50  & 1056(44) & 1236(50) & 11.60 & 12674(636) & 13371(690)  \\
 6.60  & 3399(188) & 3731(190) &     &          &               \\
 6.70  & 7973(674) & 8973(744) &     &          &              \\ \hline
\multicolumn{3}{|c|}{$SU(5)$} & \multicolumn{3}{|c|}{$SU(6)$} \\ \hline
17.22  & 189(4)   &  196(4)   & 25.05 & 1158(31)   & 1177(31)    \\
17.43  & 695(13)  &  721(13)  & 25.32 & 5992(262)  & 6105(272)   \\
17.63  & 2686(90) & 2749(93)  & 25.55 & $23.6(1.8)\times 10^3$  &  $24.4(1.8)\times 10^3$   \\
18.04  & $35.7(3.7)\times 10^3$ & $37.0(3.7)\times 10^3$ & 26.22 & $1.43(99)\times 10^6$  & $1.43(99)\times 10^6$ \\
18.375 & $40(17)\times 10^4$    & $44(19)\times10^4$    &       &  & \\ \hline
\multicolumn{3}{|c|}{$SU(8)$} & \multicolumn{3}{|c|}{} \\ \hline
 44.10 & 514(10)                & 528(10)                  &   &  &  \\
 44.85 & $26.3(3.5)\times 10^3$ & $26.3(3.5)\times 10^3$   &  &   &    \\ 
 45.50 &  $7.1(3.5)\times 10^5$ & $7.1(3.5)\times 10^5$    &  &   &   \\
 45.50 & $12.5(5.1)\times 10^5$ & $12.5(5.1)\times 10^5$   &  &   &  \\ \hline
\multicolumn{3}{|c|}{$SU(10)$} & \multicolumn{3}{|c|}{$SU(12)$} \\ \hline
  69.20  & 4793(211) & 4831(214)  & 99.86  & $4.83(58)\times 10^4$ &  $4.83(58)\times 10^4$   \\  \hline
\end{tabular}
\caption{Average number of sweeps, $\tau_Q$, between $\Delta Q= \pm 1$ changes normalised to
  a standard space-time volume of $(3/\surd\sigma)^4$, with $\Tilde{\tau}_Q$ including a correction for
`near-dislocations'.}
\label{table_tauQ_SUN}
\end{table}

\begin{table}[htb]
\centering
\begin{tabular}{|c|c|c|} \hline
\multicolumn{3}{|c|}{$\ln\{\tau_Q\} = b - c(N)\ln\{a\surd\sigma\}$} \\ \hline
$N$ & $c$ &  $11N/3 - 5$ \\ \hline
3  & 6.82(7)   &  6.0 \\
4  & 9.27(12)  &  9.6$\dot{6}$ \\
5  & 11.13(20) &  13.3$\dot{3}$ \\
6  & 13.50(30) &  17.0 \\
8  & 17.58(43) &  24.3$\dot{3}$ \\ \hline
 \end{tabular}
\caption{Fitted values of $c(N)$ from Fig.\ref{fig_tauQ_KsuN} compared to the asymptotic
  dilute gas prediction in eqn(\ref{eqn_Ia}).}
\label{table_tauQ_a}
\end{table}

\begin{table}[htb]
\centering
\begin{tabular}{|c|c|c|c|} \hline
\multicolumn{4}{|c|}{$SU(2)$ topology} \\ \hline
$\beta$  &  lattice  & $\langle Q_L^2 \rangle$ & $\langle Q_I^2 \rangle$ \\ \hline
2.2986  & $8^316$  & 2.266(9)  & 3.057(12)    \\ 
        & $12^316$ & 7.694(42) & 9.600(49)    \\   \hline
2.3714  & $10^316$ & 2.705(9)  & 3.544(12)    \\ 
        & $14^316$ & 7.343(61) & 9.220(77)    \\   \hline
2.427   & $12^316$ & 2.824(7)  & 3.551(8)     \\ 
        & $16^4$   & 6.635(56) & 8.066(64)    \\ 
        & $20^316$ & 13.09(19) & 15.41(22)   \\  
        & $24^316$ & 22.56(20) & 26.18(23)    \\   \hline
2.509   & $16^320$ & 3.411(13) & 4.067(16)    \\ 
        & $22^320$ & 8.822(58) & 10.137(65)   \\   \hline
2.60    & $22^330$ & 4.491(31)  & 5.109(33)   \\ 
        & $30^4$   & 11.48(10)  & 12.65(12)   \\  \hline
2.65    & $26^334$ & 4.702(41)  & 5.268(45)   \\  \hline
2.70    & $28^340$ & 3.786(36)  & 4.181(39)   \\ 
        & $40^4$   & 11.07(14)  & 12.09(14)   \\ \hline
2.75    & $34^346$ & 4.328(88)  & 4.711(94)   \\   \hline
2.80    & $40^354$ & 4.43(12)   & 4.76(12)   \\   \hline
\end{tabular}
\caption{Average values of $Q^2_L$, after 20 cooling sweeps, in $SU(2)$ for various values of $\beta$ and lattice sizes.
  $Q_I$ is the projection of $Q_L$ to an integer value.}
\label{table_QQ_SU2}
\end{table}

\begin{table}[htb]
\centering
\begin{tabular}{|c|c|c|c||c|c|c|c|} \hline
\multicolumn{4}{|c|}{$SU(3)$ topology} & \multicolumn{4}{|c|}{$SU(4)$ topology} \\ \hline
$\beta$  &  lattice  & $\langle Q^2_L \rangle$ & $\langle Q_I^2 \rangle$
& $\beta$  &  lattice  & $\langle Q^2_L \rangle$ & $\langle Q_I^2 \rangle$ \\ \hline
5.6924 & $8^316$  & 4.034(21) &  5.452(29)  & 10.70 & $12^316$  & 5.813(37) & 7.084(48) \\
5.80   & $10^316$ & 4.021(13) &  5.111(16)  & 10.85 & $14^320$  & 5.718(45) & 6.674(53) \\
5.8941 & $12^316$ & 3.570(35) &  4.352(43)  & 11.02 & $18^320$  & 6.010(95) & 6.78(11) \\
5.99   & $14^320$ & 3.790(48) &  4.428(56)  & 11.20 & $22^4$    & 5.97(33)  & 6.57(36) \\
6.0625 & $14^320$ & 2.312(22) &  2.649(26)  & 11.40 & $26^4$    & 6.19(67)  & 6.68(73) \\
6.235  & $18^326$ & 2.32(12)  &  2.55(13)   & 11.60 & $30^4$    & 5.52(73)  & 5.80(78) \\
6.3380 & $22^330$ & 2.72(13)  &  2.94(14)   &  &  &  &  \\
6.50   & $26^338$ & 2.09(20)  &  2.23(21)   &  &  &  &  \\
6.60   & $32^340$ & 2.60(35)  &  2.74(37)   &  &  &  &  \\
6.70   & $36^344$ & 1.54(30)  &  1.61(31)   &  &  &  &  \\ \hline
\end{tabular}
\caption{Average values of $Q_L^2$, after 20 cooling sweeps, in $SU(3)$ and in  $SU(4)$
  for the values of $\beta$ and lattices shown.
  $Q_I$ is the projection of $Q_L$ to an integer value.}
\label{table_QQ_SU3_SU4}
\end{table}

\begin{table}[htb]
\centering
\begin{tabular}{|c|c|c|c||c|c|c|c|} \hline
\multicolumn{4}{|c|}{$SU(5)$ topology} & \multicolumn{4}{|c|}{$SU(6)$ topology} \\ \hline
$\beta$  &  lattice  & $\langle Q^2_L \rangle$ & $\langle Q_I^2 \rangle$
& $\beta$  &  lattice  & $\langle Q^2_L \rangle$ & $\langle Q_I^2 \rangle$ \\ \hline
16.98   & $10^316$  & 3.273(25)  & 3.906(30) & 24.67   & $10^316$  & 3.178(45)  & 3.747(53) \\
17.22   & $12^316$  & 2.682((60) & 3.073(68) & 25.05   & $12^316$  & 2.50(13)  & 2.83(15) \\
17.43   & $14^320$  & 2.95(13)   & 3.30(14)  & 25.32   & $14^320$  & 2.76(32)  & 3.07(35) \\
17.63   & $16^320$  & 2.86(31)   & 3.15(35)  & 25.35   & $14^320$  & 2.91(60)  & 3.24(67) \\
18.04   & $20^324$  & 3.44(1.27) & 3.71(1.36)& 25.35   & $18^318$  & 5.83(78)  & 6.48(87) \\
18.375  & $24^330$  & 2.21(86)   & 2.35(92)  &    &   &   &  \\ \hline
\end{tabular}
\caption{Average values of $Q^2_L$, after 20 cooling sweeps, in $SU(5)$ and in  $SU(6)$
  for the values of $\beta$ and lattices shown.
  $Q_I$ is the projection of $Q_L$ to an integer value.}
\label{table_QQ_SU5_SU6}
\end{table}

\begin{table}[htb]
\centering
\begin{tabular}{|c|ccc|ccc|} \hline
\multicolumn{7}{|c|}{continuum topological susceptibility} \\ \hline
group  &  $\chi_L^{1/4}/\surd\sigma$ & $\beta\in$ & $\chi^2/n_{df}$ & $\chi_I^{1/4}/\surd\sigma$ & $\beta\in$ & $\chi^2/n_{df}$  \\ \hline
SU(2) & 0.4773(14) & [2.509,2.75]   & 0.46 & 0.4857(14) & [2.509,2.75]   & 0.55 \\
SU(3) & 0.4196(35) & [5.8941,6.60]  & 0.85 & 0.4246(36) & [5.8941,6.60]  & 0.89 \\
SU(4) & 0.3925(25) & [10.70,11.60]  & 0.19 & 0.3964(27) & [10.70,11.60]  & 0.26 \\
SU(5) & 0.3786(59) & [16.98,18.375] & 0.44 & 0.3818(60) & [16.98,18.375] & 0.45  \\
SU(6) & 0.386(13)  & [24.67,25.35]  & 0.29 & 0.390(12)  & [24.67,25.35]  & 0.30 \\ \hline
SU($\infty$) & 0.3655(27)  &        & 0.92 & 0.3681(28) &                & 0.93 \\ \hline
\end{tabular}
\caption{Continuum limit of the topological susceptibility in units of the string tension for the
  gauge groups shown; $\chi_L$ is from $Q^2_L$ and $\chi_I$ is from $Q^2_I$. Fitted range of $\beta$
  also shown, as is the chi-squared per degree of freedom, $\chi^2/n_{df}$.}
\label{table_khiK_SUN_cont}
\end{table}

\begin{table}[htb]
\centering
\begin{tabular}{|ll|ll|ll|ll|} \hline
\multicolumn{8}{|c|}{$\langle{Q}_{hot}\rangle_{Q=Q_I} = Z_Q(\beta) Q_I$} \\ \hline
\multicolumn{2}{|c|}{$SU(2)$} &\multicolumn{2}{|c|}{$SU(3)$} & \multicolumn{2}{|c|}{$SU(4)$} & \multicolumn{2}{|c|}{$SU(5)$} \\ \hline
$\beta$ & $Z_Q(\beta)$ & $\beta$ & $Z_Q(\beta)$ & $\beta$ & $Z_Q(\beta)$ & $\beta$ & $Z_Q(\beta)$  \\ \hline
2.452 & 0.1386(23) &  5.6924 & 0.0646(9)  & 10.70 & 0.0948(14) & 16.98  & 0.0992(15) \\
2.65  & 0.233(12)  &  5.80   & 0.0877(10) & 10.85 & 0.1146(20) & 17.22  & 0.1137(20) \\
2.70  & 0.239(17)  &  5.99   & 0.1306(26) & 11.02 & 0.1347(35) & 17.43  & 0.1348(26) \\
2.75  & 0.258(18)  &  6.0625 & 0.1462(31) & 11.02 & 0.1397(29) & 17.63  & 0.1476(31) \\
2.80  & 0.239(22)  &  6.235  & 0.1808(54) & 11.20 & 0.1524(46) & 18.04  & 0.1587(48) \\
      &            &  6.338  & 0.2044(65) & 11.40 & 0.1710(66) & 18.375 & 0.190(11) \\
      &            &  6.50   & 0.231(10)  & 11.60 & 0.184(10)  &  &  \\
      &            &  6.60   & 0.232(19)  &  &  &  &  \\
      &            &  6.70   & 0.241(28)  &  &  &  &  \\ \hline
\end{tabular}
\caption{Multiplicative renormalisation factor, $Z_Q(\beta)$, relating 
  the average lattice topological charge, $Q_{hot}$,  calculated on the rough
  Monte Carlo fields, and the integer valued topological charge $Q_I$ calculated
  after 20 `cooling' sweeps of those fields. For the gauge groups and $\beta$ shown.}
\label{table_ZQA}
\end{table}

\begin{table}[htb]
\centering
\begin{tabular}{|ll|ll|ll|ll|} \hline
\multicolumn{8}{|c|}{$\langle{Q}_{hot}\rangle_{Q=Q_I} = Z_Q(\beta) Q_I$} \\ \hline
\multicolumn{2}{|c|}{$SU(6)$} &\multicolumn{2}{|c|}{$SU(8)$} & \multicolumn{2}{|c|}{$SU(10)$} & \multicolumn{2}{|c|}{$SU(12)$} \\ \hline
$\beta$ & $Z_Q(\beta)$ & $\beta$ & $Z_Q(\beta)$ & $\beta$ & $Z_Q(\beta)$ & $\beta$ & $Z_Q(\beta)$  \\ \hline
24.67 & 0.0976(20) & 44.10  & 0.0912(19)  & 69.20 & 0.0930(19) & 99.86  & 0.0899(17) \\
25.05 & 0.1132(29) & 44.85  & 0.1194(31)  & 70.38 & 0.1165(27) & 101.55 & 0.1129(26) \\
25.32 & 0.1316(40) & 45.50  & 0.1254(35)  & 71.38 & 0.1283(37) & 103.03 & 0.1292(33) \\
25.55 & 0.1444(56) & 46.10  & 0.1491(50)  & 72.40 & 0.1525(41) & 104.55 & 0.1566(46) \\
26.22 & 0.1716(75) & 46.70  & 0.1617(54)  & 73.35 & 0.1731(52) & 105.95 & 0.1561(66) \\
26.71 & 0.170(12)  & 47.75  & 0.1866(88)  &  &  &  & \\ \hline
\end{tabular}
\caption{Multiplicative renormalisation factor, $Z_Q(\beta)$, relating 
  the average lattice topological charge, $Q_{hot}$,  calculated on the rough
  Monte Carlo fields, and the integer valued topological charge $Q_I$ calculated
  after 20 `cooling' sweeps of those fields. For the gauge groups and $\beta$ shown.}
\label{table_ZQB}
\end{table}

\begin{table}[htb]
\centering
\begin{tabular}{|c|c|c|c|} \hline
\multicolumn{4}{|c|}{$Z^{int}_Q = 1 - z_0 g^2N - z_1 (g^2N)^2$} \\ \hline
$N$ & $z_0$ & $z_1$ & $\chi^2/n_{df}$ \\ \hline
 2 & 0.190(30) & 0.023(9)   & 1.17  \\
 3 & 0.162(10) & 0.0425(31) & 0.62  \\
 4 & 0.156(20) & 0.047(7)   & 1.32  \\
 5 & 0.203(21) & 0.035(7)   & 2.76  \\
 6 & 0.205(30) & 0.036(11)  & 1.37  \\
 8  & 0.187(24) & 0.043(9)   & 1.71   \\
 10 & 0.141(44) & 0.060(16)  & 1.05 \\
 12 & 0.182(24) & 0.071(22)  & 2.22 \\ \hline
\end{tabular}
\caption{Interpolating funtions for the multiplicative renormalisation factor, $Z_Q(\beta)$,
  for our $SU(N)$ calculations, with $g^2N=2N^2/\beta$.}
\label{table_ZQint}
\end{table}

\clearpage


\clearpage


\begin{figure}[htb]
\begin	{center}
\leavevmode
\input	{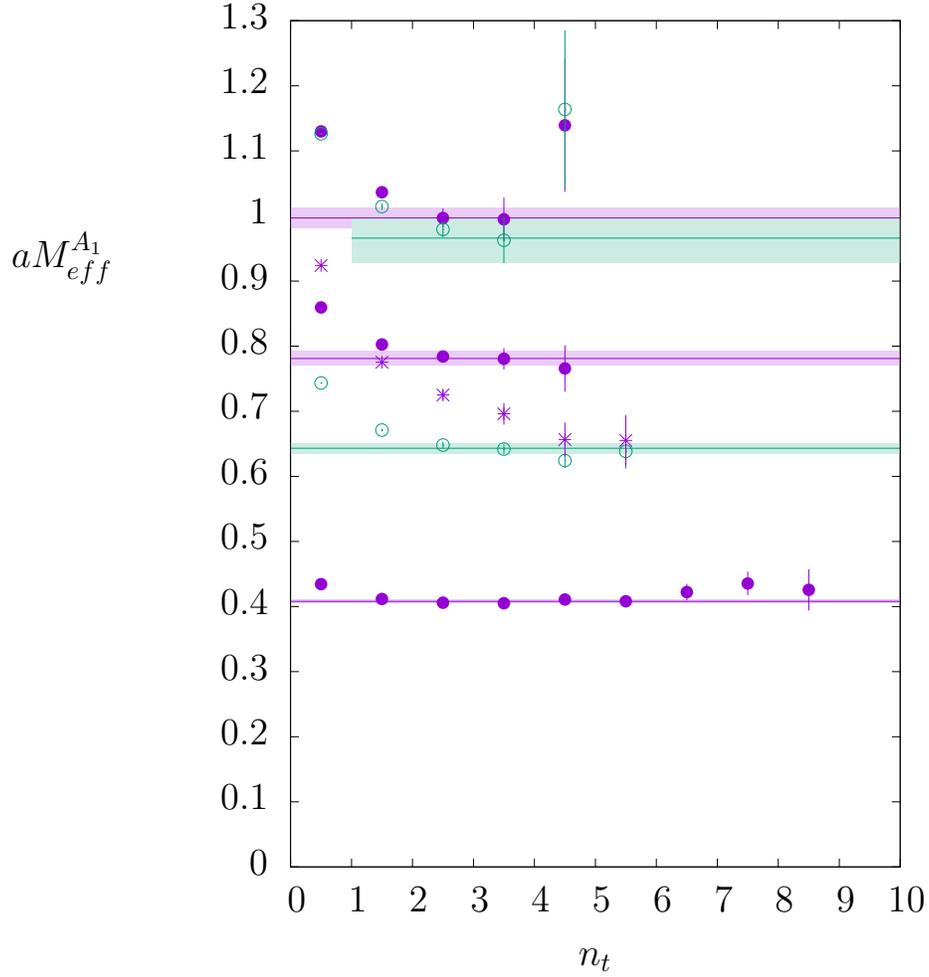}
\end	{center}
\caption{Effective masses for the lightest three $A_1^{++}$ ($\bullet$) and the lightest
  two $A_1^{-+}$ ($\circ$) glueball states, as well as the main $A_1^{++}$ ditorelon state ($\ast$).
  Lines are our best glueball mass estimates, with bands corresponding to $\pm 1$ standard deviations.
  All on a $20^330$ lattice at $\beta=47.75$ in $SU(8)$.
  In the continuum limit the lightest two glueball states in each sector become the lightest
  two $J^{PC}=0^{++}$ and $J^{PC}=0^{-+}$ glueballs. The ditorelon disappears in the thermodynamic
  limit.}
\label{fig_MeffA1_SU8}
\end{figure}

\begin{figure}[htb]
\begin	{center}
\leavevmode
\input	{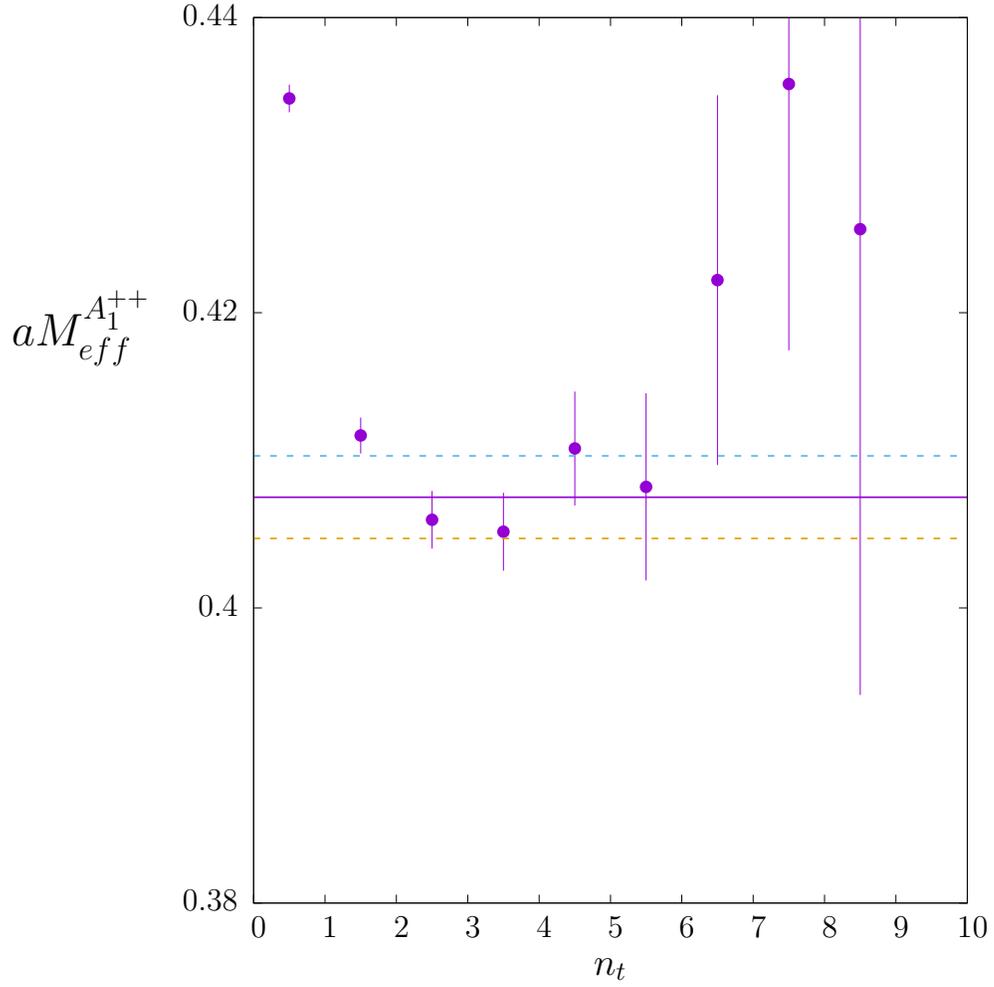}
\end	{center}
\caption{Effective mass plot for the ground state $A_1^{++}$ ($\bullet$) 
  on a $20^330$ lattice at $\beta=47.75$ in $SU(8)$, as in Fig.\ref{fig_MeffA1_SU8},
  but rescaled so as to expose the errors on the effective masses. The straight line
  is the best estimate for the mass obtained by fitting the correlation function, and
  the two dashed lines bound the $\pm 1$ standard deviation error band on this mass.}
\label{fig_MeffA1b_SU8}
\end{figure}

\begin{figure}[htb]
\begin	{center}
\leavevmode
\input	{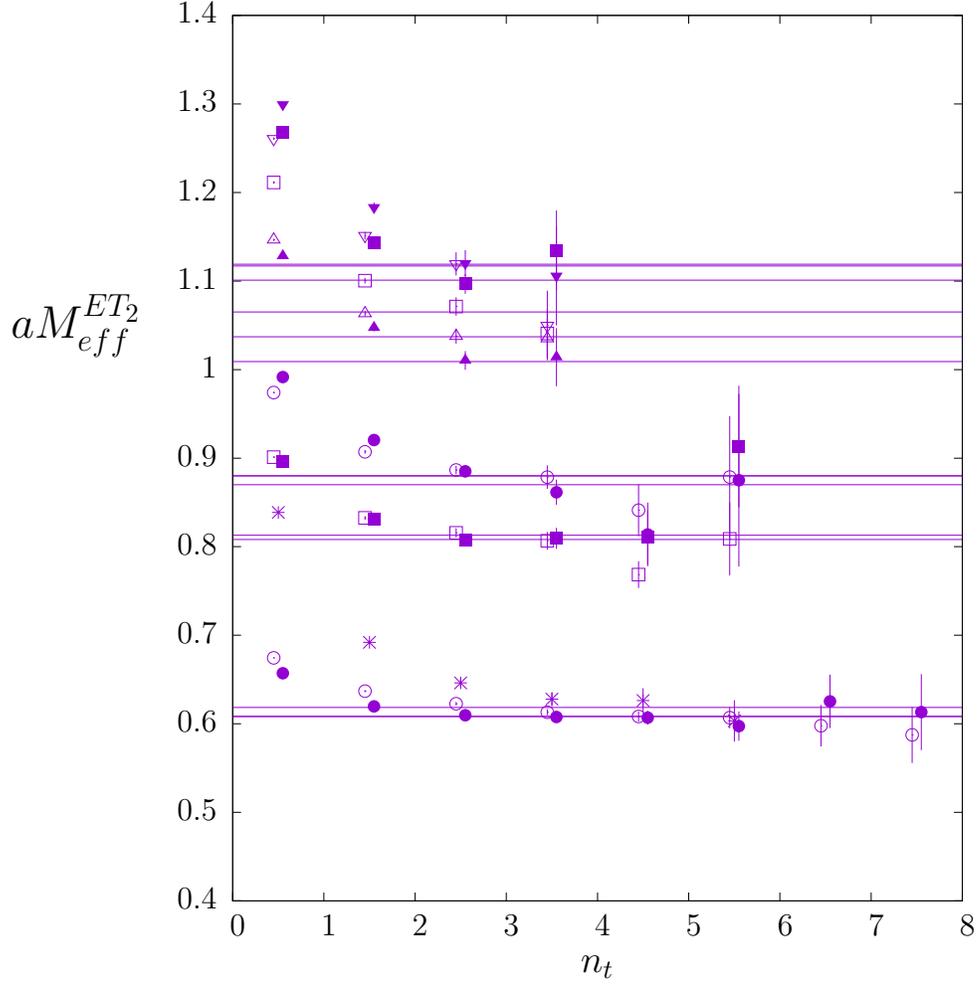}
\end	{center}
\caption{Effective masses for the lightest two $E^{++}$ ($\bullet$) and $T_2^{++}$ ($\circ$) 
  glueballs; the lightest two  $E^{-+}$ ($\blacksquare$) and $T_2^{-+}$ ($\square$) glueballs;
  the lightest $E^{+-}$ ($\blacktriangledown$) and $T_2^{+-}$ ($\triangledown$) glueballs;
  and the lightest $E^{--}$ ($\blacktriangle$) and $T_2^{--}$ ($\vartriangle$) glueballs.
  The state labelled by $\ast$ is, mainly, the $E^{++}$ ditorelon.
  Lines are mass estimates. All on a $20^330$ lattice at $\beta=47.75$ in $SU(8)$.
  In the continuum limit each of the $E$ doublets and corresponding $T_2$ triplets will
  pair up to give the five states of a $J=2$ glueball.}
\label{fig_MeffET2_SU8}
\end{figure}



\begin{figure}[htb]
\begin	{center}
\leavevmode
\input	{plot_MJ02ppK_cont_SU4.tex}
\end	{center}
\caption{Lightest two glueball masses in the $A_1^{++}$ ($\bullet$), $E^{++}$ ($\blacklozenge$)
  and $T_2^{++}$ ($\lozenge$) sectors, in units of the string tension. Lines are linear
  extrapolations to the continuum limit. In that limit the  $A_1^{++}$ states become the
  lightest two $J^{PC}=0^{++}$ scalar glueballs while the doublet $E^{++}$ and triplet $T_2^{++}$
  pair up to give the five components of each of the lightest two $J^{PC}=2^{++}$ glueballs.
  All in $SU(4)$.}
\label{fig_MJ02ppK_cont_SU4}
\end{figure}

\begin{figure}[htb]
\begin	{center}
\leavevmode
\input	{plot_MJ02mpK_cont_SU4.tex}
\end	{center}
\caption{Lightest two glueball masses in the $A_1^{-+}$ ($\bullet$), $E^{-+}$ ($\blacklozenge$)
  and $T_2^{-+}$ ($\lozenge$) sectors, in units of the string tension. Lines are linear
  extrapolations to the continuum limit. In that limit the  $A_1^{-+}$ states become the
  lightest two $J^{PC}=0^{-+}$ pseudoscalar glueballs while the doublet $E^{--+}$ and triplet $T_2^{-+}$
  pair up to give the five components of each of the lightest two $J^{PC}=2^{-+}$ glueballs.
  All in $SU(4)$.}
\label{fig_MJ02mpK_cont_SU4}
\end{figure}

\begin{figure}[htb]
\begin	{center}
\leavevmode
\input	{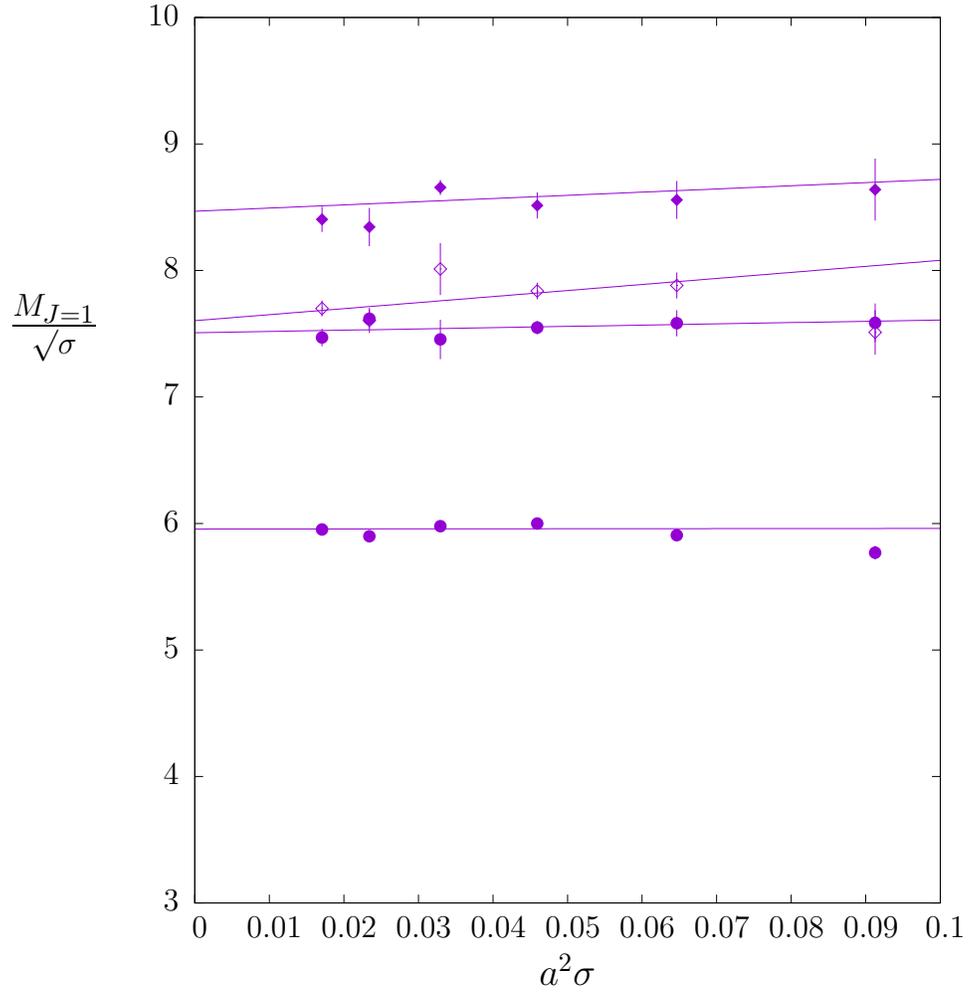}
\end	{center}
\caption{Lightest two glueball masses in the $T_1^{+-}$ ($\bullet$) representation
  and the lightest ones in the $T_1^{-+}$ ($\blacklozenge$)
  and $T_1^{--}$ ($\lozenge$) representations, in units of the string tension. Lines are linear
  extrapolations to the continuum limit. In that limit the  $T_1^{+-}$ states become the
  lightest two $J^{PC}=1^{+-}$ glueballs while the other two becomes the
  $1^{-+}$ and $1^{--}$ ground state glueballs.  All in $SU(4)$.}
\label{fig_MJ1K_cont_SU4}
\end{figure}


\begin{figure}[htb]
\begin	{center}
\leavevmode
\input	{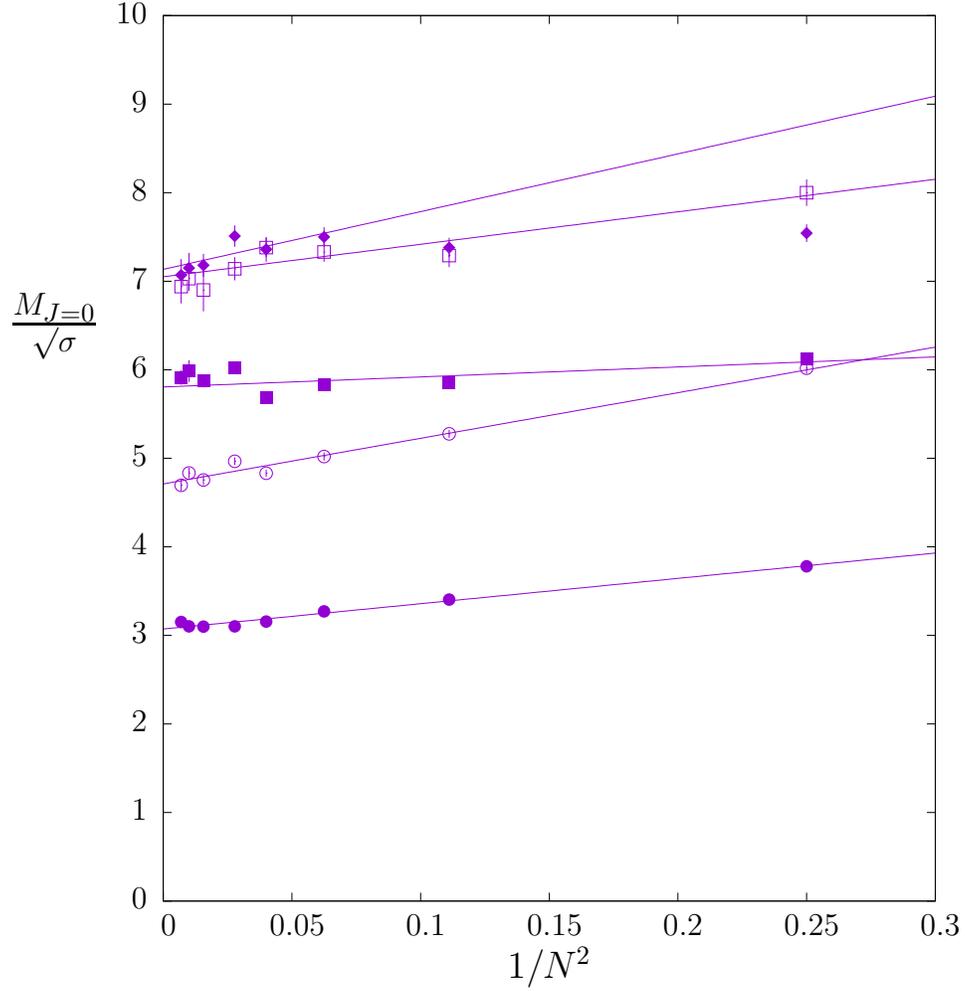}
\end	{center}
\caption{Continuum masses of the lightest ($\bullet$) and first excited ($\blacksquare$)
  $J^{PC}=0^{++}$ scalars and of the lightest ($\circ$) and first excited ($\square$)
  $0^{-+}$ pseudoscalars, in units of the string tension. The state denoted by
  $\blacklozenge$ is either the $4^{++}$ ground state or the second excited $0^{++}$.
  With extrapolations from values in the range $N\in[2,12]$ to $N=\infty$.}
\label{fig_M0pp0mpK_N}
\end{figure}

\begin{figure}[htb]
\begin	{center}
\leavevmode
\input	{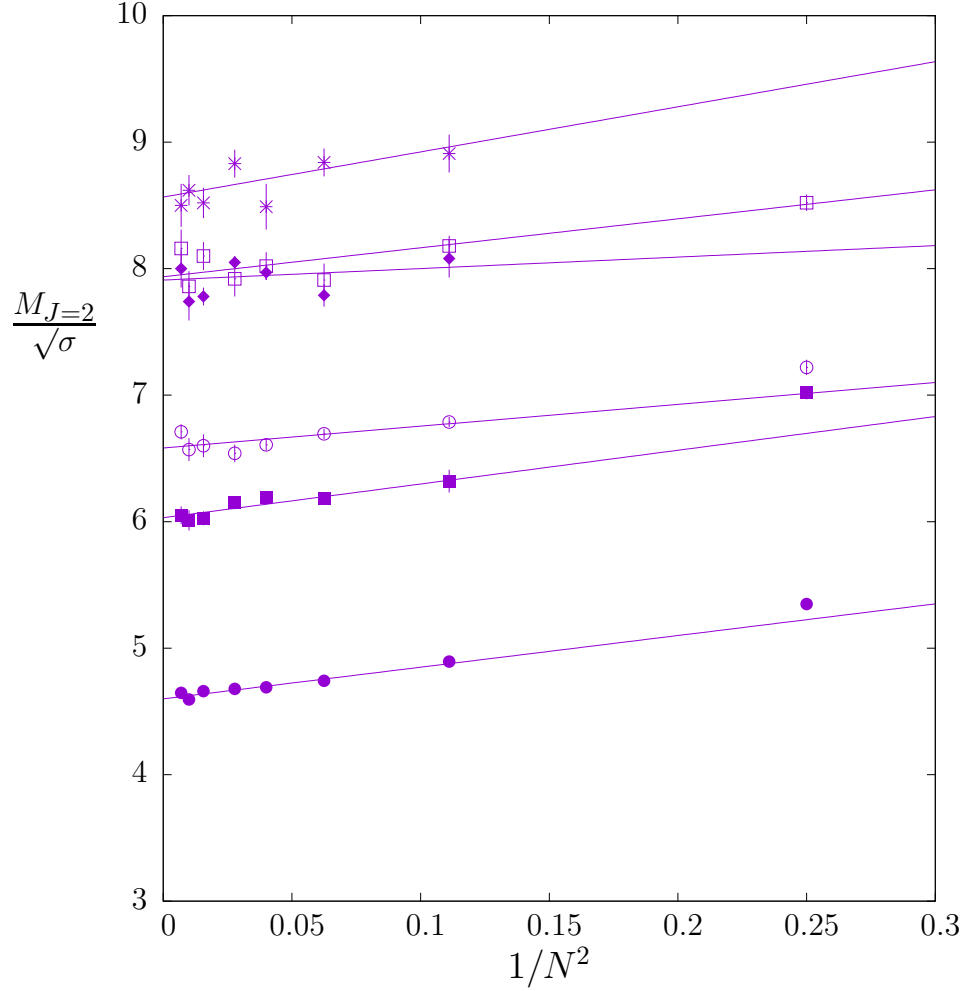}
\end	{center}
\caption{Continuum masses of the lightest ($\bullet$) and first excited ($\circ$)
  $J^{PC}=2^{++}$ tensors, the lightest ($\blacksquare$) and first excited ($\square$)
  $2^{-+}$ pseudotensors, the lightest $2^{+-}$ ($\ast$), and the lightest
  $2^{--}$ ($\blacklozenge$), all in units of the string tension.
  With extrapolations to $N=\infty$ from $N\leq 12$.}
\label{fig_MJ2PCK_N}
\end{figure}

\begin{figure}[htb]
\begin	{center}
\leavevmode
\input	{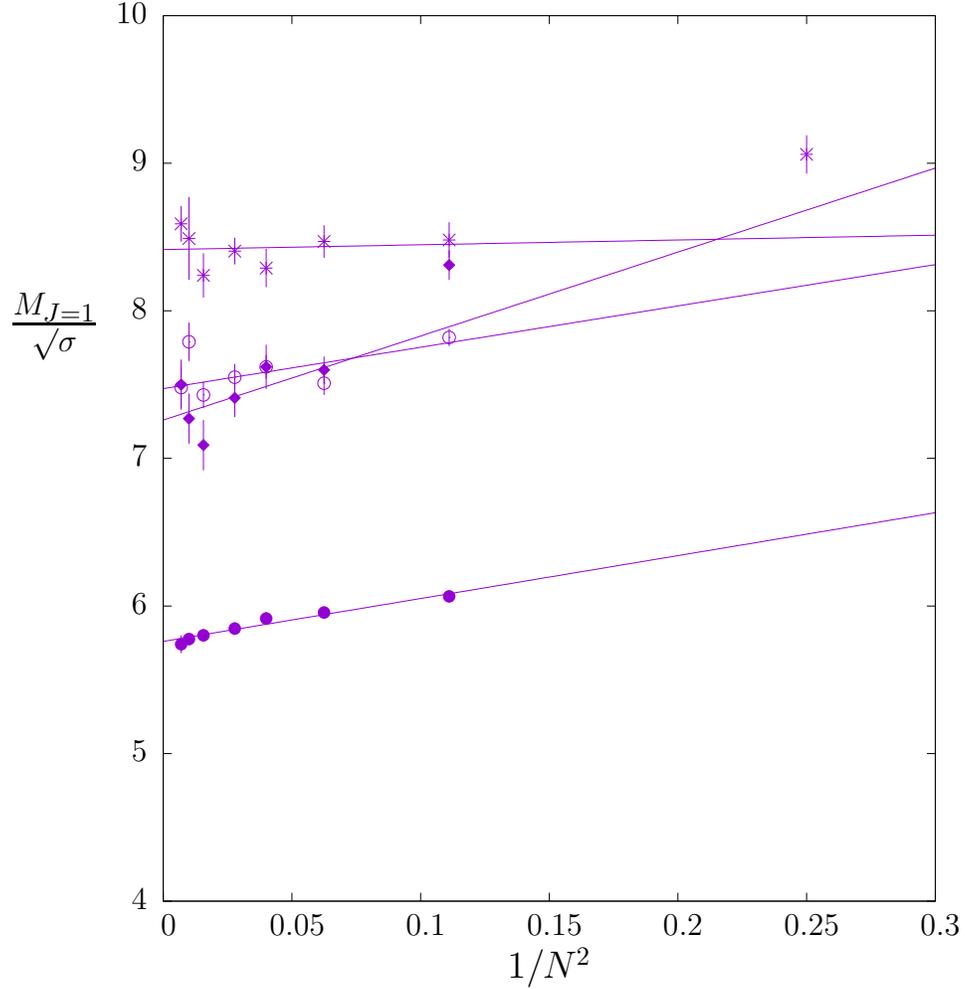}
\end	{center}
\caption{Continuum masses of the lightest ($\bullet$) and first excited ($\circ$) $J^{PC}=1^{+-}$
  glueballs, as well as the lightest $1^{-+}$ ($\ast$) and $1^{--}$ ($\blacklozenge$)
  glueballs. in units of the string tension, with extrapolations to $N=\infty$.}
\label{fig_MJ1K_N}
\end{figure}


\begin{figure}[htb]
\begin	{center}
\leavevmode
\input	{plot_MeffG+GGA1++l26n_SU3.tex}
\end	{center}
\caption{Effective masses for the lightest few glueballs in the `scalar' $A_1^{++}$
  representation, for the single trace operators ($\circ$) and for the same set
  augmented with double trace operators (filled points), with points shifted for clarity.
  The extra `scattering' state amongst the latter is shown as $\blacklozenge$. 
  Horizontal line indicates twice the mass of the lightest glueball.
  On the $26^326$ lattice at $\beta=6.235$ in SU(3).}
\label{fig_MeffG+GGA1++l26n_SU3}
\end{figure}

\begin{figure}[htb]
\begin	{center}
\leavevmode
\input	{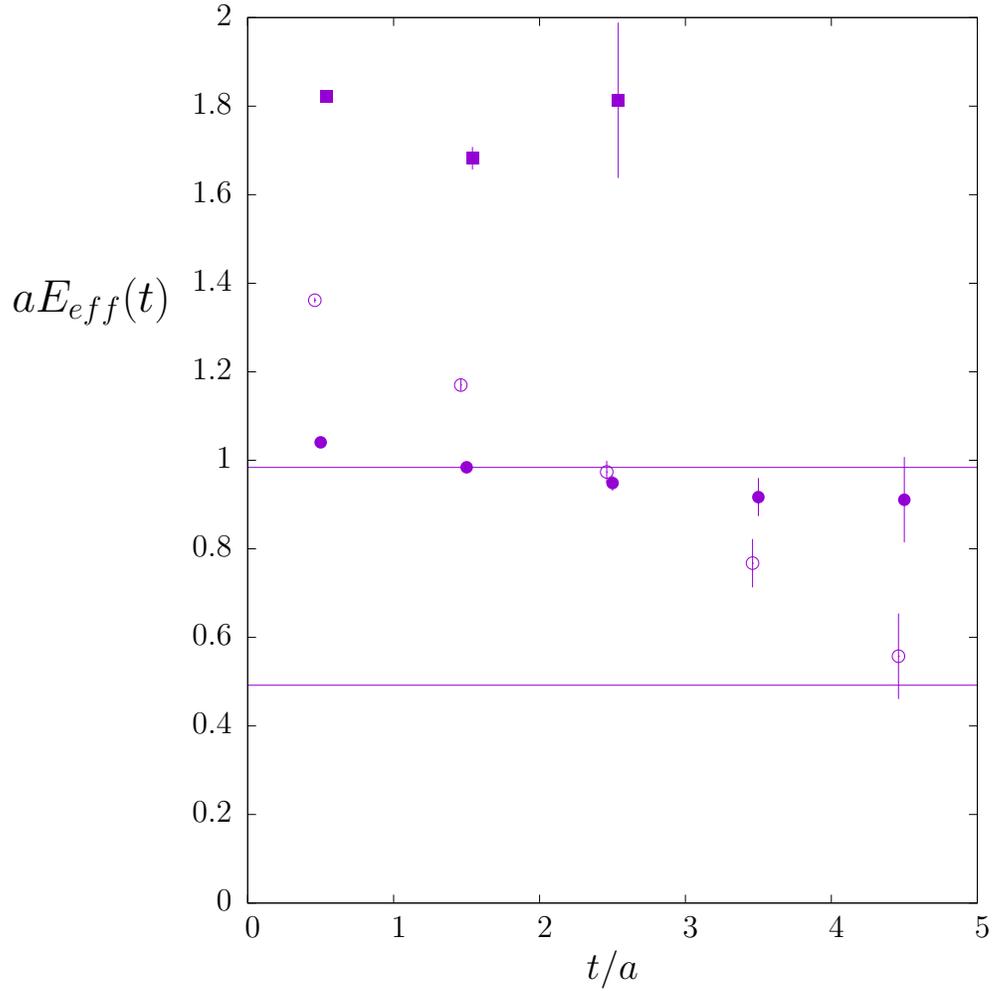}
\end	{center}
\caption{Effective masses for the lightest three states
  in the `scalar' $A_1^{++}$ representation, for the double trace operators.
  Lower horizontal line indicates the mass of the lightest glueball, and
  upper  horizontal line indicates twice the mass of the lightest glueball.
  On the $26^326$ lattice at $\beta=6.235$ in SU(3).}
\label{fig_MeffGGA1++l26n_SU3}
\end{figure}

\clearpage

\begin{figure}[htb]
\begin	{center}
\leavevmode
\input	{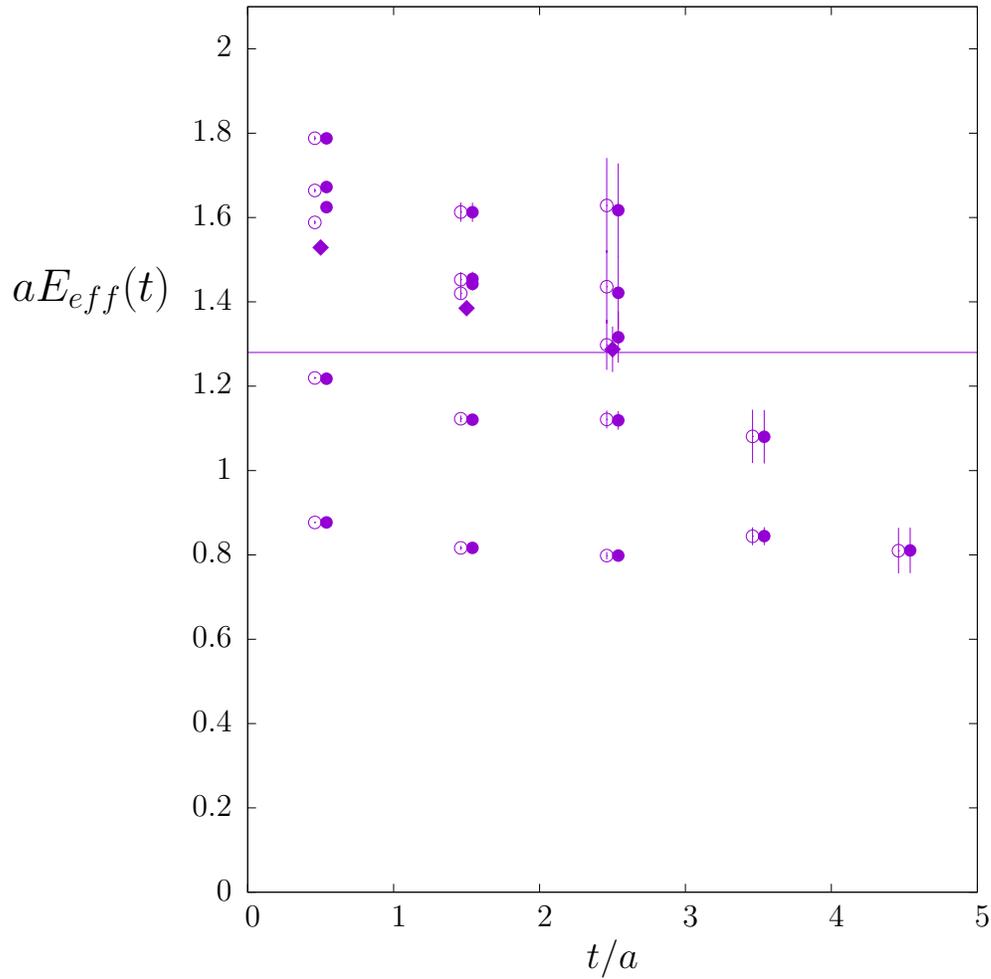}
\end	{center}
\caption{Effective masses for the lightest few glueballs in the `pseudoscalar' $A_1^{-+}$
  representation, for the single trace operators ($\circ$) and for the same set
  augmented with double trace operators (filled points), with points shifted for clarity.
  The likely extra `scattering' state amongst the latter is shown as $\blacklozenge$. 
  Horizontal line indicates the sum of the lightest $A_1^{++}$ and $A_1^{-+}$ glueball masses.
  On the $26^326$ lattice at $\beta=6.235$ in SU(3).}
\label{fig_MeffG+GGA1-+l26n_SU3}
\end{figure}

\begin{figure}[htb]
\begin	{center}
\leavevmode
\input	{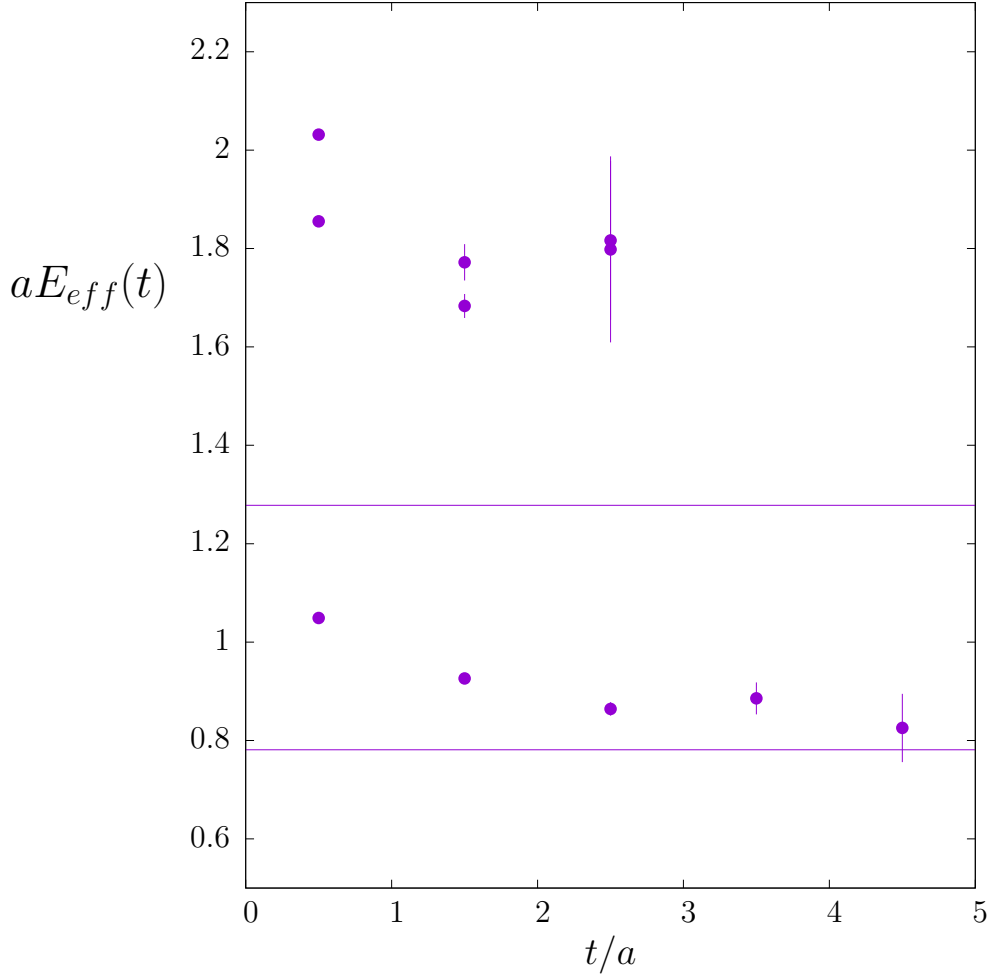}
\end	{center}
\caption{Effective masses for the lightest three states
  in the `pseudoscalar' $A_1^{-+}$ representation, for the double trace operators.
  Lower horizontal line indicates the mass of the lightest $A_1^{-+}$ glueball, and
  upper  horizontal line indicates the sum of the masses of the lightest
  $A_1^{++}$ and $A_1^{-+}$ glueballs.
  On the $26^326$ lattice at $\beta=6.235$ in SU(3).}
\label{fig_MeffGGA1-+l26n_SU3}
\end{figure}



\clearpage

\begin{figure}[htb]
\begin	{center}
\leavevmode
\input	{plot_Qcool_su5b17.63.tex}
\end	{center}
\caption{The number of lattice fields with topological charge $Q_L$ after 2 ($\circ$) and after 20 ($\bullet$)
cooling sweeps, from sequences of $SU(5)$ fields generated at $\beta=17.63$. $N(Q_L)=0$ points suppressed.}
\label{fig_Qcool20_su5}
\end{figure}

\begin{figure}[htb]
\begin	{center}
\leavevmode
\input	{plot_Qcool_su8b47.75.tex}
\end	{center}
\caption{The number of lattice fields with topological charge $Q_L$ after 2 ($\circ$) and after 20 ($\bullet$)
cooling sweeps, from sequences of $SU(8)$ fields generated at $\beta=47.75$. $N(Q_L)=0$ points suppressed.}
\label{fig_Qcool20_su8}
\end{figure}

\begin{figure}[htb]
\begin	{center}
\leavevmode
\input	{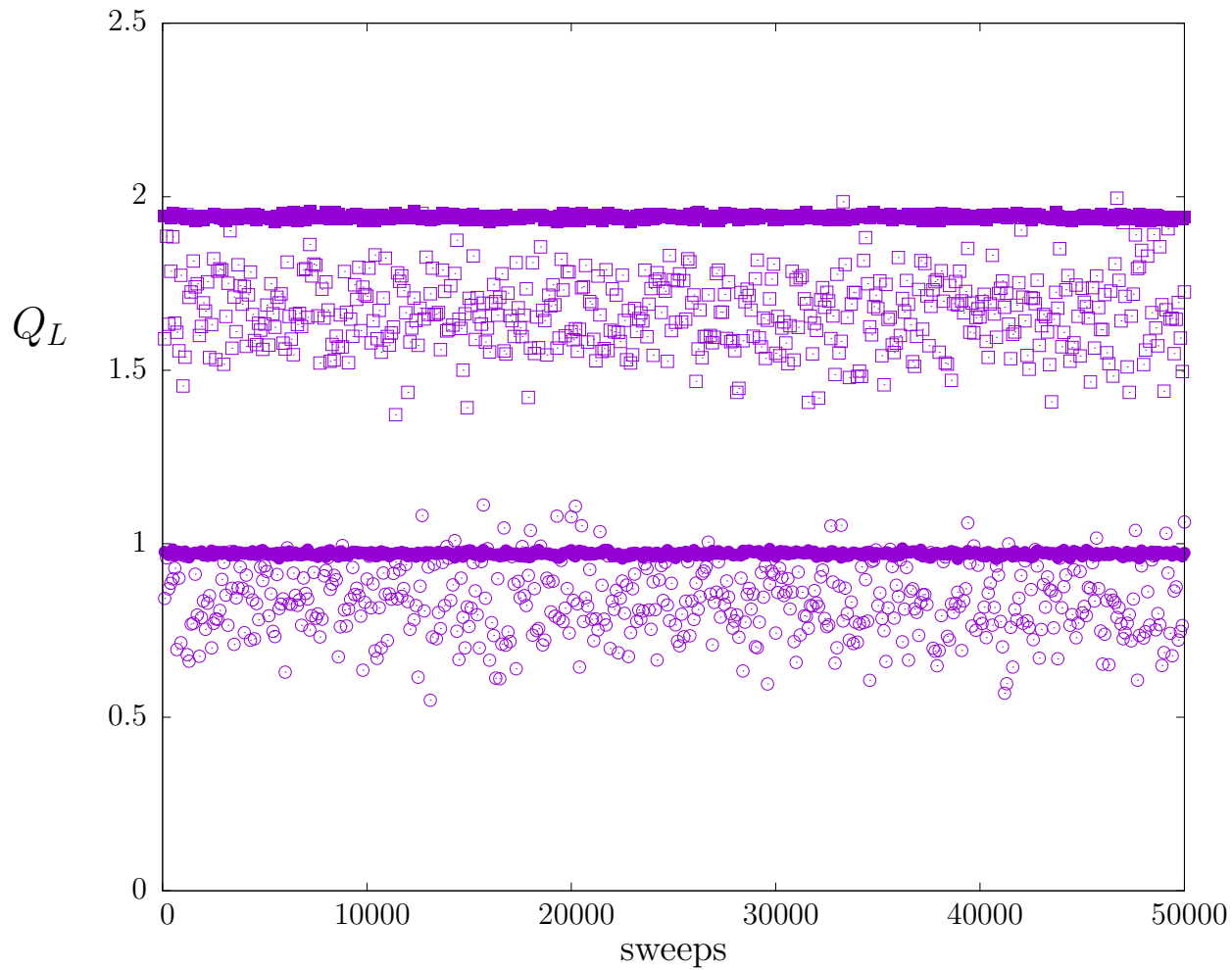}
\end	{center}
\caption{The lattice topological charge $Q_L$ for two sequences of $SU(8)$ lattice fields,
  calculated after 2 ($\circ,\square$) and 20 ($\bullet,\blacksquare$) cooling sweeps.
  Calculations of $Q_L$ made every 100 Monte Carlo sweeps for each sequence of 50000 sweeps,
  at $\beta=47.75$ on a $20^330$ lattice.}
\label{fig_Qseq_su8}
\end{figure}

\begin{figure}[htb]
\begin	{center}
\leavevmode
\input	{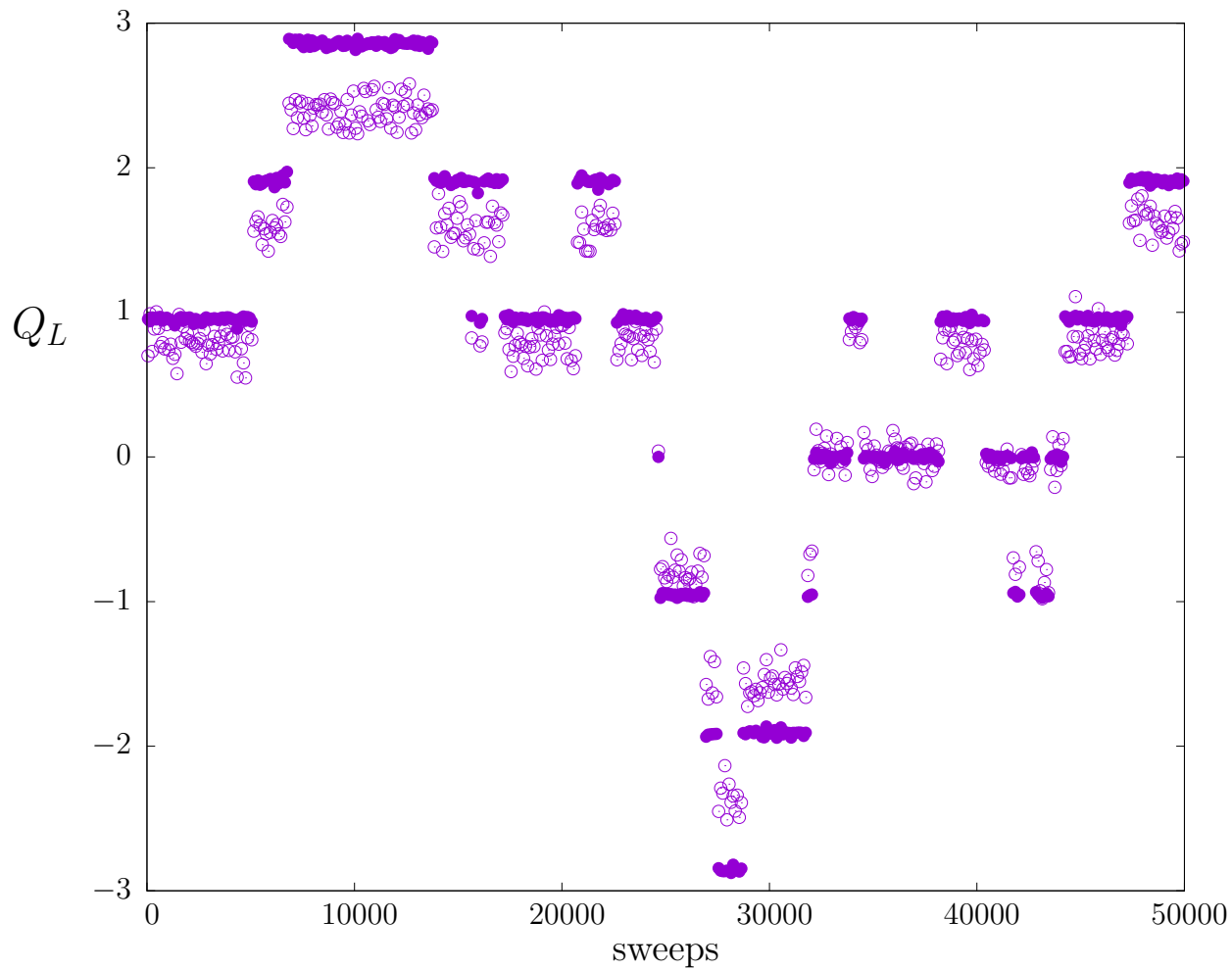}
\end	{center}
\caption{The lattice topological charge $Q_L$ for a sequence of $SU(5)$ lattice fields after
2 ($\circ$) and 20 ($\bullet$) cooling sweeps. Calculations of $Q_L$ every 100 Monte Carlo sweeps
over a sequence of 50000 sweeps, at $\beta=17.63$ on a $16^320$ lattice.}
\label{fig_Qseq_su5}
\end{figure}

\begin{figure}[htb]
\begin	{center}
\leavevmode
\input	{plot_Qcool20c2_su8b47.75.tex}
\end	{center}
\caption{The distribution in the topological charge $Q_L$ as obtained after 2 cooling sweeps,
for fields that after 20 cooling sweeps have topological charges $Q=0$ ($\circ$),
$Q=1$ ($\blacksquare$) and $Q=2$ ($\square$). $N(Q_L)=0$ points suppressed. 
From the same sequences of $SU(8)$ fields generated 
at $\beta=47.75$ plotted in Fig.\ref{fig_Qcool20_su8}.} 
\label{fig_Qcool20c2_su8}
\end{figure}

\begin{figure}[htb]
\begin	{center}
\leavevmode
\input	{plot_Qcool20c1_su8b47.75.tex}
\end	{center}
\caption{The distribution in the topological charge $Q_L$ as obtained after only 1 cooling sweep,
for fields that after 20 cooling sweeps have topological charges $Q=0$ ($\circ$),
$Q=1$ ($\blacksquare$) and $Q=2$ ($\square$).  $N(Q_L)=0$ points suppressed. 
From the same sequence of $SU(8)$ fields generated 
at $\beta=47.75$ plotted in Fig.\ref{fig_Qcool20_su8}.} 
\label{fig_Qcool20c1_su8}
\end{figure}

\begin{figure}[htb]
\begin	{center}
\leavevmode
\input	{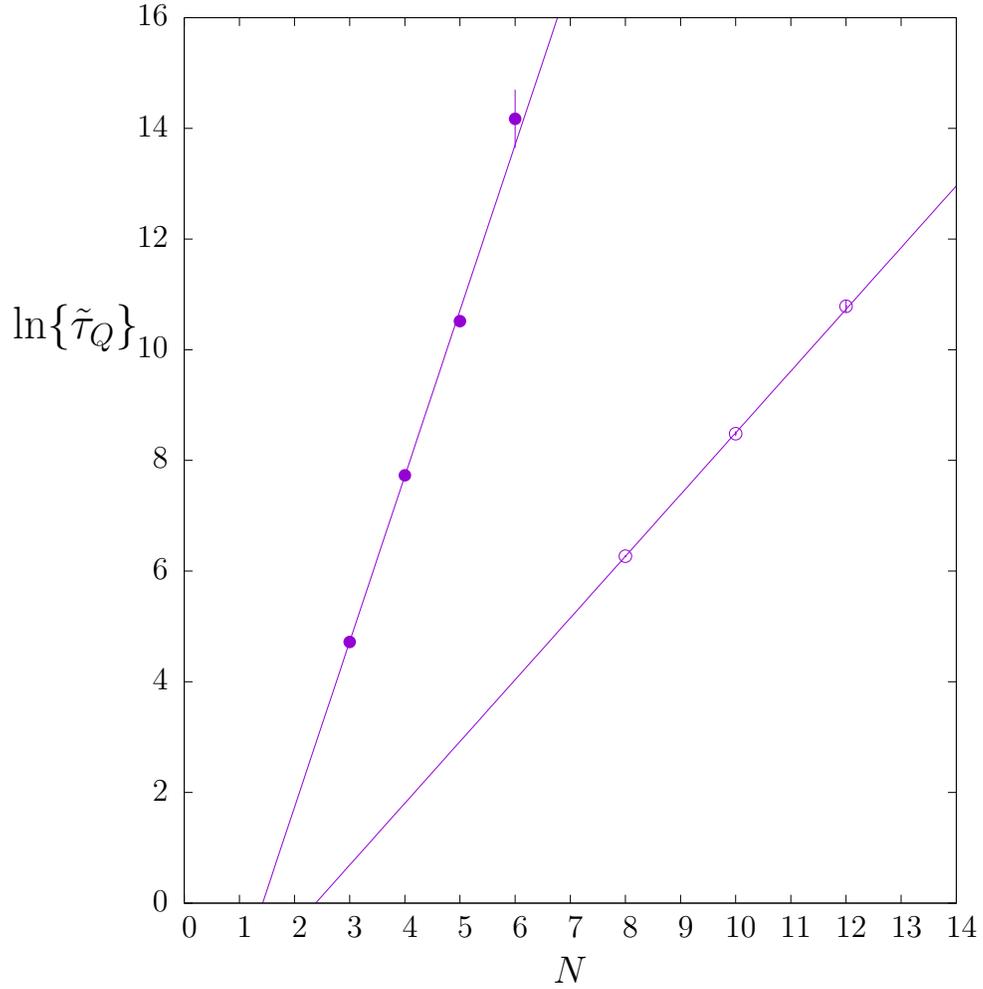}
\end	{center}
\caption{Correlation length $\Tilde{\tau}_Q$, the average number of sweeps between changes of $Q$ by $\pm 1$,
  measured for the $SU(N)$ lattice topological charge
  for $a\surd\sigma \sim 0.15$ ($\bullet$) and for $a\surd\sigma \sim 0.33$ ($\circ$).
  Lines are fits $\Tilde{\tau}_Q = b\exp\{cN\}$.}
\label{fig_tauQ_suN}
\end{figure}

\begin{figure}[htb]
\begin	{center}
\leavevmode
\input	{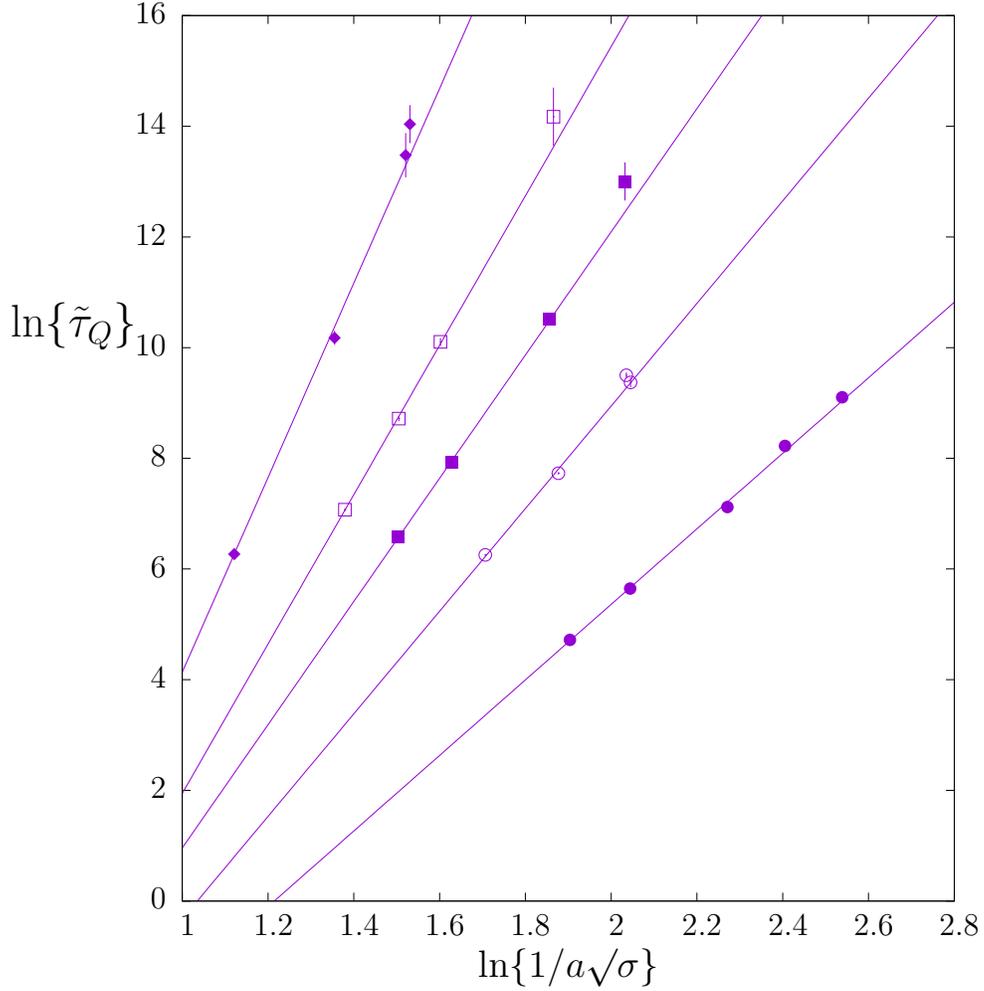}
\end	{center}
\caption{Variation of $\Tilde{\tau}_Q$, the average number of sweeps between changes of $Q$ by $\pm 1$,
  against the lattice spacing and normalised to our standard space-time volume $V=(3/\surd\sigma)^4$.
  For $SU(3)$ ($\bullet$), $SU(4)$ ($\circ$), $SU(5)$ ($\blacksquare$),
  $SU(6)$ ($\square$), $SU(8)$ ($\blacklozenge$). The two pairs of points nearly overlapping
  are obtained at same $a$ but from different volumes. Lines are fits
  $\Tilde{\tau}_Q = b\{1/a\surd\sigma\}^c$.}
\label{fig_tauQ_KsuN}
\end{figure}

\begin{figure}[htb]
\begin	{center}
\leavevmode
\input	{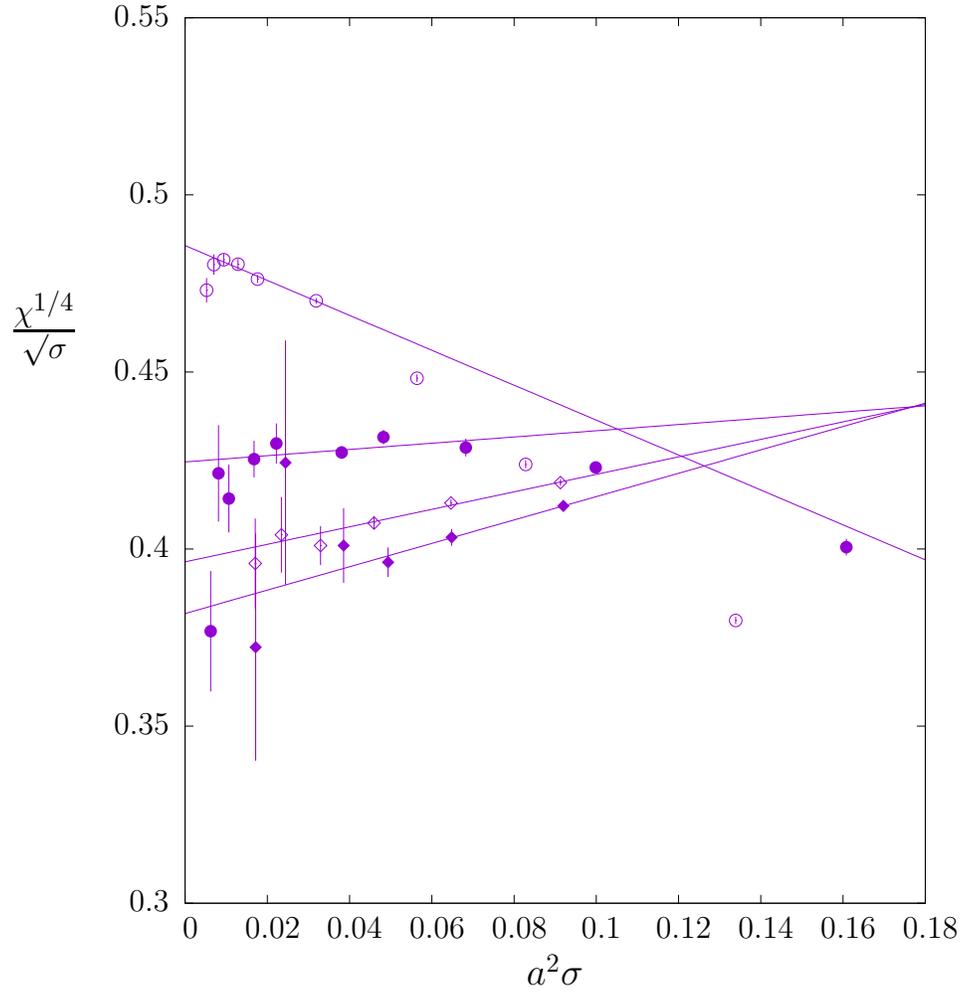}
\end	{center}
\caption{Continuum extrapolations of the topological susceptibility in units of the string tension for
  continuum $SU(2)$, $\circ$, $SU(3)$, $\bullet$, $SU(4)$, $\lozenge$, and $SU(5)$, $\blacklozenge$,
  gauge theories.}
\label{fig_khiIK_cont}
\end{figure}

\begin{figure}[htb]
\begin	{center}
\leavevmode
\input	{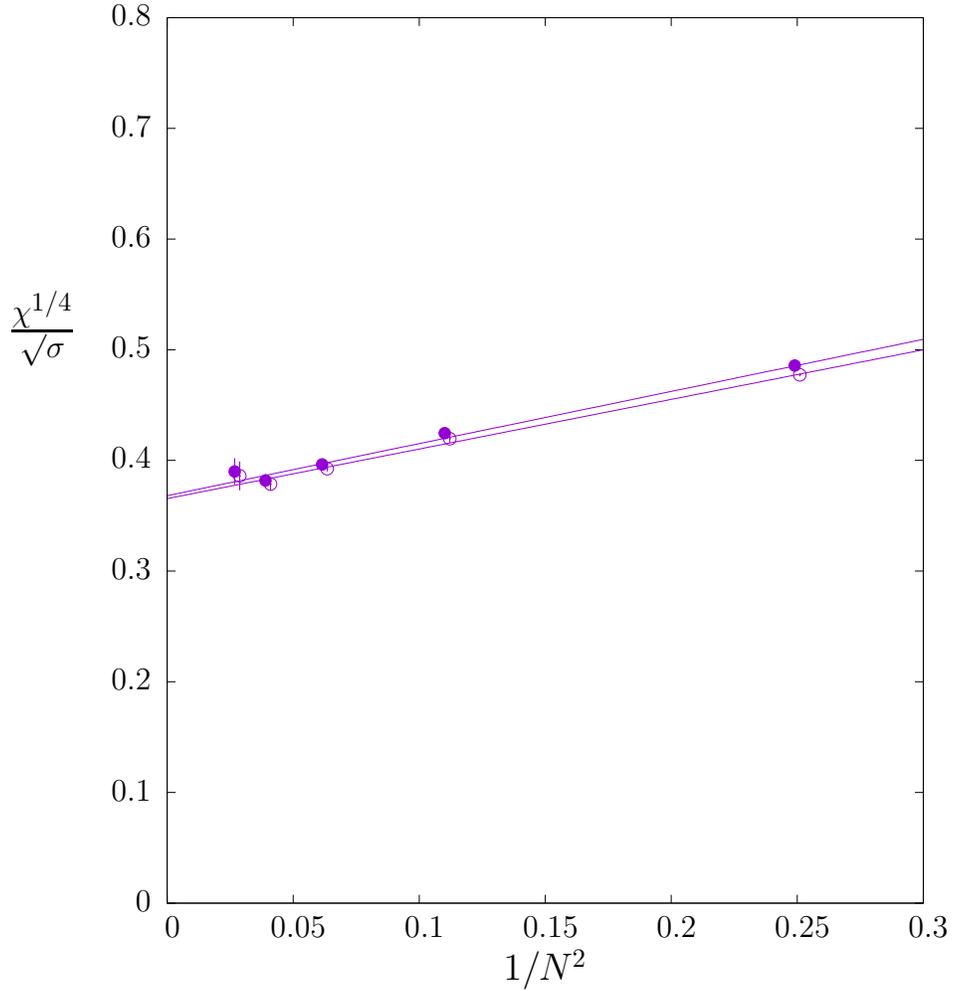}
\end	{center}
\caption{Topological susceptibility in units of the string tension for
  continuum $SU(N)$ gauge theories with $N=2,3,4,5,6$. For integer valued
  charge, $\bullet$, and for non-integer lattice charge, $\circ$. Points
  shifted slightly for clarity. Lines are extrapolations to $N=\infty$. }
\label{fig_khiK_N}
\end{figure}

\begin{figure}[htb]
\begin	{center}
\leavevmode
\input	{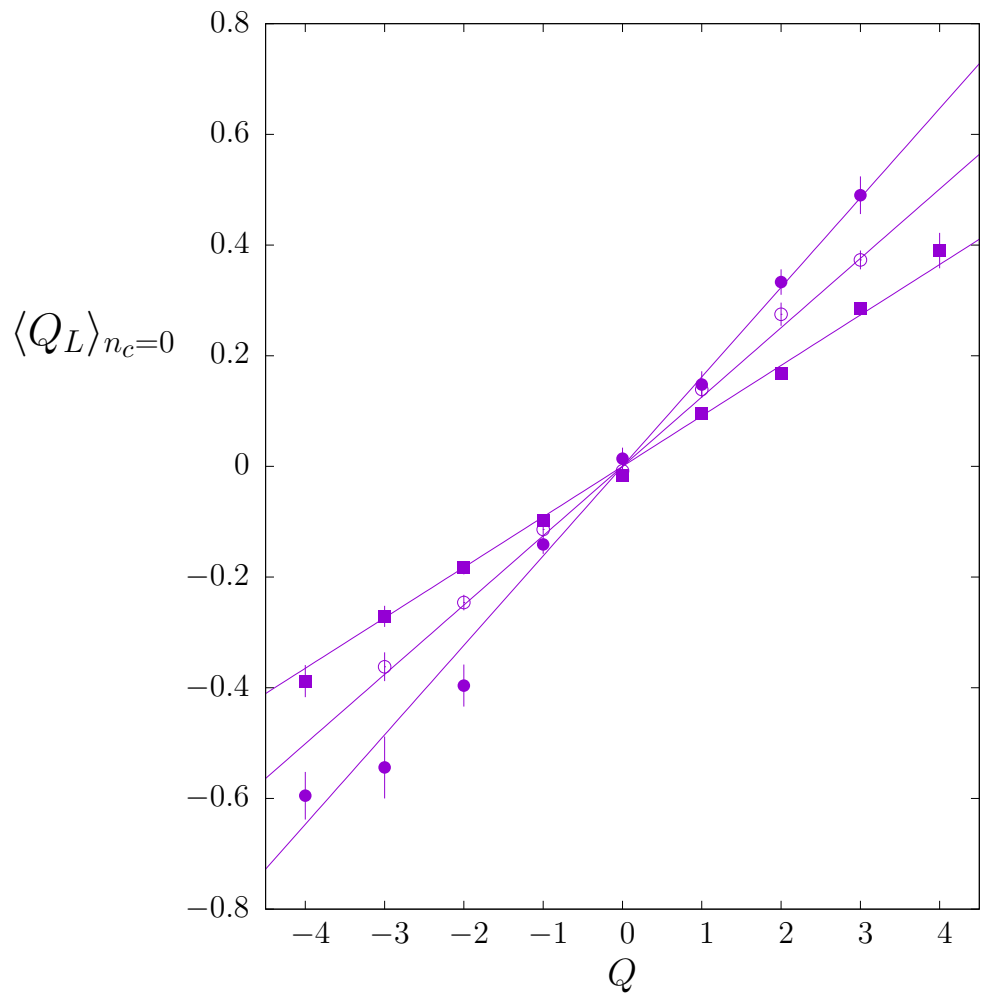}
\end	{center}
\caption{Average value of topological charge on lattice fields that
  have charge $Q$ after 20 cooling sweeps. In $SU(8)$ at $\beta =44.10$, $\blacksquare$,
  $\beta =45.50$, $\circ$, and $\beta =46.70$, $\bullet$. Slope of fits gives the
  renormalisation factor $Z_Q(\beta)$.} 
\label{fig_ZQ_su8}
\end{figure}

\end{document}